\documentstyle[11pt]{article}

\date{October 13, 1999}

\long\def\comment#1{\ifdim\overfullrule>0pt{\sf[{#1}]}\fi}
\long\def\micsays#1{\comment{#1---M}}

\long\def\zzsaid#1{} \long\def\micsaid#1{}

\let\originallabel\label
\def\label#1{\originallabel{#1}\comment{#1}}

\newenvironment{proof}{\noindent
{\bf Proof:} \rm}{\mbox{} \hfill $\Box$\par\vspace{1ex}}

\def\case#1{\smallskip\par\noindent{\em Case #1:~}}

\newtheorem{theorem}{Theorem}[section]
\newtheorem{lemma}[theorem]{Lemma} 
\newtheorem{corollary}[theorem]{Corollary} 
\newtheorem{claim}[theorem]{Claim}
\newtheorem{fact}[theorem]{Fact}
\newtheorem{definition}[theorem]{Definition}

\newtheorem{assume}{Assumption}

\makeatletter
\global\def\thefigure{\thesection.\@arabic\c@figure}
\@addtoreset{figure}{section}
\makeatother

\input{epsf}  \let\figbox\epsfbox		

\def\fig[#1]#2#3{\begin{figure}[htp]
	\centerline{\figbox{#1.ps}}
	\caption{\comment{fig:#2} #3}
	\originallabel{fig:#2}
	\end{figure}}

\title{Map Graphs\thanks
	{A preliminary version appeared with the title
	 ``Planar Map Graphs''~\protect\cite{CGP98}.
	}}

\author{Zhi-Zhong Chen\thanks
        {Department of Mathematical Sciences, Tokyo Denki University, Hatoyama,
        Saitama 350-0394, JAPAN.
        Part of work done while visiting University of California at Berkeley.
        E-mail: {\tt  chen@r.dendai.ac.jp}.}
        \and
        Michelangelo Grigni\thanks
        {Department of Mathematics and Computer Science, Emory University,
	Atlanta, GA 30322. 
        E-mail: {\tt mic@mathcs.emory.edu}.}
        \and
        Christos H. Papadimitriou\thanks
        {Computer Science Division, University of California at Berkeley,
	Berkeley, CA 94720. Supported by NSF Grant number CCR-9626361.
        E-mail: {\tt christos@cs.berkeley.edu}.}
	}

\begin{document}
\maketitle

\micsaid{pass22: Changed ``fact'', ``claim'', and ``definition''
to follow same numbering scheme as ``theorem''; I think this makes things
easier to find (also it is the style of some journals, for example SIAM).
Note that ``assumption'' is still exceptional.}

\begin{abstract}
  We consider a modified notion of planarity, in which two nations of a
map are considered adjacent when they share any {\em point} of their
boundaries (not necessarily an {\em edge}, as planarity requires).
  Such adjacencies define a {\em map graph}.
  We give an NP characterization for such graphs, and an $O(n^3)$-time 
recognition algorithm for a restricted version: given a graph,
decide whether it is realized by adjacencies in a map without holes,
in which at most four nations meet at any point.
\end{abstract}

\section{Introduction}\label{sec:intro}

\subsection{Motivation and Definition}

  Suppose you are told that there are four planar regions, and
for each pair you are told their topological relation:
$A$ is inside $B$, $B$ overlaps $C$, $C$ touches $D$ on
the outside, $D$ overlaps $B$, $D$ is disjoint from $A$, and $C$
overlaps $A$.
\micsaid{pass18: Merged ``each pair'' sentence into the first.}
  All four planar regions are ``bubbles'' with no holes; more
precisely, they are closed disc homeomorphs.
  Are such regions possible?
  If so, we would like a model, a picture of four
regions so related; if not, a proof of impossibility.

This extension of propositional logic is known as the {\em topological
inference problem}~\cite{GPP95}.
  No decision algorithm or finite axiomatization is known, although
the problem becomes both finitely axiomatizable and polynomial-time
decidable in any number of dimensions other than two~\cite{A83,R98},
for some reasonable vocabularies of topological relations.
\micsaid {pass30: was ``a reasonably chosen vocabulary'' above.}
\micsaid {pass18: Renz only addresses the vocabulary RCC-8, which is
the same as the ``high-resolution'' vocabulary in~\cite{GPP95}.  The
argument also applies to other vocabularies such as those defined
in~\cite{GPP95}.  But I cannot find references for such results, so I just
added the extra clause at the end of the last sentence.}
  In fact, the following special case has been open since the 
1960's~\cite {EET76}: for every pair of regions, we are only told
whether the regions intersect or not.
  This is known as the {\em string graph problem}, because the input
is a simple graph, and we may assume that the regions are in fact
planar curves (or slightly fattened simple curves, if we insist on disc
homeomorphs).
\micsaid{pass30: added previous ``simple''.}
  In other words, we are seeking a recognition algorithm for the
intersection graphs of planar curves.
  It is open whether this problem is decidable; it is known that there
are infinitely many forbidden subgraphs, that recognition is at least
NP-hard~\cite {K91}, and that there are string graphs that require
exponentially many string intersections for their realization~\cite
{KM91}.

  The difficulty of the string graph problem stems to a large extent
from the complex overlaps allowed between regions.
  But many practical applications are so structured that no two
regions in them overlap arbitrarily.
  For example, consider maps of political regions: such regions either
contain one another or else they have disjoint interiors; general
overlaps are not allowed.

In this paper, we consider the following special case: for every pair
of regions, they are either disjoint, or they intersect only on their
boundaries.
\micsaid{pass18: changed next ``on'' to ``of''.}
Since the nations of political maps intersect in this way, we call
such regions {\em nations}.
\micsaid{pass18: rewrote next.}
In this paper we study {\em map graphs}, the intersection graphs of nations.

\micsaid{pass18: I merged the definition into the first subsection
	(with expanded title), before discussing planarity.
	This arrangement avoids repetition.}
\zzsaid{pass17: Changed this subsection's title.}
\micsaid{pass16: added this subsection defining map, map-graph, k-map,
	k-map graph, and hole-free.  The definition of hole is delayed
	until needed.  Possibly, this entire subsection could move to
	the first subsection of Chapter~2, but this way we have less
	trouble with repetition and forward-reference.}
\begin{definition}\label{def:map}
Suppose $G$ is a simple graph. 
\micsaid{pass18: removed ``...with vertex set $V$''.}
A {\em map} of $G$ is a function ${\cal M}$ taking each vertex $v$ of $G$
to a closed disc homeomorph ${\cal M}(v)$ in the sphere, such 
that:
\begin{enumerate}
\item For every pair of distinct vertices $u$ and $v$, the interiors 
	of ${\cal M}(u)$ and ${\cal M}(v)$ are disjoint.
\item Two vertices $u$ and $v$ are adjacent in $G$ if and only if
	the boundaries of ${\cal M}(u)$ and ${\cal M}(v)$ 
	intersect.
\end{enumerate}
\micsaid{pass18: Moved next sentence up from end.}
If $G$ has a map, then it is a {\em map graph}.
\zzsaid{pass17: Rewrote the next sentence, and added hole's definition
	because we need to mention ``hole" in the ``Summary of Results"
	subsection.}
\micsaid{pass18: changed next ``on'' to ``of'', added next ``such''.}
The regions ${\cal M}(v)$ are the {\em nations} of ${\cal M}$.
The uncovered points of the sphere fall into open connected regions;
the closure of each such region is a {\em hole} of ${\cal M}$.
\end{definition}

Can we recognize map graphs?
\micsaid{pass18: removed ``or in other words....''}
This problem is closely related to planarity, one of the most basic
and influential concepts in graph theory.
\micsaid{pass18: rewrote rest of paragraph.}
Usually, planarity is defined as above, but with adjacency only for
those pairs of nations sharing a curve segment.
Planarity may also be defined in terms of maps: specifically, a planar
graph has a map {\em such that no four nations meet at a point}.

\micsaid{Deleted ``interesting philosophical question.''}

We consider two natural restrictions on maps and map graphs.
First, suppose we restrict our map so that {\em no more than $k$
nations meet at a point}; we call this a {\em $k$-map}, and the
corresponding graph is a {\em $k$-map graph}.
\micsaid{pass12: rewrote next sentence, which trivially claimed
``2-map graphs are also 3-map graphs'' (oops).}
  In particular the ordinary planar graphs are the 2-map graphs;
in fact all 3-map graphs are also 2-map graphs, as argued below.

Second, suppose that every point of the sphere is covered by some
nation.  Then we say that this is a {\em hole-free map}, and the
corresponding graph is a {\em hole-free map graph}.  \micsaid{pass14:
Added ``4-'' in next sentence.}  For our algorithm starting in
Section~\ref{sec:algorithm}, we consider hole-free 4-map graphs.

In our figures we draw a map 
by projecting one point of the sphere to infinity; we always choose a
point that is not on a nation boundary.

\subsection{Examples}\label{subsec:examples}

We give three examples.
First, consider the adjacency graph of the United States in
Figure~\ref{fig:usa}; this is a 4-map graph.  It is not planar, since
the ``four corners'' states (circled)
form a $K_4$, which is part of a $K_5$-minor.
\micsaid{pass18: added next.}
  Since removing one nation from a hole-free map leaves a connected
set of nations, all hole-free map graphs are 2-connected;
consequently, this example is not a hole-free map graph.

\micsaid{pass40: added two missing edges in {\tt usa.eps}.
 The one remaining induced 4-cycle is correct; it is
 four states around a lake.}
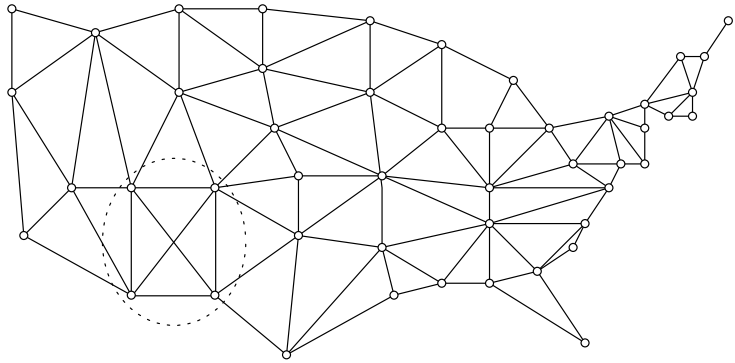
\fig[figure101]{usa}{The USA map graph.}

\zzsaid{pass19: Rewrote the next paragraph more.}
\micsaid{pass20: I mentioned the number of nations so it is clearer that 
  both letters and numbers denote the same thing (nations).
  Also, the graph $H''$ is not necessarily an induced subgraph of
every witness $H$, for example it is easy to find a witness where the
only point joining 1 and 2 is the 4-point of 1,2,3,4.
but I think my modified statement is ok.
Also in the caption, I changed ``$G$'s'' to ``its''.}
\zzsaid{pass21: By ``a witness graph of $G$", I here meant a plane graph
whose faces are cycles of length 4. In this meaning, $H''$ must be an
induced subgraph of every witness graph of $G$. }
\zzsaid{pass21: Moved most part of the next paragraph to the end of 
		Section 2.1.}
\micsaid{pass22: Ok, this example is fine now.}

Second, consider the 17-nation hole-free 4-map in Figure~\ref{fig:ex1}(1).
Let $G$ be its map graph. At the end of Section~\ref{sec:witness}, we show
that after deleting the edge $\{6,7\}$ from $G$, the result is not a map 
graph. This example demonstrates that the 4-map graph property is not 
monotone.

\fig[figure102]{ex1}%
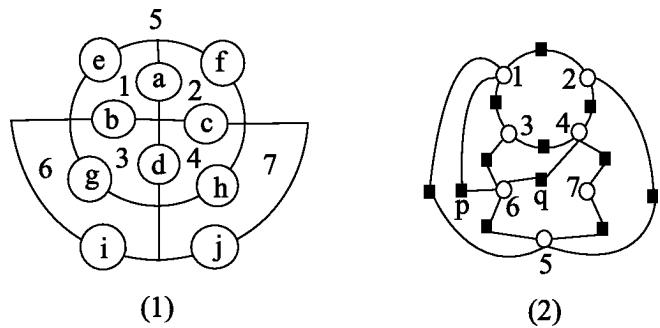
{An example hole-free 4-map, and a subgraph of its witness.}

\micsaid{pass16: changed ``order-2'' to ``mirror'', added ``hole-free''.
	Slightly revised flipflop.eps, adding a missing square and
	removing the unnecessary 2-point between $g$ and~$h$.}
Third, consider Figure~\ref{fig:flipflop}: part (1) is a map graph,
part (2) is a hole-free 4-map of the graph, and part (3) is a
corresponding witness (as defined in Section~\ref{sec:witness}).
  The graph has a mirror symmetry exchanging $a$ with $c$, but the map and
the witness do not.
  In fact a careful analysis shows that no map or witness of
this graph has such a symmetry, and so a layout algorithm must somehow
``break symmetry'' to find a map for this graph.

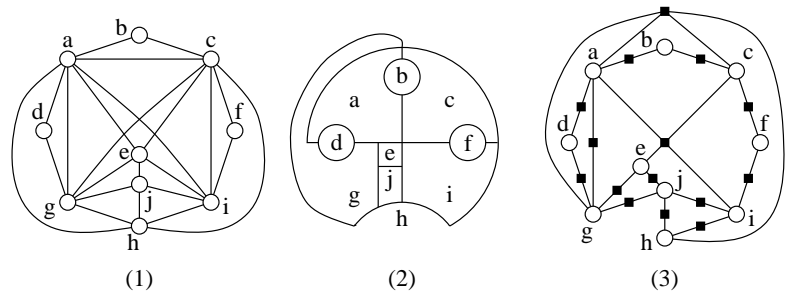
\fig[figure103]{flipflop}{A symmetric map graph, a map, and a witness.}

\micsaid{pass30: made cosmetic change to {\tt flipflop.eps}.}
\micsaid{pass14: Adopted your new example, get updated flipflop.eps.}
\zzsaid{pass11: Suppose we modify your example map graph by adding
the two edges $\{a,h\}, \{c,h\}$, and a new nation $j$ connected to
$e, g, h, i$. That is, the original internal hole is now
the new nation $j$. Then, the graph becomes a hole-free 4-map graph
(see my home page for file ``ff.eps").
I rewrote your next comment to explain why the modification works.}
\micsaid{pass10: Ok, why is there no such symmetry?

If there is a map with a geometric symmetry exchanging $a$ and $c$,
then there is a witness with a graph symmetry exchanging $a$ and $c$;
so for contradiction, we assume the latter.  
There must be a simple 16-cycle $C$ in the witness graph on which 
nations $a,b,c,f,i,h,g,d$ appear in this cyclic order. 
Let $p_{g,h}$ be the common neighbor of nations $g$ and $h$ on $C$.
Similarly, let $p_{i,h}$ be the common neighbor of nations $i$ and $h$ 
on $C$. Consider a planar embedding of the witness with nation $e$ 
inside the interior of $C$. 

Some point $p$ connects $g$ and $c$. The graph symmetry takes $p$ to
another point $p'$ connecting $a$ and $i$; $p$ and $p'$ are distinct
since $g$ and $i$ are non-adjacent. By symmetry again, $p=p_{g,h}$ 
if and only if $p'=p_{i,h}$. 

ZZ, pass11: Not sure about the last sentence, because I do not
	know the precise definition of "symmetry" here.

Mic, pass12: you asked about symmetry, I am really only
  considering witness graph symmetries; that is, a permutation of the
  vertices which preserves edges (and also the bipartite nation/point
  coloring).  I think this is at least as general as any reasonable
  definition of symmetry in a map.  In that context, there is the
  following fine point: a map with holes may have a homeomorphism
  (defined on the union of nations, and taking nations to nations)
  which cannot extend to the entire sphere.

First, suppose $p=p_{g,h}$ and $p'=p_{i,h}$. Then, 
to avoid a non-planar crossing, one of the witness edges 
$\{p_{g,h},c\}$ and $\{a,p_{i,h}\}$ must be embedded inside 
the interior of $C$ and the other outside; suppose
$\{p_{g,h},c\}$ is inside (the other case is similar). 
Then, the witness edge $\{p_{g,h},c\}$ split the interior of $C$ into 
two regions; no matter $e$ is embedded in which of the regions,
either the edge $\{a,e\}$ or the edge $\{e,i\}$ cannot be witnessed.

Second, suppose $p\not=p_{g,h}$ and $p'\not=p_{i,h}$. Then, 
to avoid a non-planar crossing, one of $p$ and $p'$ must be embedded 
inside the interior of $C$ and the other outside; suppose
$p$ is inside (the other case is similar). 
Nation $e$ is not adjacent to $p'$ because $p'$ is outside the cycle.
Since $e$ is fixed by the symmetry, $e$ is not adjacent to $p$ either.

The witness edges $g-p-c$ divide our original cycle into two halves,
one sub-cycle through $a$ and the other through $i$.  If $e$ is
inside the first sub-cycle, then there is no way for it to be adjacent
to $i$; likewise if $e$ is inside the second sub-cycle, it cannot be
adjacent to $a$.  But this contradicts our assumption that $e$ is
inside the original cycle.

}

\subsection{Summary of Results}

In the preliminary version of this paper~\cite{CGP98} we gave an
NP-characterization of map graphs, and we also sketched a
polynomial-time recognition algorithm for 4-map graphs.
  In this paper we prove the first result, but for reasons of brevity
we present a simpler variant of the second result.
  Specifically, we present a polynomial-time recognition algorithm for
{\em hole-free} 4-map graphs.
  We believe that our argument for the case with holes is correct,
however it is a very long case analysis which would double the length
of the present paper.

We left general map graph recognition as an open
problem; Thorup~\cite{T98} has recently presented a polynomial-time 
algorithm for recognizing map graphs. 
  Thorup's result does not necessarily imply ours, since even if we
are given a map realizing a map graph, it is not clear that it
helps us to find a map with the additional restrictions we want
(a hole-free 4-map). 
  Also, Thorup's algorithm is complex and the exponent of its
  time bound polynomial is about $120$, while our algorithm is more
  understandable and its time bound is $O(n^3)$.

\micsaid{pass10: some Europeans prefer the ``na{\"\i}ve'' spelling.}
\zzsaid{pass11: You make the choice.}
There is an obvious naive approach for all these recognition problems:
for each maximal clique in the given graph, assert a point in the map
where the nations of the clique should meet.
 This approach fails because there are other maps which realize a
clique, as we will see in Figure~\ref{fig:cliques}.

In Section~\ref{sec:fundamentals} we characterize map graphs as the
``half-squares'' of planar bipartite graphs
(Theorem~\ref{thm:witness}); this implies that map graph
recognition is in NP (Corollary~\ref{cor:np}).
  Using this, we list all possible clique maps
(Theorem~\ref{thm:cliques}), and we show that a map graph has a
linear number of maximal cliques (Theorem~\ref{thm:linearcliques}).

In Section~\ref{sec:algorithm} we give a high-level presentation of
our algorithm for recognizing hole-free 4-map graphs; the
algorithm produces such a map, if one exists.
\zzsaid{pass9: Rewrote the next sentence.}
  In Sections~\ref{sec:conn} through~\ref{sec:four} we
present the structural results needed to prove the correctness of the
algorithm; these sections are the technical core of our paper.
We give a time analysis in Section~\ref{sec:time}, and concluding
remarks in Section~\ref{sec:conclude}.

\section{Map Graph Fundamentals}\label{sec:fundamentals}

\subsection{A Characterization}\label{sec:witness}

We characterize map graphs in terms of planar bipartite graphs.  
  For a graph $H$ and a subset $A$ of its vertices, let $H[A]$ denote
the subgraph of $H$ induced by $A$.  Let $H^2$ denote the square of
$H$, that is the simple graph with the same vertex set, where vertices
are adjacent whenever they are connected by a two-edge path in $H$.
We represent a bipartite graph as $H=(A,B;E)$, where the vertices
partition into the independent sets $A$ and $B$, and $E$ is the set of
edges.

\begin{definition}
  Suppose $H=(A,B;E)$ is a bipartite graph.
  Then $H^2[A]$ is the {\em half-square} of $H$.
  That is, the half-square has vertex set $A$, where two vertices are
adjacent exactly when they have a common neighbor in~$B$.
\end{definition}

\zzsaid{pass8: Changed the witness graph $G'$ to $H$ below.}

\begin{theorem}\label{thm:witness}
A graph $G$ is a map graph if and only if it is the half-square of 
some planar bipartite graph $H$.
\end{theorem}
\begin{proof}
For the ``only if" part, suppose $G$ is a map graph.
Let ${\cal R}$ be the set of nations in a map of~$G$; for convenience
we identify the $n$ vertices of $G$ with the corresponding nations in
${\cal R}$.
\micsaid{pass18: explicitly identify ${\cal R}$ with $V(G)$, above.}

\micsaid{pass42: consulted a grammar book, and changed next ``so'' to
	``and so'' (or else I should change the comma to a semicolon).
	Made similar changes elsewhere without comment.}
Consider a single nation $R$.  Clearly at most $n-1$ boundary points
will account for all the adjacencies of $R$ with other nations, and so a
finite collection ${\cal P}$ of boundary points witnesses all the
adjacencies among the nations in ${\cal R}$.

In each nation $R$ we choose a representative interior point, and
connect it with edges
\zzsaid{pass19: The last ``edges" were ``arcs" [ok - M]}
through the interior of $R$ to the points of ${\cal P}$ bounding $R$.
\micsaid{pass22: reworded next sentence slightly.}
  In this way we construct 
a planar embedding of the bipartite graph
 $H=({\cal R}, {\cal P}; E')$,
such that any two nations $R_1$ and $R_2$ overlap if 
and only if they have distance two in $H$.  In other words, 
$G$ is the half-square $H^2[{\cal R}]$.

For the ``if'' part, given a bipartite planar graph $H=({\cal R},
{\cal P}; E')$, we embed it in the plane.  By drawing a sufficiently
thin star-shaped nation around each $R\in{\cal R}$ and its edges in
$E'$, we obtain a map for $H^2[{\cal R}]$.
\end{proof}

When graphs $G$ and $H$ are related as above, $H$ acts as a proof
that $G$ is a map graph.
\micsaid{pass10: defined witness ``points'' in next sentence.}
We call $H$ a {\em witness} for $G$, and we call the vertices in
${\cal P}$ the {\em points} of the witness; such points are displayed
as squares in the example Figure~\ref{fig:flipflop}(3).
\zzsaid{pass17: Previous Figure~\ref{fig:ex1}(2) was
	Figure~\ref{fig:flipflop}(3).}
\micsaid{pass20: mentioned both figures.}
\zzsaid{pass21: changed back to one figure.}
The above argument shows that $H$ has at most quadratic size, but we
can do better.

\begin{lemma}\label{lem:linearsize}
If $G$ is a map graph with $n$ vertices, then it has a witness
$H$ with $O(n)$ vertices and edges.
\end{lemma}

\begin{proof}
Construct $H$ as above.  A point $p \in {\cal P}$ is {\em redundant}
if all pairs of its neighbors are also connected through other points
of ${\cal P}$.  Deleting a redundant point does not change the half-square;
we repeat this until $H$ has no redundant points.

Consider a drawing of $H$.  For each $p \in {\cal P}$, we choose a pair of
nations $R_1$ and $R_2$ connected only by $p$.  Remove each $p$ and
its arcs, and replace them by a single arc from $R_1$ to $R_2$.  
In this way, we draw a simple planar subgraph $H'$ of $G$ with 
edge set ${\cal P}$ and vertex set ${\cal R}$.
Hence $|{\cal P}| \leq 3n-6$, and $H$ has less than $4n$ vertices.

\micsaid{pass10: by considering 2-points separately, I think I can
improve this to $|{\cal P}| \leq 2.5n-5$.  But this is probably not
tight, since the best examples I can find (red squares of a large
checkerboard) has a leading constant of~2.  In fact for hole-free map
graphs, maybe the leading constant is~1.}
\zzsaid{pass11: Please don't waste your time on such an inessential point.}

\micsaid{pass22: Moved $4n$ remark into previous paragraph.  Inserted
edge bound here in terms of $|{\cal P}|$, since we need it later.}
Since $H$ is simple and bipartite, 
by Euler's formula it has at most $2(n+|{\cal P}|)-4$ edges, which is
less than~$8n$.
\end{proof}
In particular, a map graph has a witness which may be checked in
linear time~\cite{HT74}, and so we have:
\begin{corollary}\label{cor:np}
The recognition problem for map graphs is in NP.
\end{corollary}

\micsaid{pass44: moved arboricity definition, bound, and argument here.}
Let $\alpha(G)$ denote the {\em arboricity} of a graph $G$, the
minimum number of forests whose union is $G$.
The next result is useful for the time analysis in Section~\ref{sec:time}.
\begin{corollary}\label{cor:edgebound}
A $k$-map graph with $n$ vertices has $O(kn)$ edges and arboricity $O(k)$.
\end{corollary}
\begin{proof}
By Lemma~\ref{lem:linearsize}, the map graph has a witness $H$ with
less than $8n$ edges.  So, some nation $R$ has degree at most~7 in
$H$, and consequently $R$ has degree less than $7k$ in the map graph.
Now we delete $R$, and prove our edge bound by induction on~$n$.

Since $\alpha(G)=\max_U\lceil\frac{|E(G[U])|}{|U|-1}\rceil$ where $U$
ranges over all subsets of $V(G)$ containing at least two
vertices~\cite{N61}, and each $G[U]$ is again a $k$-map graph, the
edge bound implies the arboricity bound. 
\zzsaid{pass45: changed previous citation. [ok-M]}
\end{proof}

For the simplicity of our figures, we prefer to draw maps rather than
planar bipartite witness graphs.  The arguments in
Theorem~\ref{thm:witness} show efficient transformations back and
forth between them; so nothing will be lost by using maps.
We conclude with some further simple consequences of
Theorem~\ref{thm:witness}:
\zzsaid{pass21: Re-ordered the following items.}
\begin{itemize}

\item In the witness graph, a point of degree three may be replaced by
three points of degree two.  Consequently, 3-map graphs are 2-map
graphs (planar graphs).

\item For $k\geq3$, $k$-map graphs are those with witnesses $H$
such that every point has degree at most $k$.

\item Hole-free map graphs are those with witnesses $H$ such that
every face has exactly four sides.  Since a six-sided face may always
be partitioned into three four-sided faces (by adding a redundant
point), we may also allow six-sided faces.
\micsaid{pass30: added ``redundant point'' remark.}

\item
  If $G$ has no clique of size four,
  then it is a map graph if and only if it is a planar graph.  

\item A map graph may contain cliques of arbitrary size.

\item {From} the previous two remarks, it is clear that the ``map graph'' 
	property is not monotone, and hence cannot be characterized by
	forbidden subgraphs or minors. 

\micsays{pass30: The planarity remark also implies that there are
infinitely many forbidden induced subgraphs; unlike the argument for
string graphs, this is a relative triviality.  Just consider the
infinite family of ``M\"{o}bius ladder'' graphs ($P_n \times K_2$,
with the first and last copies of $K_2$ identified with a twist); each
graph (with $n\geq3$) is not a map graph, and also not a subgraph of
another.}

\end{itemize}
\micsaid{pass22: this is a much better place than
Section~\ref{subsec:examples}, thanks!  I moved the argument out of
the itemization, into a claim.  Minor changes: added two ``let''s and
a ``the''.}  Regarding the last point, we can also show a stronger
example:
\begin{claim}
There is a hole-free 4-map graph $G$ and edge $e$ such that $G-e$ has no map.
\end{claim}
\begin{proof}
Let $G$ be the graph realized by the hole-free 4-map in
Figure~\ref{fig:ex1}(1), and let $H$ be a witness of $G$.
  We may assume that $H$ has no point of degree~1.
  Let $H'$ be the bipartite graph in Figure~\ref{fig:ex1}(2), and let
$H''$ be the graph obtained from $H$ by deleting points $p$ and $q$
and their incident edges.
  Since each $v\in\{a,\ldots,j\}$ has exactly two neighbors in $G$,
the degree of $v$ in $H$ is either 1 or 2; we may assume the former
case because in the latter case, the two points adjacent to nation $v$
in $H$ can be identified. By this assumption, $H''$ is an induced
subgraph of~$H$.
 In turn, by the existence of edges $\{1,6\}$ and $\{4,6\}$ in $G$ and
the planarity of $H$, $H'$ must be a (not necessarily induced)
subgraph of $H$.
  Now, by Figure~\ref{fig:ex1}(2) and planarity, point $q$ must also
be adjacent to nations 3 and~7 in $H$.
  Our argument so far did not depend on edge $e=\{6,7\}$ in $G$, but
now this edge has been forced by considering other edges.
  So in other words, $G-e$ is not a map graph.
\end{proof}

\subsection{Cliques in Map Graphs}

Suppose $G$ is the clique $K_n$, then it may be realized in the
following ways, corresponding to the four parts of
Figure~\ref{fig:cliques}:
\micsaid{pass30: cosmetic changes to {\tt cliques.eps}}
\fig[figure201]{cliques}{Cliques in map graphs.}
\begin{enumerate}
\def\labelenumi{(\theenumi)}  

\item The $n$ nations share a single boundary point.
  We call this the {\em pizza}.

\item Some $n-1$ nations share a single boundary point, and
the one remaining nation is arbitrarily connected to them at other
points.
  We call this the {\em pizza-with-crust}.

\item If $n \geq 6$, there may be three points supporting
all adjacencies in the clique, with at most $n-2$ nations at any one
point.
  In particular, there are at most two nations adjacent to all three
of the points.
  We call this the {\em hamantasch}.

\item A clique with all boundary points of degree two; that
is, an ordinary planar clique.  Since the planar $K_2$ (edge) and
$K_3$ (triangle) are a pizza and pizza-with-crust respectively, the
only new clique to list here is the planar $K_4$, which we call the
{\em rice-ball}.

\end{enumerate}
\micsaid{pass10: although it works, it seems unnatural to make the
four classes disjoint.  In particular, the extra restrictions in the
hamantasch definition are just for that purpose; also it should not
matter that some planar cliques are also pizzas or pizza-with-crust.
An example from geometry: it would be poor style to insist that a
rectangle has unequal sides, just to avoid a square.}

\begin{theorem}\label{thm:cliques}
  A map graph clique must be one of the above four types.
\end{theorem}

\begin{proof}
By Theorem~\ref{thm:witness}, we have a bipartite planar witness graph
$H=({\cal R}, {\cal P}; E')$ such that the half-square
$H^2[{\cal R}]$ is the clique $K_n$, where $n=|{\cal R}|$.
Let $d$ be the maximum degree of all points $p \in {\cal P}$.

\micsaid{pass10: rewrote this paragraph to avoid modifying the map.}
If $n=d$, we have a pizza.  If $n=d+1$, we have a pizza-with-crust.
So we may assume $n \geq d+2$. If $d \leq 3$, then the map graph is 
planar; since $K_5$ is not planar, this forces $n=4$ and $d=2$,
the rice-ball. \micsaid{pass30: added previous $K_5$ remark}
We now assume $d \geq 4$.

\micsaid{pass32: rewrote the rest. I delayed the introduction of
$R_1,\ldots,R_6$ until an explicit choice can be made.  This should
explain why I put those dots where I did.  Also, I permuted some of
the $R$'s, get the revised {\tt p1p2.eps}.}

Pick point $p_1$ of maximum degree $d$, and nations $x$ and $y$
not adjacent to $p_1$.  Consider the set ${\cal P}'$ of all points
connecting $x$ or $y$ to the nations around $p_1$.  We claim
that there is a point $p_2 \in {\cal P}'$ connecting $x$, $y$,
and at least two nations adjacent to $p_1$;
otherwise, by drawing arcs through the points of ${\cal P}'$, we
could get a planar $K_{d,2}$ with the $d$ nations on
\zzsaid{pass31: last word was ``around"}
a common face, which is impossible.
Since $p_1$ has maximum degree, there are also two nations
adjacent to $p_1$ but not $p_2$.  In summary, the following
three disjoint sets each contain at least two nations:
\begin{eqnarray*}
{\cal R}_1 & = & \{R \in {\cal R}|
 \mbox{ $R$ is adjacent to $p_2$ but not $p_1$}\} \\
{\cal R}_2 & = & \{R \in {\cal R}|
 \mbox{ $R$ is adjacent to $p_1$ but not $p_2$}\} \\
{\cal R}_3 & = & \{R \in {\cal R}|
 \mbox{ $R$ is adjacent to both $p_1$  and $p_2$}\}
\end{eqnarray*}

We will choose six distinct nations $R_1,R_2 \in {\cal R}_1$, $R_3,R_4
\in {\cal R}_2$, and $R_5,R_6 \in {\cal R}_3$; no matter how we
choose, the graph $H$ will contain the induced subgraph in
Figure~\ref{fig:p1p2}(1), with the cycle $C= p_1 R_5 p_2 R_6$.
  The graph $H'=H[\{p_1,p_2\} \cup {\cal R}_3]$ is a complete
bipartite graph, with a planar embedding inherited from $H$.
  Each face of $H'$ is a 4-cycle; furthermore all nations in ${\cal
R}_1 \cup {\cal R}_2$ must lie inside one face, in order to be
connected by other points.
  So, we choose $R_5, R_6 \in {\cal R}_3$ on this face.
\zzsaid{pass33: rewrote next sentence.}
Then this face is bounded by $C$; we have an embedding with ${\cal R}_1
\cup {\cal R}_2$ inside~$C$, and ${\cal R}_3-\{R_5,R_6\}$ outside~$C$.
  By an appropriate choice of nations $R_1, R_2 \in {\cal R}_1$ and
$R_3, R_4 \in {\cal R}_2$, we arrive at Figure~\ref{fig:p1p2}(2), the
embedding of $p_1$, $p_2$, and all their edges to adjacent nations.
  In this figure, the three occurrences of ``$\ldots$'' locate any other
nations in ${\cal R}_1 \cup{\cal R}_2 \cup{\cal R}_3$.

\fig[figure202]{p1p2}{A subgraph of $H$, and its embedding.}


\zzsaid{pass31: rewrote the rest of proof, because the last sentence
	in your original argument is false when both $R_3$ and $R_4$
	are adjacent to $p_3$.
	You need to modify Figure~\ref{fig:p1p2}(2) and (3) by changing
	each $R_i$ to $R'_i$, adding a ``$\ldots$" to the right of
	$R'_3$, and moving the ``$\ldots$" on the right of $R'_4$ to
	the left.} 

There must exist a third point $p_3$ inside $C$ connecting $R_1$
and $R_4$.  These edges now separate ${\cal R}_1-\{R_1\}$ from ${\cal
R}_2-\{R_4\}$, so all these nations are adjacent to $p_3$ as well,
yielding Figure~\ref{fig:p1p2}(3).
  This figure is not necessarily an induced subgraph, since the edges
$\{R_5,p_3\}$ and $\{R_6,p_3\}$ may occur in $H$.
\zzsaid{pass33: rewrote the rest of this paragraph. [ok-M]} 
 But by the maximality of the degree of $p_1$, if exactly $i\in\{1,2\}$ of
these edges exist, then there exist $i$ other nations $R_7 \in {\cal R}_3$,
necessarily outside~$C$. So, no matter whether these edges exist or not,
the points $p_1$, $p_2$, and $p_3$ support a hamantasch on
${\cal R}_1 \cup {\cal R}_2 \cup {\cal R}_3$. Hence, we are done
if ${\cal R}={\cal R}_1 \cup {\cal R}_2 \cup {\cal R}_3$.

For contradiction, suppose ${\cal R}$ contains some nation $R$ not
adjacent to $p_1$ or~$p_2$.
\zzsaid{pass33: rewrote the rest. [ok-M]}
We need to place $R$, and some new points and edges, in
Figure~\ref{fig:p1p2}(3) so that $R$ has neighbor points connected to
the other nations. However by planarity of $H$, if $\{R_5,p_3\}$ or
$\{R_6,p_3\}$ is an edge in $H$, then $R$ cannot be placed so that
both $R_1$ and $R_7$ have neighbor points connected to $R$.
Similarly, if neither $\{R_5,p_3\}$ nor $\{R_6,p_3\}$ is an edge in $H$,
then $R$ cannot be placed so that all of $R_1$, $R_5$, and~$R_6$ have
neighbor points connected to $R$. 
\end{proof}

By a careful analysis of each kind of clique, we can now show:

\micsaid{pass22: No explicit use found for the ``dual map graph''
$H^2[{\cal P}]$.  Added ``less than'' in theorem statement.}
\zzsaid{pass27: Wrong! It is used in the first step of our main
	algorithm to bound the enumeration of maximal cliques.
	It is odd to say ``if $G$ has more than $O(n)$ maximal cliques,
	then reject". So, I restored the constant.}
\micsaid{pass28: Regarding {\tt\string\ConsFac}, I simply observed
 that it is not used elsewhere in the paper (although it could be).
 This relates to the ``time bound'' argument for Assumption~1; it is
 the general argument usable whenever proposing a polynomial-time
 algorithm for an NP search problem.  I think the general argument is
 more immediately acceptable by our typical reader, who probably knows
 about general algorithmic theory but not much about map graphs.
 Otherwise, at various points in our algorithmic description we take
 back Assumption~1, which seems inconsistent.  I think that we would
 only need to check for two things: too many edges, and too many
 cliques.  But for a first-time reader, it is not obvious that these
 two checks suffice.
 
 I inserted ``at most'' in the theorem statement.}

\def\ConsFac{27} 
\begin{theorem}\label{thm:linearcliques}
A map graph $G$ with $n$ vertices has at most $\ConsFac n$ maximal cliques.
\end{theorem}
\micsaid{pass26: changed proof statement to $O(n)$, and removed
all uses of {\tt ConsFac}.}

\begin{proof} We may assume that $G$ is connected.
As in Theorem~\ref{lem:linearsize}, we choose a planar witness
$H=({\cal R}, {\cal P}; E')$ for $G$ where ${\cal R}$ is the set of
nations, ${\cal P}$ is the set of at most $3n-6$ points,
and $E'$ is the set of edges.
\micsaid{pass26: added next sentence.}
\zzsaid{pass27: commented out it, because it is confusing.}

Fix a plane embedding of $H$.  If vertices $u_1,u_2,v_1,v_2$ appear in
that cyclic order as distinct neighbors of some vertex $w$ in $H$, then
we say that the pairs $\{u_1,v_1\}$ and $\{u_2,v_2\}$ {\em cross} at $w$.

Each point can contribute to at most one maximal pizza, and so there
are at most $3n-6$ maximal pizzas in $G$.  Note that each MC$_2$ is a
pizza.

Next, let $C_1,\ldots,C_\ell$ be the maximal cliques in $G$ that are
either non-pizza MC$_3$'s or hamantaschen. For each hamantasch $C_i$,
we may choose three points $p_i,q_i,r_i \in {\cal P}$ and three nations
$a_i,b_i,c_i\in C_i$ such that
$T_i=p_i a_i q_i b_i r_i c_i$ is an induced cycle in $H$
and $C_i$ consists of all nations adjacent to at least two
of the points $p_i,q_i,r_i$ in $H$.
For each pair $\{a,b\}$ of vertices in $G$
such that some non-pizza MC$_3$ contains both $a$ and $b$, let $s_{a,b}$
be a point in ${\cal P}$ that is adjacent to both $a$ and $b$ in $H$.
For each non-pizza MC$_3$ $C_i$, let $C_i=\{a_i,b_i,c_i\}$,
$p_i=s_{a_i,c_i}$, $q_i=s_{a_i,b_i}$, $r_i=s_{b_i,c_i}$, and $T_i$ be
the induced cycle $p_i a_i q_i b_i r_i c_i$ in $H$.
In either case, we define $f(C_i)=\{p_i,q_i,r_i\}$ and note that
$C_i$ consists of all nations adjacent to at least two points of $f(C_i)$.
This implies $f(C_i)\not=f(C_j)$ for distinct $C_i$ and~$C_j$, because
otherwise $C_i\cup C_j$ would be a larger clique.

Define $H'$ as the simple graph with vertex set ${\cal P}$ and edge set 
$\{\{p,q\}\;|\ \{p,q\} \subset f(C_i)$ for some $i\}$.
We claim that $H'$ is planar. To see this, we embed $H'$ in $H$ by drawing 
the edge $\{p_i,q_i\}$ of $H'$ through their neighbor $a_i$ in $H$, and 
similarly for the other two edges $\{q_i,r_i\}$ and $\{p_i,r_i\}$. Towards 
a contradiction, assume that two edges of $H'$ cross in the embedding. 
Then, by cycle symmetries, we may assume that for some distinct $C_i$ and 
$C_j$, pairs $\{p_i,q_i\}$ and $\{p_j,q_j\}$ cross at nation $a_i=a_j$ 
(call it~$a$) in $H$. Since the cycles $T_i$ and $T_j$ cross at $a$ in $H$,
they must cross again, sharing either another nation or a point.

\case{1} $T_i$ and $T_j$ share another nation but no point. 
By symmetry, it suffices to consider the case $b_i=b_j=b$. 
If $C_i$ and $C_j$ were both non-pizza MC$_3$'s, then we would have 
$q_i=q_j=s_{a,b}$, contradicting the crossing. 
So at least one of $C_i$ and $C_j$ is a hamantasch, we suppose $C_i$. 
Then $C_i$ has another nation $a'$ also adjacent to 
$p_i$ and $q_i$ or else $C_i$ would be a pizza-with-crust. 
Because of $T_j$, it must be $a'=c_j$.  Now since $q_i$ is adjacent to 
$a,b,c_j$, in order for $C_j$ to be a non-pizza, $C_j$ must also be 
a hamantasch. Then $C_j$ has another nation $a''$ adjacent to $p_j$ 
and $q_j$, but planarity of $H$ makes this impossible.

\case{2} $T_i$ and $T_j$ share a point. 
Since $T_i$ and $T_j$ are induced, $a$ is adjacent to neither $r_i$ nor 
$r_j$, so the only possible shared point is $r_i=r_j$. In turn, 
$C=\{a,b_i,c_i,b_j,c_j\}$ is a clique of $G$. So, neither $C_i$ nor $C_j$ 
is an MC$_3$, and both are hamantaschen. Now as in the previous case, 
we find it is impossible to add a nation $a'\not\in C$ between $p_i$ and 
$q_i$ and a nation $a''\not\in C$ between $p_j$ and $q_j$. By this, 
both $C_i$ and $C_j$ are pizza-with-crusts, a contradiction.

By the above case-analysis, the claim holds, and $H'$ is a planar graph 
with at most $3n-6$ vertices and at least $\ell$ distinct triangles. 
An easy exercise shows that any simple planar graph with $h$ vertices
has at most $3h$ triangles. So, $\ell\leq 9n-18$.   
\zzsaid{pass21: commented out the rest of this paragraph.} 

There are at most $n$ rice-balls, since they all have different center
nations.

\micsaid{pass26: accepted the $2d_p-2$ bound, and checked your changes.}
\zzsaid{pass27: rewrote the rest of proof.}
\micsaid{pass28: checked, and rewrote.  I changed $u,w,v$ to $a,b,c$ to
get same alphabetic cycle ordering as before; also changed the
introduction of your two cases.  Enumerated the cliques, so I can use
``$q_i$'' rather than ``$q_{C_i}$''.}

It remains to bound the number of maximal pizza-with-crusts of size 4
or more. Fix a point $p$ in $H$, let $V_p$ denote the set of nations
adjacent to $p$ in $H$, and let ${\cal C}_p =\{C_1,\ldots,C_{\ell_p}\}$ 
be the maximal pizza-with-crusts with center $p$ and size~4 or more.  
We claim that $\ell_p = |{\cal C}_p| \leq 2(2|V_p|-3)$.  
This claim implies that $G$ has at most 
$\sum_p (4|V_p|-6)$ maximal pizza-with-crusts of size~4 or more. This
sum equals $4|E'|-6|{\cal P}|$; since $|E'| \leq 2(n+|{\cal P}|)-4$
and $|{\cal P}|\leq 3n-6$, the sum is less than $14n$.

Now we prove the claim. The embedding gives a cyclic clockwise order 
on the nations of $V_p$ around~$p$; this order defines 
``consecutive'' nations and ``intervals'' of nations around~$p$.
For nations $u,v\in V_p$, let $[u,v]$ denote the circular interval of 
nations starting at $u$, proceeding clockwise around~$p$, and ending 
at~$v$. For each clique $C_i$ in ${\cal C}_p$, let $b_i$ be the crust 
of $C$. Since $C_i$ is not a pizza, we can choose distinct nations 
$a_i, c_i \in C_i-\{b_i\}$ and distinct points $q_i,r_i\not=p$ 
satisfying the following three conditions:
\begin{itemize}

\item $T_i=p a_i q_i b_i r_i c_i$ is a simple cycle in $H$.

\item If a nation of $C_i-\{a_i,b_i,c_i\}$ is adjacent to $q_i$ or $r_i$,
	then it lies outside $T_i$, otherwise it lies inside.

\item All nations of $C_i$ lying inside the cycle $T_i$ lie in 
	the interval $[a_i,c_i]$.

\zzsaid{pass29: commented out next item for not requiring uniqueness.}
\end{itemize}
Denote the unordered pair $\{a_i,c_i\}$ by $g(C_i)$,
	and $\{q_i,r_i\}$ by $h(C_i)$.
By considering such 6-cycles $T_i$ and the planarity of $H$, there are no
cliques $C_i,C_j \in {\cal C}_p$ such that $g(C_i)$ and $g(C_j)$ cross at
$p$; consequently the graph $G_p=(V_p, \{g(C_i)\;|\ C_i\in{\cal C}_p\})$ 
is simple outerplanar, where we use the same cyclic order on $V_p$ 
for the outerplanar embedding. Since $G_p$ is simple outerplanar, 
it can have at most $2|V_p|-3$ edges.  Thus, to prove
$|{\cal C}_p| \leq 2(2|V_p|-3)$, it suffices to prove that each edge of 
$G_p$ equals $g(C_i)$ for at most two $C_i \in {\cal C}_p$.

\zzsaid{pas29: rewrote this paragraph.} 
For contradiction, assume that there exist three distinct cliques
$C_i,C_j,C_k \in {\cal C}_p$ with $g(C_i)=g(C_j)=g(C_k)$.
Say that two of these cliques are {\em nested} if the crust of one 
is inside the cycle of the other.  We consider two cases.

\case{I} There are two non-nested cliques.  We may suppose that they
are $C_i$ and $C_j$.  By planarity, the interiors of $T_i$ and $T_j$
are disjoint, with $a_i=b_j$ and $a_j=b_i$.  
\zzsaid{pass29: rewrote next sentence.} 
Moreover, no matter whether $h(C_i)\cap h(C_j)=\emptyset$ or not, 
no nation of $V_p - g(C_i)$ is adjacent to both $p$ and at least 
one point of $h(C_i)\cup h(C_j)$ in $H$. Thus, the set of nations 
lying inside $T_i$ and the set of nations lying inside $T_j$ form 
a partition of $V_p-g(C_i)$. By this and the maximality of cliques 
in ${\cal C}_p$, the crust of $C_k$ must lie inside $T_i$ or $T_j$; 
by symmetry we suppose it lies inside $T_i$.  On the other hand, 
since $C_i$ is a maximal clique of size 4 or more, there exists a
nation $x_i\in C_i-g(C_i)$ lying inside $T_i$.  Since $x_i$ cannot
be adjacent to $q_i$ or $r_i$, there is another point $s_i$ lying
inside $T_i$ that connects $x_i$ with $b_i$.  So, we have a path
$P_i=p x_i s_i b_i$ sharing only its endpoints with $T_i$, and
bisecting the interior of $T_i$.  Now the crust $b_k$ must lie inside
$T_i$, to one side or the other of $P_i$.  To achieve $g(C_k)=g(C_i)$,
$b_k$ must be adjacent to $s_i$ in $H$.  But then we would see that
$s_i \in h(C_k)$, and so either $a_k$ or $c_k$ was chosen incorrectly.

\case{II} All three cliques nest. 
\zzsaid{pass29: rewrote next two sentences.} 
Again by planarity, the cycles cannot cross. So, their interiors nest
in some order; we may assume that $b_k$ lies inside $T_j$, and $b_j$
lies inside $T_i$. We have $a_i=a_j=a_k$ and $c_i=c_j=c_k$.
Since $C_j$ is a maximal clique, it contains some nation $x_j$ not
adjacent to $b_i$ in $G$. By planarity, $x_j$ must lie inside $T_j$,
and there is some point $s_j$ connecting $b_j$ with $x_j$ in $H$.
\zzsaid{pass29: rewrote next sentence.}
We cannot have $s_j \in h(C_j)$, by the choice of $g(C_j)$; so again
we have a path $P_j=p x_j s_j b_j$, sharing only its endpoints with
$T_j$ and bisecting its interior. Now the crust $b_k$ must lie inside
$T_j$, to one side or the other of $P_j$; 
the rest of the argument proceeds as in the last case.
\end{proof}


\section{Recognizing Hole-Free 4-Map Graphs}\label{sec:algorithm}

The rest of this paper is devoted to an $O(n^3)$-time algorithm for 
deciding whether a given graph is a hole-free 4-map graph.  It follows 
from Theorem~\ref{thm:cliques} that 4-map graphs have no 7-cliques, that 
all 6-cliques are hamantaschen, and that all 5-cliques are pizza-with-crusts.
\zzsaid{pass9: Added the next sentence.}
\micsaid{pass18: modified next sentence.}
Also, as observed in Section~\ref{subsec:examples}, all hole-free map
graphs are 2-connected.
Unfortunately, these simple observations do not lead to a 
polynomial-time algorithm. Indeed, our algorithm is very sophisticated 
and too long to be included in a single section. This section only gives 
a high-level sketch of the algorithm. The correctness and implementation
details are given in subsequent sections. 

\subsection{Definitions}\label{subsec:algdefs}
\micsaid{pass18: Changed title back (removed ``More'').}

Before presenting the algorithm, we define some vocabulary for the
remainder of the paper.
\zzsaid{pass13: Commented out the next two sentences.}
%
A {\em marked graph} is a simple graph in which each edge is either
marked or not marked. Throughout the remainder of this paper, $G$ is a
marked graph with vertex set $V(G)=V$ and edge set $E(G)=E$.
  For a vertex $v$ in $G$, $N_G(v)$ denotes the set of vertices
adjacent to $v$ in $G$. For a subset $U$ of $V$, $N_G(U)$ denotes
$\cup_{u\in U}N_G(u)$.
  Let $F$ be a subset of $E$.
  $G-U-F$ denotes the marked graph obtained from $G$ by deleting the
edges in $F$ and the vertices in $U$ together with the edges incident
to them.
  When $U$ or $F$ is empty, we drop it from the notation $G-U-F$.
\micsays{pass22: removed $G[U]$ definition, since it is already given
 and used in the last section.}

\begin{definition}\label{def:layout}
Let $U$ be a subset of $V$. A {\em layout} ${\cal L}$ of $G[U]$ is a 4-map
${\cal L}$ of $G[U]$ such that for every marked edge $\{u,v\}$ in $G$ with
$u,v\in U$, the boundaries of nations ${\cal L}(u)$ and ${\cal L}(v)$ 
share a curve segment (not just one or more isolated points). 
\micsaid{pass36: defined ``well-formed'' here.} 
\zzsaid{pass37: reworded it.} 
${\cal L}$ is {\em well-formed} if for every edge $\{u,v\}$ in $G[U]$, 
the intersection of ${\cal L}(u)$ and ${\cal L}(v)$ is either a point 
or a curve segment (but not both). 
\end{definition}

\begin{definition}\label{def:atlas}
If a layout ${\cal L}$ of $G$ covers every point of the sphere,
we call it an {\em atlas} of $G$. 
\end{definition}

Our goal is to design an efficient algorithm to decide whether a given
$G$ has an atlas; it will either return an atlas or report failure.
\micsaid{pass36: added promise here, note ``3-connected'' not yet defined.}
Furthermore, the algorithm returns a well-formed atlas
whenever possible (see Corollary~\ref{cor:well}).

If $G$ has an atlas, then by Lemma~\ref{lem:linearsize} it has a
witness graph checkable in linear time~\cite{HT74}.
So in fact we describe an algorithm that makes the following assumption:
\begin{assume}\label{ass:map}
  $G$ has an atlas. 
\end{assume}
If $G$ does not have an atlas, we will discover this when our algorithm
either fails, returns an invalid atlas, or takes more time than allowed
by our analysis in Section~\ref{sec:time}.
\micsaid{pass22: Added ``time'' above.  Otherwise, we should say
   something about early termination our the clique enumeration.}
\zzsaid{pass23: I dislike this, because you did not mention how
	much time. So, instead of this, we may just say ``our 
	algorithm rejects if $G$ has too many maximal cliques."}
\micsaid{pass28: expanded previous ``time'' clause, since I think the
   general argument is easier (it is not as practical, but it is more
   familiar and immediately acceptable to CS theorists).}
\micsaid{pass12: moved next sentence from before
	Definition~\ref{def:layout}; it makes more sense here.}
Since we assume that $G$ has an atlas, we will call the vertices in $G$
{\em nations}; we will use lower-case letters to denote them.

Throughout the rest of this subsection, fix a subset $U$ of $V$ and 
a layout ${\cal L}$ of $G[U]$. 
A vertex $u\in U$ {\em touches} a hole ${\cal H}$ of ${\cal L}$ 
if ${\cal L}(u)$ intersects ${\cal H}$; 
they necessarily intersect on their boundaries.  Vertex
$u$ {\em strongly touches}
${\cal H}$ if the boundaries of ${\cal L}(u)$ and ${\cal H}$ share a curve 
segment. A {\em 2-hole} is a hole strongly touched by exactly two vertices. 
{\em Erasing a 2-hole} ${\cal H}$ in ${\cal L}$ is the operation of 
modifying ${\cal L}$ by extending ${\cal L}(u)$ to occupy ${\cal H}$, 
where $u$ is one of the vertices strongly touching ${\cal H}$. 
Figure~\ref{fig:def1}(1) depicts the operation. 

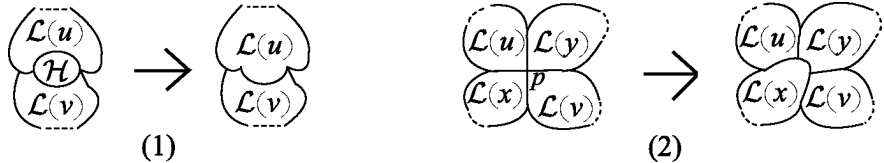
\fig[figure301]{def1}{Erasing a 2-hole ${\cal H}$, and a $(u,v)$-point $p$.
Dashed curves may intersect.}

A {\em $k$-point} in ${\cal L}$ is a point shared by exactly $k$ nations. 
Let $u\in U$ and $v\in U$. A {\em $(u,v)$-point} in ${\cal L}$ is a 4-point 
$p$ at which ${\cal L}(u)$ and ${\cal L}(v)$ together with two other nations 
${\cal L}(x)$ and ${\cal L}(y)$ meet cyclically in the order 
${\cal L}(u)$, ${\cal L}(x)$, ${\cal L}(v)$, ${\cal L}(y)$. 
{\em Erasing the $(u,v)$-point} $p$ in ${\cal L}$ is the operation of 
modifying ${\cal L}$ by extending nation ${\cal L}(x)$ to occupy a small disc
around $p$ touching only ${\cal L}(u)$, ${\cal L}(v)$, ${\cal L}(x)$, and 
${\cal L}(y)$. Figure~\ref{fig:def1}(2) depicts the operation. 
\zzsaid{pass13: Commented out the following sentence.} 

A {\em $(u,v)$-segment} in ${\cal L}$ is a curve segment $S$ shared by the 
boundaries of ${\cal L}(u)$ and ${\cal L}(v)$ such that each endpoint of $S$ 
is a 3- or 4-point. Note that two $(u,v)$-segments must be disjoint. 
An edge $\{u,v\}$ of $G$ is {\em good} in ${\cal L}$ if there is either 
exactly one $(u,v)$-segment or exactly one $(u,v)$-point, but not both.
An edge that is not good in ${\cal L}$ is {\em bad} in ${\cal L}$. 
\micsaid{pass36: reworded next, since it is no longer a definition.}
Note that ${\cal L}$ is well-formed if and only if every edge of
$G[U]$ is good in ${\cal L}$.

\begin{definition}
If ${\cal M}$ is an atlas of $G$ and $W$ is a subset of $V$, then ${\cal
M}|_W$ denotes the layout of $G[W]$ obtained by restricting
${\cal M}$ to $W$. 
Such a layout is called an {\em extensible} layout of $G[W]$.
\micsaid{pass18: rewrote next, closer to your original.}
${\cal L}$ is {\em transformable} to another layout ${\cal L}'$ of $G[W]$ 
if whenever ${\cal L}$ is extensible, so is~${\cal L}'$.
\end{definition}

Note that an extensible layout never has a 4-point on a hole boundary, 
since filling the hole would create an illegal 5-point. Similarly, 
if two holes touch in an extensible layout, it must be at a 2-point. 

\micsaid{old: Since we will seek well-formed layouts, this helps
justify the operation of ``erasing'' 2-holes.  That is, we could point
out that if a layout is well-formed, connected, and has at least three
nations, then it has no 2-hole.}

\subsection{Making Progress}

\micsaid{pass22: Changed title and reworded first paragraph to give
   meaning to phrase ``make progress'' (which we use often).}
\zzsaid{pass35: rewrote next sentence to mention the ``promise".}
\micsaid{pass36: I took back your change.  We want a well-formed atlas
  for each 3-connected smaller graph, even if $G$ is 2-connected.}
To find an atlas for $G$, our algorithm may ``make progress'' by
producing one or more smaller marked graphs, so that finding an atlas
for $G$ is reduced to finding an atlas
for each of these smaller graphs. 
\micsaid{pass34: replaced two old sentences with the next.}
Here we define the graph features that our algorithm may identify in
order to make progress; subsequent sections show how to make progress
for each.
\micsaid{pass36: added last two words.}

A {\em $k$-cut} of $G$ is a subset $U$ of $V$ with $|U|=k$ whose removal 
disconnects $G$. $G$ is {\em $k$-connected} if it has no $i$-cut with 
$i\leq k-1$. Section~\ref{sec:conn} shows that the algorithm can 
always make progress when $G$ is not 4-connected. On the other hand, 
under the assumption that $G$ is 4-connected, Corollary~\ref{cor:well} 
guarantees that $G$ has a well-formed atlas. 

\begin{definition}
A clique consisting of $k$ vertices is called a {\em $k$-clique}. 
A clique $C$ in $G$ is {\em maximal} if no clique in $G$ properly contains $C$.
A maximal $k$-clique is called an MC$_k$.
\zzsaid{pass33: moved next two sentences to Section~\ref{sec:five}.}
\end{definition}

If $G$ is 4-connected and has at most 8 vertices, our algorithm will 
construct a well-formed atlas for $G$ by exhaustive search. On the other hand, 
under the assumptions that $G$ is 4-connected and has at least 9 vertices, 
Lemma~\ref{lem:noMC6} guarantees that $G$ has no 6-clique. 

\begin{definition}
A {\em correct 4-pizza} is a list $\langle a, b, c, d\rangle$ of four nations 
in $G$ such that $G$ has a well-formed atlas in which nations $a$, $b$, $c$,
$d$ meet at a point cyclically in this order.
\zzsaid{pass33: commented out next line.}
 {\em Removing a correct 4-pizza} 
$\langle a, b, c, d\rangle$ from $G$ is the operation of modifying $G$ 
as follows: Delete the edge $\{a, c\}$ from $G$ and mark the edges $\{b,d\}$, 
$\{a, b\}$, $\{b, c\}$, $\{c, d\}$, and $\{d, a\}$. 
\zzsaid{pass33: added next line.}
\end{definition}
\zzsaid{pass33: moved next sentence from below.}
\micsaid{pass34: rewrote next sentence to use the (just defined)
``removing'' operation.}
Under the assumption that $G$ is 4-connected, Lemma~\ref{lem:remove4}
allows our algorithm to make progress by removing a correct 4-pizza in
$G$, whenever we identify one.

\zzsaid{pass33: lem:remove4 was here before.}

\micsaid{pass32: the following ``correct center'' material could all
move to Section~\ref{sec:five}.}
\zzsaid{pass33: I do not agree, because the moving makes the algorithm
	sketch less clear.}

\zzsaid{pass33: added next subsentence.}
\micsaid{pass34: made next ``pizza'' singular.}
To see a particular type of correct 4-pizza,
consider an extensible layout of an MC$_5$ $C$ in $G$.
Since a 5-clique is not planar, the layout contains at least one 4-point.
Inspection shows there is exactly one 4-point in a well-formed layout.
\micsaid{Otherwise there would be a 2-hole by pigeon-hole argument.}
This ``pizza-with-crust'' layout motivates the following definition.  
\begin{definition}
A {\em correct center} of $C$ is a list $\langle a, b, c,
d\rangle$ of four nations in $C$, such that $C$ has a well-formed
extensible layout in which nations $a$, $b$, $c$, $d$ meet at a point
in this cyclic order.  The unique nation in $C - \{a, b, c, d\}$
is the corresponding {\em correct crust} of $C$.  
\end{definition}

\begin{fact}\label{fac:center}
Let $C$ be an MC$_5$ in $G$. 
Then, every correct center of $C$ is a correct 4-pizza in $G$. 
\end{fact}
\micsaid{pass12: moved next sentence down from above the ``Fact.''}
Note that $C$ may have multiple correct centers, each from a different
extensible layout. 
\zzsaid{pass33: moved next sentence to right after Definition 3.6.}

\micsaid{pass22: moved this text up from before def:separators, for a
better transition from the previous text on 4-points.}
\micsaid{pass34: changed next ``here'' to ``below''.}
  Besides the $k$-cuts mentioned above, we also consider the more
specialized separators introduced below in
Definition~\ref{def:separators}.
  Section~\ref{sec:adv} will show how the algorithm may make progress
as long as $G$ contains one of these.

\micsaid{Here I define ``crossable'' edges, and define ${\cal E}$ in
terms of that.  This changes ${\cal E}$ slightly: now ${\cal E}[a,b]$
is empty if $\{a,b\}$ is marked.  But this is probably what you
intended anyway.}
\zzsaid{pass5: All right.} 
\begin{definition}
Edges $\{a,b\}$ and $\{x,y\}$ in $G$ are {\em crossable}
if they are both unmarked and $\{a,b,x,y\}$ is an MC$_4$ in $G$.
For an edge $\{a,b\}$, let ${\cal E}[a,b]$ denote the set of all edges
$\{x,y\}$ crossable with $\{a,b\}$.
\end{definition}

\begin{fact}\label{fac:Eabcde}
If $\{a,b\}$ is an edge and $G-\{a,b\}$ has a triangle $\{c,d,e\}$,
then at most one edge of that triangle is in ${\cal E}[a,b]$.
\end{fact}
\begin{proof}
Two edges would imply two MC$_4$'s,  sitting inside the
5-clique $\{a,b,c,d,e\}$.
\end{proof}

\begin{definition}\label{def:separators} 
We define the following separators in the marked graph $G$:
\begin{enumerate} 
\item\label{def:sepedge} 
      A {\em separating edge} of $G$ is an edge $\{a,b\}$ such that 
	$G - \{a,b\} - {\cal E}[a,b]$ is disconnected.
\item\label{def:sep4cyc} 
      An {\em induced 4-cycle} in $G$ is a set $C$ of four vertices in $G$ 
	such that $G[C]$ is a cycle. A {\em separating 4-cycle} of $G$ is 
	an induced 4-cycle $C$ in $G$ such that $G-C$ is disconnected. 
\item\label{def:septrip} 
      A {\em separating triple} of $G$ is a list $\langle a, b, c\rangle$ 
	of three vertices in $G$ such that $C=\{a,b,c\}$ is a clique in $G$ 
	and $G - C - {\cal E}[a,b]$ is disconnected.
\item\label{def:sepquad} 
      A {\em separating quadruple} is a list $\langle a,b,c,d\rangle$ of 
	four vertices in $G$ such that (i) $G[\{a,b,c,d\}]$ is a cycle and 
	(ii) $G - \{a, b,c,d\}-{\cal E}[a,b]$ is disconnected. 
\item\label{def:septria}
      A {\em separating triangle} of $G$ is a list $\langle a,b,c\rangle$ 
	of three vertices in $G$ such that (i) $C=\{a,b,c\}$ is a clique 
	in $G$ and (ii) $G'=G-C-({\cal E}[a,b] \cup {\cal E}[a,c])$ is 
	disconnected. If in addition, $G'$ has a connected component 
	consisting of a single vertex, then $\langle a,b,c\rangle$ is 
	a {\em strongly separating triangle} of $G$. 
\end{enumerate}
\end{definition}

\micsaid{pass18: removed next sentence, redundant with next paragraph.}

\subsection{Sketch of the Algorithm}\label{subsec:sketch}

Given a marked graph $G=(V,E)$, our algorithm rejects if $G$ is not 
2-connected, and it solves the problem by exhaustive search 
when $|V|\leq 8$. When $|V|\geq 9$, it searches $G$ for 
a 2-cut (cf.\ Lemma~\ref{lem:3-con}), 3-cut (cf.\ Lemma~\ref{lem:3-cut}), 
separating edge (cf.\ Lemma~\ref{no2hole}), separating 4-cycle (cf.\ 
Lemma~\ref{noS4-cycle}), separating triple (cf.\ Lemma~\ref{noStriple}), 
separating quadruple (cf.\ Lemma~\ref{noSquadruple}), 
strongly separating triangle (cf.\ Lemma~\ref{noSStriangle}), or 
separating triangle (cf.\ Lemma~\ref{noStriangle}), {\em in this order}. 
In each case, as the lemmas show, the algorithm makes progress by 
either (1) removing a correct 4-pizza or (2) reducing the problem for 
$G$ to the problems for certain marked graphs smaller than $G$
whose total size is that of $G$ plus a constant.
\zzsaid{pass8: Added ``whose total size ... constant."} 

If none of the above separators exists in $G$, then $G$ has no 6-clique 
(cf.\ Lemma~\ref{lem:noMC6}) and the algorithm searches $G$ for an 
MC$_5$ or MC$_4$, {\em in this order}. If an MC$_5$ $C$ is found, 
it tries to find an extensible layout of $C$ by doing a case-analysis 
based on the neighborhood of $C$ in $G$ (cf.\ Section~\ref{sec:five}). 
The absence of the above separators guarantees that only 
a few cases needed to be analyzed.
\zzsaid{pass35: rewrote next sentence to mention the ``promise".}
The case-analysis either yields an extensible layout of $C$ whose
center is then removed to make progress, or produces a marked graph
$G'$ smaller than $G$ such that finding a well-formed atlas for $G$
can be reduced to finding a well-formed atlas for $G'$.
\zzsaid{pass35: commented out next sentence.}

\zzsaid{pass43: strangely, this paragraph appeared in mg41.tex. 
Similarly, if an MC$_4$ $C$ is found in $G$, the algorithm tries to 
find an extensible layout of $C$ by doing a case-analysis based on 
the neighborhood of $C$ in $G$ (cf.\ Section~\ref{sec:four}).
\zzsaid{pass35: rewrote next sentence.} 
The analysis consists of only a few cases due to the absence of the 
above separators, and yields a correct 4-pizza or a marked graph $G'$ 
smaller than $G$ such that finding a well-formed atlas for $G$ can be
reduced to finding a well-formed atlas for $G'$.
In either case, it makes progress.
}
\micsaid{pass44: The paragraph you just commented was there since
subsec:sketch appeared in mg4.tex.  But you are right, it no
longer belongs.}
 
\zzsaid{pass43: did you check this paragraph?}
\micsaid{pass44: checked; minor changes. I slashed some spaces, removed
  an extraneous instance of ``rice-ball'', and fixed one verb.}
If no MC$_5$ but an MC$_4$ is found in $G$, the algorithm scans all MC$_4$'s 
of $G$ in an arbitrary order. While scanning an MC$_4$ $C$, it decides whether 
$C$ has a rice-ball layout (cf.\ Lemma~\ref{mc4:riceball}). If $C$ has a 
rice-ball layout, the algorithm quits the scanning and makes progress by 
removing a correct 4-pizza obtained from the rice-ball layout of $C$. 
On the other hand, if no rice-ball is found after scanning all MC$_4$'s, 
the algorithm scans all MC$_4$'s of $G$ in an arbitrary order, once again.
But this time, while scanning an MC$_4$ $C$, it decides whether $C$ has a 
non-pizza layout, by doing a case-analysis based on the neighborhood of 
$C$ in $G$ (cf.\ Section~\ref{subsec:dist}). The analysis consists of only 
a few cases due to the absence of the above separators. If $C$ has a 
non-pizza layout, the algorithm quits the scanning and makes progress by 
removing a correct 4-pizza obtained from the layout of $C$. 
Otherwise, all MC$_4$'s are pizzas; the algorithm finds their centers 
(cf. Section~\ref{subsec:pizzas}), and removes all of them so that 
$G$ no longer has an MC$_4$. 

\micsaid{pass42: rewrote next paragraph with ``triangulated''
 condition.  Removed this issue from sec:time.  Perhaps this is too
 long; maybe it should just be stated as ``obvious'', instead of
 explicitly argued like this.  Also, 3-connectivity would suffice.}
\zzsaid{pass43: rewrote it again. Please check.}
\micsaid{pass44: it is ok, added end sentence stating what the
  algorithm actually does.}
If neither MC$_5$ nor MC$_4$ is found in $G$, then this is a base case.
As observed in Section~\ref{sec:witness}, $G$ must be planar, or else 
we reject. When $G$ is planar, then by its 4-connectivity it has a 
unique planar embedding. We claim that $G$ has a well-formed atlas if 
and only if all its faces are triangles. The ``if'' direction is obvious 
because the planar dual of $G$ is an atlas, which is well-formed by the
connectivity of~$G$. Conversely, suppose $G$ has a well-formed atlas 
${\cal M}$. Since ${\cal M}$ has no $k$-point for $k>3$, all adjacent 
pairs of nations strongly touch in ${\cal M}$, and so the 3-points and
boundaries in ${\cal M}$ define a 3-regular planar graph $G'$, whose
dual is $G$.
So, it suffices for the algorithm to check that $G$ is planar and has
a 3-regular dual; if so, it returns the dual as an atlas.

\micsaid{pass42: added next ``recursive''}
\zzsaid{pass43: rewrote this paragraph. [ok-M]}
In all the recursive cases, the smaller graphs that we generate have
total size at most the size of $G$ plus a constant, and we spend 
quadratic time on generating them. A simple argument 
(cf.\ Section~\ref{sec:time}) shows that the overall time is cubic. 

\micsaid{pass3: We could also argue that each recursive step is
charged to some unmarked edge which is either marked or disappears.}

\subsection{Figures}

For our arguments of the algorithm's correctness, we need a convenient 
graphical notation for the possible extensible layouts of $G[U]$, where $U$ 
is some small subset of $V(G)$.  First, as is very natural, we consider 
two layouts equivalent when they are homeomorphic.  But beyond this, 
we also introduce a convenient graphic notation for partially determined
layouts of $G[U]$.  In particular, we introduce contractible segments
and permutable labels. 

\begin{definition}
A {\em figure} of $G[U]$ is a list ${\cal D} = 
\langle {\cal L}$, ${\cal S}$, $L_1$, \ldots, $L_k\rangle$, where ${\cal L}$ 
is a layout of $G[U]$, ${\cal S}$ is a set of curve segments in ${\cal L}$, 
and $L_1$, \ldots, $L_k$ are disjoint lists of 
vertices in $U$. We call ${\cal L}$ the {\em layout} in ${\cal D}$, call 
the curve segments in ${\cal S}$ the {\em contractible segments} in ${\cal 
D}$, and call $L_1$, \ldots, $L_k$ the {\em permutable lists} in ${\cal D}$. 

We say ${\cal D}$ {\em displays} a layout ${\cal L}'$ of $G[U]$
if ${\cal L}'$ can be obtained from ${\cal L}$ by:
\begin{enumerate}
\item contracting parts of some contractible segments to points,
\item  erasing all resulting 2-holes,
 and
\item for each permutable list $L_i$, selecting a permutation 
$\pi$ of $L_i$ and renaming each nation $u\in L_i$ as $\pi(u)$. 
\end{enumerate}
We say ${\cal D}$ {\em displays} $G[U]$,
or ${\cal D}$ is a {\em display} of $G[U]$,
if ${\cal D}$ displays an extensible layout of $G[U]$.
${\cal D}$ is {\em transformable} to another figure ${\cal D}'$ of $G[U]$ 
if whenever ${\cal D}$ displays $G[U]$, so does ${\cal D}'$. 
\end{definition}

\micsaid{After contraction and 2-hole removal, ${\cal L}'$ may still
not be well-formed (consider an atlas of $K_4$, and then contract
one edge).  Should we disqualify such layouts?}
\zzsaid{pass5: Yes, if a resulting map is not well-formed, then
we just ignore it.}

To illustrate a figure ${\cal D}$ we draw ${\cal L}$, emphasize
the contractible segments in bold,
and then for each permutable list $L_i$, we label each nation $u \in L_i$ as
$u^i$. Since we may contract any part of a contractible segment to a point,
the endpoints do not matter, and they are not emphasized.
The holes are unlabeled, and should be regarded as ``optional'' if a
contraction could reduce it to a 2-hole.

\fig[figure302]{intro1}{A display of MC$_5$ $\{a,b,c,d,e\}$.}
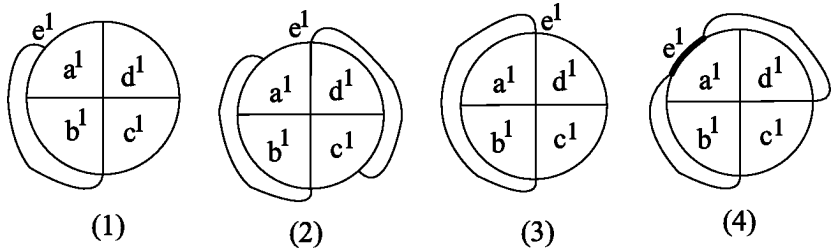
\fig[figure303]{intro2}{Possible displays of MC$_5$ $\{a,b,c,d,e\}$.}

For example, when $G$ has a well-formed atlas ${\cal M}$ and at least 
6 vertices but has no separating triangle, Figure~\ref{fig:intro1} displays 
$G[U]$ for an MC$_5$ $U=\{a,b,c,d,e\}$ of $G$. To see this, we contract 
a set of contractible segments in the figure to obtain 
Figure~\ref{fig:intro2}(1) through (4). We argue that one of them 
must display ${\cal M}|_U$ as follows. First, $U$ is a pizza-with-crust in 
${\cal M}$. Suppose that the four non-crust nations $a^1,b^1,c^1,d^1$ meet 
at a 4-point $p$ in ${\cal M}$ in this order. Let $q_{a,b}$ be the endpoint 
of the $(a^1,b^1)$-segment in ${\cal M}$ other than $p$. Define $q_{b,c}$, 
$q_{c,d}$, and $q_{d,a}$ similarly. Let $k$ be the number of points among 
$q_{a,b}$, $q_{b,c}$, $q_{c,d}$, $q_{d,a}$ that are touched by the crust 
$e^1$ of $U$ and another nation of $V-U$ in ${\cal M}$. 
Since ${\cal M}$ is well-formed, $k \leq 2$. On the other hand, 
since $G\not=G[U]$ and has no separating triangle, $k\geq 1$. 
If $k=1$, then Figure~\ref{fig:intro2}(1) displays ${\cal M}|_U$.
If $k=2$, then Figure~\ref{fig:intro2}(2), (3) or (4) displays ${\cal M}|_U$. 
We note that 
Figure~\ref{fig:intro1} has a unique permutable set, namely, $U$ itself.

\section{Establishing Connectivity}\label{sec:conn} 

Our goal here is to reduce the algorithmic problem to the case when
$G$ is 4-connected.
\zzsaid{pass9: Commented out next sentence:}
Since $G$ is already 2-connected, we first show how to
reduce to the 3-connected case.
\micsaid{pass22: changed ``next'' to ``first'', above.}

\begin{lemma}\label{lem:3-con} Let ${\cal M}$ be an atlas of $G$. 
Let $u$ and $v$ be two distinct vertices of $G$. Then,
the following statements hold: 
\begin{enumerate}
\item $G-\{u,v\}$ is disconnected if and only if
	there are at least two $(u,v)$-segments in ${\cal M}$. 
\item Suppose that $G-\{u,v\}$ is disconnected and its connected components 
are $G_1$, \ldots, $G_k$. Then for each $i\in\{1,\ldots,k\}$, 
the marked graph $G'_i$ obtained from $G[V(G_i) \cup \{u,v\}]$ 
by marking edge $\{u,v\}$ has an atlas. Moreover, given an atlas 
${\cal M}_i$ for each $G'_i$, we can easily construct an atlas of $G$.
\end{enumerate}
\end{lemma}

\begin{proof} 
We first prove Statement 1. If $\{u,v\}\not\in E$, 
then nations $u$ and $v$ are disjoint disc homeomorphs in ${\cal M}$, and 
hence removing them from ${\cal M}$ leaves exactly one connected region. 
So we know $\{u,v\}\in E$. Let $k$ be the number of $(u,v)$-segments 
in ${\cal M}$. Consider the following three cases.

\case{1} $k=0$. We erase the $(u,v)$-points in ${\cal M}$. Then, 
${\cal M}$ becomes an atlas of $G-\{\{u,v\}\}$ and nations $u$ and $v$ are 
disjoint disc homeomorphs in it. So, removing $u$ and $v$ from ${\cal M}$ 
leaves exactly one connected region. 

\case{2} $k=1$. We erase the $(u,v)$-points in ${\cal M}$. 
${\cal M}$ remains an atlas of $G$. Moreover, edge $\{u,v\}$ becomes good 
in ${\cal M}$. Since the union of nations $u$ and $v$ is a disc homeomorph in 
${\cal M}$, removing them from ${\cal M}$ leaves exactly one connected region. 

\case{3} $k\geq 2$. We erase the $(u,v)$-points in ${\cal M}$. 
${\cal M}$ remains an atlas of $G$. Moreover, there are exactly $k$ disjoint
 holes in ${\cal M}|_{\{u,v\}}$. So, removing nations $u$ and $v$ from ${\cal 
M}$ leaves exactly $k$ connected regions. Each of these regions forms a 
connected component of $G-\{u,v\}$. This completes the proof of Statement 1. 

We next prove Statement 2. 
For each $i$, let $U_i=V(G_i)$.
Each hole in ${\cal M}|_{U_i\cup\{u,v\}}$ is a 2-hole and is touched only by 
$u$ and $v$, and hence erasing the holes in ${\cal M}|_{U_i\cup\{u,v\}}$ 
yields an atlas of $G'_i$. On the other hand, given an atlas ${\cal M}_i$ of 
each $G'_i$, we erase any $(u,v)$-points in ${\cal M}_i$. ${\cal M}_i$ remains 
an atlas of $G'_i$, because edge $\{u,v\}$ is marked in $G'_i$ and so there 
exists a $(u,v)$-segment in ${\cal M}_i$.
Since $G'_i-\{u,v\}=G_i$ is connected, Statement 1 implies there is exactly one
$(u,v)$-segment.
Thus removing nations $u$ and $v$ from ${\cal M}_i$ leaves 
exactly one connected region, and the closure ${\cal L}_i$ of this region is 
a disc homeomorph. The boundary of ${\cal R}_i$ can be divided into two curve 
segments $S_{i,u}$ and $S_{i,v}$ such that $S_{i,u}$ (respectively, $S_{i,v}$) 
is a portion of the boundary of nation $u$ (respectively, $v$) in ${\cal M}$. 
Now, we can obtain an atlas of $G$ as follows. First, put ${\cal R}_1$, \ldots,
${\cal R}_k$ on the sphere in such a way that no two of them touch and each 
$S_{i,u}$ appears on the upper half of the sphere while each $S_{i,v}$ appears 
on the lower half. Second, draw nation $u$ (respectively, $v$) to occupy the 
area of the upper (respectively, lower) half of the sphere that is occupied 
by no ${\cal R}_i$. This gives an atlas of $G$. 
\end{proof}

\begin{corollary}\label{cor:well}
$G$ is 3-connected if and only if $G$ has a well-formed atlas. 
\end{corollary}

\begin{proof}
By Lemma~\ref{lem:3-con}, the ``if" part is obvious.
For the other direction, suppose $G$ is 3-connected. 
Let ${\cal M}$ be an atlas of $G$. If no edge of $G$ is bad in ${\cal M}$, 
then ${\cal M}$ is well-formed and we are done. So, suppose that some edge 
$\{u,v\}$ is bad in ${\cal M}$. Since $G$ is 3-connected, there is at most 
one $(u,v)$-segment in ${\cal M}$. If there is no $(u,v)$-segment in ${\cal 
M}$, we erase all but one $(u,v)$-points in ${\cal M}$; otherwise, we erase 
all the $(u,v)$-points in ${\cal M}$. In both cases, ${\cal M}$ remains 
an atlas of $G$ and edge $\{u,v\}$ becomes good in ${\cal M}$ while 
no good edge becomes bad in ${\cal M}$. Consequently, we can  
make all bad edges good in ${\cal M}$. 
\end{proof}

Using Statement 2 of Lemma~\ref{lem:3-con}, our algorithm may reduce
the given graph to its 3-connected components, and so we assume:
\begin{assume}\label{ass:3-con}
$G$ is 3-connected, and it has a well-formed atlas denoted by ${\cal M}$.
\end{assume}

We say that two nations $u$ and $v$ {\em strongly touch} in ${\cal M}$
if there is a $(u,v)$-segment in ${\cal M}$; they {\em weakly touch}
in ${\cal M}$ if there is a $(u,v)$-point in ${\cal M}$.
To simplify the sequel, we also suppose that small graphs are handled
by exhaustive methods.  We assume:
\begin{assume}\label{ass:big}
$G$ has $|V|\geq 9$ vertices.
\end{assume}

\micsaid{pass32: the next lemma was in fact already cited in
Lemma~\ref{lem:remove4} and Section~\ref{subsec:sketch}.  It is probably
not worth pointing out, but there was no circular reasoning.}
\zzsaid{pass33: Of course, we need the next lemma, because it is cited
	in many places in the sequel. But to avoid forward citing in
	Lemma~\ref{lem:remove4}, I moved lem:remove4 to the end of
	Section~\ref{sec:conn}.}

\begin{lemma}\label{lem:3-cut}
Let $C=\{a,b,c\}$ be a set of three distinct vertices in $G$. 
Then, the following statements hold: 
\begin{enumerate}

\item When $C$ is not a clique in $G$, $G-C$ is connected. 

\item When $C$ is a clique in $G$, $G-C$ is disconnected if and only if
 (i) the nations in $C$ do not meet at a point in ${\cal M}$ and 
 (ii) each pair of nations in $C$ strongly touch in ${\cal M}$. 

\item Suppose that $G-C$ is disconnected. Then, (i) $G-C$ has exactly two 
connected components $G_1$ and $G_2$, and (ii) both $G'_1$ and $G'_2$ have 
a well-formed atlas, where $G'_1$ (respectively, $G'_2$) is the marked graph 
obtained from $G[V(G_1)\cup C]$ (respectively, $G[V(G_2)\cup C]$) by marking 
the edges in $E(G[C])$. Moreover, given a well-formed atlas of $G'_1$ and 
one of $G'_2$, we can easily construct one of $G$. 

\end{enumerate}
\end{lemma}

\begin{proof}
To prove Statement 1, suppose that $C$ is not a clique. 
For each edge $\{u,v\}\in E(G[C])$, if nations $u$ and $v$ weakly touch 
in ${\cal M}$, then we erase the $(u,v)$-point in ${\cal M}$. 
Then, ${\cal M}|_{V-C}$ is a layout of $G-C$ and the holes in 
${\cal M}|_{V-C}$ are disjoint disc homeomorphs. So, $G-C$ must be connected. 

To prove Statement 2, suppose that $C$ is a clique. The ``if" part
is clear. To prove the ``only if" part, suppose that (i) or (ii) in 
Statement 2 does not hold. In case (i) is false, $a$, $b$ and $c$ meet at a
point in ${\cal M}$, and the well-formedness of ${\cal M}$ ensures that 
their union is a disc homeomorph, and so $G-C$ is connected. 
Otherwise, suppose (i) is true and (ii) is false.
For each edge $\{u,v\}\in E(G[C])$, if nations $u$ and $v$ weakly touch 
in ${\cal M}$, then we erase the $(u,v)$-point to get atlas ${\cal M}'$. 
Then ${\cal M}'|_{V-C}$ is a layout of $G-C$ and the holes in 
${\cal M}'|_{V-C}$ are disjoint disc homeomorphs. So, $G-C$ is connected. 

Next, we prove Statement 3. Since $G-C$ is disconnected, (i) and (ii) in 
Statement 2 hold. By this, there are exactly two holes ${\cal H}_1$ and 
${\cal H}_2$ in ${\cal M}|_C$ and they are disjoint. For $i\in\{1,2\}$, 
let $U_i$ be the set of nations that occupy ${\cal H}_i$ in atlas ${\cal M}$. 
Each $G[U_i]$ is a connected component of $G$. Let $G'_i$ be the marked graph 
obtained from $G[U_i\cup C]$ by marking the edges in $E(G[C])$. There is a 
unique hole in ${\cal M}|_{U_1\cup C}$ and it is (strongly) touched only by 
the nations of $C$. So, modifying ${\cal M}|_{U_1\cup C}$ by extending nation 
$a$ to occupy its unique hole yields a well-formed atlas of $G'_1$. 
Similarly, we can obtain a well-formed atlas of $G'_2$. 

On the other hand, suppose that we are given a well-formed atlas ${\cal M}_1$ 
of $G'_1$ and one ${\cal M}_2$ of $G'_2$. Let $i\in\{1,2\}$. Since the edges 
in $E(G[C])$ are marked in $G'_i$, each pair of nations of $C$ strongly touch 
in ${\cal M}_i$. Note that $G'_i - C$ is connected. 
Then by Statement 2 and the well-formedness of ${\cal M}_i$, the nations 
in $C$ meet at a 3-point $p_i$ in ${\cal M}_i$. Let $D_i$ be a disc that is 
centered at $p_i$ and touches no nation of $U_i$ in atlas ${\cal M}_i$. 
To obtain a well-formed atlas of $G$, we remove each $D_i$ from ${\cal 
M}_i$ to obtain a connected region ${\cal R}_i$, and then glue ${\cal R}_1$ 
and ${\cal R}_2$ together by identifying nations $a$, $b$, $c$ in ${\cal R}_1$
 with those in ${\cal R}_2$, respectively. 
\end{proof}

Statement 3 above gives an effective algorithm to reduce the given
graph to its 4-connected components.  So we now assume:
\begin{assume}\label{ass:4-con}
Graph $G$ is 4-connected.
\end{assume}

\micsaid{It may also be useful
to state this corollary of Statement 2, above: if $G$ is 4-connected
and has a marked triangle, then those three regions meet at a point in
any atlas.  The reason is that this would be a more transparent
replacement for most uses of {\tt lem:3-cut}.}
\zzsaid{pass5: Not a big deal, because only one place uses this.}

\begin{lemma}\label{lem:noMC6}
$G$ has no 6-clique. 
\end{lemma}

\zzsaid{pass2: Added the following figure.}
\fig[figure401]{conn1}{A display of MC$_6$ $\{a,\ldots,f\}$.}

\begin{proof}
First observe that no hole-free 4-map graph has a 7-clique, by
Theorem~\ref{thm:cliques}.  Assume, on the contrary, that $G$ has an
MC$_6$ $C$. Then it must be a hamantasch, and by
Assumption~\ref{ass:4-con}, Figure~\ref{fig:conn1} displays
${\cal M}|_C$.
\zzsaid{pass6: Changed the next sentence.}
Thus, $V=C$, contradicting Assumption~\ref{ass:big}. 
\end{proof}

\micsaid{pass34: in the next lemma (and many subsequent) we require
that the atlas for $G'$ is well-formed, in order to produce that for
$G$.  Of course here $G'$ is 3-connected, and so it will always have an
well-formed atlas if it has one at all.  But algorithmically, we still
need a promise: if the input $G$ is 3-connected, then the output atlas
is well-formed.  Where to make this promise?
Such a promise could be mentioned in {\tt subsec:sketch}.  It could
maybe go earlier (``Our goal...'' paragraph of {\tt subsec:algdefs}),
but at that point we have not defined ``well-formed''.}
\zzsaid{pass35: I modified the meaning of ``making progress", and
	rewrote several sentences in the algorithm-sketch section.
	Hope these reflect your above opinion. }
\micsaid{pass36: I put promise as early as possible, it is ok now.}

\begin{lemma}\label{lem:remove4}
Let $G'$ be the marked graph obtained from $G$ by removing a correct 4-pizza 
$\langle a, b, c, d\rangle$. Then, $G'$ has a well-formed atlas. Moreover, 
given a well-formed atlas of $G'$, we can easily construct one of $G$.
\end{lemma}

\begin{proof}
Let ${\cal M}$ be a well-formed atlas of $G$ in which 
nations $a$, $b$, $c$, $d$ meet at a point $p$ in this order. After
erasing the $(a,c)$-point $p$ in ${\cal M}$, we obtain a well-formed atlas 
of $G'$ in which nations $a$, $b$, and $d$ meet at a 3-point and nations 
$b$, $c$, and $d$ meet at a 3-point. Thus, by Lemma~\ref{lem:3-cut}, both 
$G' - \{a, b, d\}$ and $G' - \{b, c, d\}$ are connected.

Let ${\cal M}'$ be 
a well-formed atlas of $G'$. Since $G'-\{a, b, d\}$ is connected and the edges 
$\{a, b\}$, $\{a, d\}$, and $\{b, d\}$ are marked in $G'$, nations $a$, $b$, 
and $d$ must meet at a 3-point $p_1$ in ${\cal M}'$ according to 
Lemma~\ref{lem:3-cut}. Similarly, nations $b$, $c$, and $d$ must meet at 
a 3-point $p_2$ in ${\cal M}'$. Thus, the boundaries of $b$ and $d$ in 
${\cal M}'$ share a curve segment $S$ with endpoints $p_1$ and $p_2$. 
We modify ${\cal M}'$ by contracting $S$ to a point, obtaining 
a well-formed atlas of $G$. 
\end{proof}

\micsaid{pass36: added this paragraph.}  In some of our reductions we
will discover that $G$ has a well-formed map with several 4-pizzas.
In those situations we may remove all the 4-pizzas at once.  This is
because the resulting graph still has a well-formed atlas, and
therefore is 3-connected, and so the above argument applies for each
removed 4-pizza.

\section{Advanced Separations}\label{sec:adv}

In this section we prove the necessary properties of the separators
introduced in Definition~\ref{def:separators}.

\subsection{Separating Edges}

\begin{definition}
A {\em shrinkable segment} $S$ in ${\cal M}$ is a $(u,v)$-segment in
${\cal M}$ such that (i) $\{u,v\}$ is an unmarked edge in $G$, (ii)
both endpoints of $S$ are 3-points, and (iii) the endpoints of $S$ are
touched by distinct nations $a$ and $b$ which are adjacent in $G$.
Nations $a$ and $b$ are called the {\em ending nations} of $S$. 
\end{definition}

In the next two results, we show a close relationship between
separating edges and shrinkable segments.

\begin{lemma}\label{no2hole}
Assume that 
$G$ has a separating edge $\{a,b\}$. Let $G'=G-\{a,b\}-{\cal E}[a,b]$. Then, 
for every $\{x,y\}\in E$ such that $x$ and $y$ belong to different connected 
components of $G'$, $\langle a,x,b,y\rangle$ is a correct 4-pizza in $G$. 
\end{lemma}

\begin{proof}
Let ${\cal M}'$ be the atlas of $G$ obtained from ${\cal M}$ by 
contracting those shrinkable segments whose ending nations are $a$ and $b$. 
All edges except $\{a,b\}$ are good in ${\cal M}'$. 

First, we claim that for every $\{u,v\} \in E$ such that $u$ and $v$ belong 
to different connected components of $G'$, there is a point in ${\cal M}'$ at 
which nations $u$, $a$, $v$, $b$ meet at $p$ cyclically in this order. Towards 
a contradiction, assume that such a point does not exist in ${\cal M}'$. 
By the definition of $G'$, $\{u,v\}$ is in ${\cal E}[a,b]$.
There is no nation $w \in V-\{a,b,u,v\}$ adjacent to both $u$ and $v$;
otherwise, $w$ would connect $u$ and $v$ in $G'$ by Fact~\ref{fac:Eabcde}.
By the fact that ${\cal M}'$ has no hole, nations $u$ and $v$ only share 
a unique curve segment $S$ in ${\cal M}'$ and the endpoints of $S$ can be 
touched only by $a$ or $b$ in ${\cal M}'$. Nation $a$ cannot touch both 
endpoints of $S$; otherwise, since the edges $\{a,u\}$ and $\{a,v\}$ are still 
good in ${\cal M}'$, nations $a$, $u$, and $v$ would have to occupy the whole 
sphere, a contradiction. Similarly, $b$ cannot touch both endpoints of $S$. 
So, both endpoints of $S$ are 3-points. In summary, one endpoints of $S$ is 
touched by $a$ and the other is touched by $b$. Then $S$ would be 
a shrinkable segment whose ending nations are $a$ and $b$, contradicting
the choice of ${\cal M}'$.

Second, we claim that there is no $(a,b)$-segment in ${\cal M}'$. Towards 
a contradiction, assume that an $(a,b)$-segment $S$ exists in ${\cal M}'$. 
By the first claim, there is an $(a,b)$-point $p$ in ${\cal
M}'$. Note that $p$ is not on $S$. Let $x$ and $y$ be the two nations
of $V-\{a,b\}$ that meet at $p$. Since ${\cal M}'$ has no hole, $G$ is 
4-connected and $|V|\geq 9$, there is a nation $z\in V-\{a,b,x,y\}$ that 
touches either $x$ or $y$ in ${\cal M}'$. If $z$ touches $x$ (respectively,
$y$) in ${\cal M}'$, then $z$ is not reachable from $y$ (respectively,
$x$) in $G-\{a,b,x\}$ (respectively, $G-\{a,b,y\}$), contradicting 
Assumption~\ref{ass:4-con}. 

Third, we claim that there are at least two $(a,b)$-points in ${\cal
M}'$.  Otherwise, if there is only one $(a,b)$-point, then by the
first claim, ${\cal E}[a,b]$ would have at most one edge, and 
the 4-connectivity of $G$ would prevent the separation of $G'$.

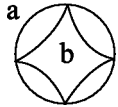
\fig[figure501]{adv1}{Layout ${\cal M}'|_{\{a,b\}}$ when $\ell=4$.}

Let $\ell$ be the number of $(a,b)$-points.  Since $\ell\geq 2$ and
there is no $(a,b)$-segment, we see that the atlas ${\cal M}'$ has a
cyclic sequence of $(a,b)$-points $q_0$, \ldots, $q_{\ell-1}$.  These
points alternate with $\ell$ 2-holes in ${\cal M}'|_{\{a,b\}}$;
Figure~\ref{fig:adv1} displays ${\cal M}'|_{\{a,b\}}$ when $\ell=4$.

For each $j$, let $x_j$ and $y_j$ be the nations meeting $a$ and $b$.
We claim that $\{a,b,x_j,y_j\}$ is an MC$_4$; otherwise to form a
containing 5-clique would force $\ell\leq3$ and ${\cal E}[a,b] =
\emptyset$, contradicting the separation of $G'$.  So in fact each
$q_j$ corresponds to an edge $(x_j,y_j)$ in ${\cal E}[a,b]$, and the
components of $G'$ correspond to the set of nations in each hole.

Now consider a particular edge $(x_j,y_j)$.  To show that $\langle a,
x_j, b, y_j\rangle$ is a correct 4-pizza, we must find a well-formed
atlas of $G$ where they meet as they do in ${\cal M}'$.  This is easy to
do: we simply erase all $(a,b)$-points in ${\cal M}'$ except $q_j$,
and the resulting atlas is well-formed.
\end{proof}

\begin{corollary}\label{cor:noSedge} 
Suppose $G$ has an edge $\{a,b\}$, not inside a 5-clique.
Then, $\{a,b\}$ is a separating edge if and only if there is a
shrinkable segment in ${\cal M}$ with ending nations $a$ and~$b$.
\end{corollary}

\micsaid{We can almost do without the 5-clique assumption; there is
essentially only one large counter-example I can think of, a 5-clique
with one hole.}

\begin{proof}
The ``if" part is obvious from the proof of Lemma~\ref{no2hole}.
To prove the ``only if" part, suppose 
there is a shrinkable segment $S$ in ${\cal M}$ with ending nations $a$ and 
$b$. Let ${\cal M}'$ be the atlas of $G$ obtained from 
${\cal M}$ by contracting $S$ to a single point $p$. Similarly to the second 
claim in the proof of Lemma~\ref{no2hole}, we can claim that there is no 
$(a,b)$-segment in ${\cal M}'$. Thus, besides $p$, there is exactly one 
$(a,b)$-point $q$ in ${\cal M}'$, inherited from ${\cal M}$.
Now, ${\cal M}'|_{\{a,b\}}$ has exactly two holes ${\cal H}_0$ 
and ${\cal H}_1$. Let $Z_0$ (respectively, $Z_1$) be the set of nations of 
$V-\{a,b\}$ occupying ${\cal H}_0$ (respectively, ${\cal H}_1$) in atlas ${\cal
 M}'$. Let $x\in Z_0$ and $y\in Z_1$ be the two nations that meet at $p$ in 
${\cal M}'$.  Similarly, let $x'\in Z_0$ and $y'\in Z_1$ be the two nations that 
meet at $q$ in ${\cal M}'$. By ${\cal M}'$, edges $\{x,y\}$ and $\{x',y'\}$ are
 not marked in $G$ and they are all the edges connecting nations of $Z_0$ to 
those of $Z_1$. Moreover, since no 5-clique contains $\{a,b\}$, $\{a,b,x,y\}$ and 
$\{a,b,x',y'\}$ are MC$_4$'s of $G$. Therefore, no connected component of 
$G-\{a,b\}-{\cal E}[a,b]$ contains both the nations of $Z_0$ and those of 
$Z_1$. In other words, $\{a,b\}$ is a separating edge of $G$. 
\end{proof}

\subsection{Separating 4-Cycles}

\begin{lemma}\label{noS4-cycle}
Suppose that $G$ has a separating 4-cycle $C$ with vertices $a,b,c,d$ 
appearing on $G[C]$ in this order. Then, $G-C$ has exactly two connected 
components $G_1$ and $G_2$; and for each $i\in\{1,2\}$, the marked graph 
$G'_i$ obtained from $G[V(G_i)\cup C]$ by adding edge $\{a,c\}$ and marking 
edges $\{a,b\}$, $\{b,c\}$, $\{c,d\}$, $\{d,a\}$, $\{a,c\}$ has a well-formed 
atlas. Moreover, given a well-formed atlas of $G'_1$ and one of $G'_2$, 
we can easily construct one of $G$. 
\end{lemma}

\begin{proof}
Since $G[C]$ is a cycle and ${\cal M}$ is well-formed, there are
exactly two holes ${\cal H}_1$ and ${\cal H}_2$ in ${\cal M}|_C$. 
For $j\in\{1,2\}$, let $U_j$ be
the set of nations that occupy ${\cal H}_j$ in atlas ${\cal M}$. Clearly, 
the nations in $U_j$ are connected together in $G-C$. By this and the 
assumption that $G-C$ is disconnected, both $G[U_1]$ and $G[U_2]$ are 
connected components of $G-C$ and $G-C$ has no other connected component. 
So, ${\cal H}_1$ and ${\cal H}_2$ must be disjoint. In turn, for each edge 
$\{u,v\}$ in $G[C]$, there is a $(u,v)$-segment in ${\cal M}$. 

For each $j\in\{1,2\}$, there is a unique hole in ${\cal M}|_{U_j\cup C}$ 
and it is (strongly) touched only by the nations of $C$. So, modifying 
${\cal M}|_{U_j\cup C}$ by extending nation $a$ to cover its unique hole 
yields a well-formed atlas of $G'_j$ in which nations $a$, $b$, and 
$c$ meet at a 3-point and nations $a$, $c$, and $d$ meet at a 3-point. 
So, by Lemma~\ref{lem:3-cut}, 
both $G'_j-\{a,b,c\}$ and $G'_j-\{a,c,d\}$ are connected.

\micsaid{pass32: this figure could be more symmetric.}
\zzsaid{pass33: I do not see the necessity.}
\fig[figure502]{adv2}{Layout ${\cal M}_j|_C$.}

Suppose that we are given an atlas ${\cal M}_j$ for each $G'_j$. 
Since $G'_j-\{a,b,c\}$ is connected and the three 
edges $\{a,b\}$, $\{a,c\}$, and $\{c,b\}$ are marked in $G'_j$, nations $a$, 
$b$, and $c$ meet at a 3-point in ${\cal M}_j$, by Lemma~\ref{lem:3-cut}.
Similarly, nations $a$, $c$, and $d$ must meet at a 3-point in ${\cal M}_j$.
Thus, by the 
well-formedness of ${\cal M}_j$, Figure~\ref{fig:adv2} displays ${\cal 
M}_j|_C$. By the figure, we can cut off a small area inside ${\cal M}_j$ 
around the $(a,c)$-segment; the nations in the other atlas can be embedded 
in the resulting open area. 
\end{proof}

\subsection{Separating Triples}

\begin{lemma}\label{noStriple}
Suppose $G$ has no 
separating edge but has a separating triple $\langle a, b, c\rangle$. Then, 
$G-\{a,b,c\}-{\cal E}[a,b]$ has exactly two connected components $G_1$ and 
$G_2$. Moreover, 
$\langle a, u, b, v\rangle$ is a correct 4-pizza, where $\{u\} = 
V(G_1)\cap N_G(V(G_2))$ and $\{v\} = V(G_2)\cap N_G(V(G_1))$. 
\end{lemma}

\fig[figure503]{adv3}{A display of $G[\{a,b,u,v\}]$.}

\begin{proof}
Let $C=\{a,b,c\}$ and $G'=G-C-{\cal E}[a,b]$.
Since $G$ is 4-connected, $G-C$ is connected.  
So ${\cal E}[a,b]$ is  non-empty to disconnect $G'$,
and we may choose $\{u,v\} \in {\cal E}[a,b]$ such that $u$ and $v$
belong to different components of $G'$.
By definition of ${\cal E}[a,b]$, $\{a, b, u, v\}$ is an MC$_4$ in $G$.

We claim that nations $u$ and $v$ do not strongly touch in ${\cal M}$. 
Assume, on the contrary, that a $(u,v)$-segment $S$ exists in ${\cal M}$. 
Since ${\cal M}$ has no hole, there are nations $w_1$, $w_2$ in $V-\{u,v\}$ 
such that $w_1$ touches one endpoint of $S$ and $w_2$ touches the
other.  If $w_1$ were not in $\{a, b\}$, then $w_1$ would connect
$u$ and $v$ in $G'$ in $G'$ by Fact~\ref{fac:Eabcde}.
Thus $w_1\in \{a,b\}$, and similarly $w_2\in \{a,b\}$. 
By the well-formedness of ${\cal M}$ and the fact that $|V| > 3$, we can 
verify that there is no way for nation $a$ or $b$ to touch  both endpoints of 
$S$. So, nation $a$ touches one endpoint of $S$ and $b$ touches the other; 
Figure~\ref{fig:adv3} displays ${\cal M}|_{\{a,b,u,v\}}$. 
By Figure~\ref{fig:adv3} and the fact that edge $\{u, v\}$ is not marked 
in $G$, $\{a,b\}$ is a separating edge of $G$, a contradiction. 
Thus, the claim holds. 

By the claim, nations $u$ and $v$ weakly touch at a point $p$ in ${\cal M}$. 
Since ${\cal M}$ has no hole, there are two distinct nations $w_1$, $w_2$ 
in $V-\{u,v\}$ which meet at $p$ in ${\cal M}$. As before, we can 
show that $\{w_1, w_2\} = \{a,b\}$. Thus, the four nations a, $u$, b, $v$ 
appear around $p$ cyclically in this order in ${\cal M}$. In turn, by the 
well-formedness of ${\cal M}$, $\langle a, u, b, v\rangle$ is a correct 
4-pizza of $G$. 

The discussions above actually prove that for every pair of adjacent nations 
$x$ and $y$ of $G$ that belong to different connected components of $G'$, 
nations $a$, $x$, $b$, $y$ must meet at a 4-point cyclically in 
this order in ${\cal M}$. Since $p$ is the unique point in ${\cal M}$ at which 
$a$ and $b$ meet, $(u, v)$ is the unique pair of adjacent nations of $G$ that 
belong to different connected components of $G'$. We now claim that 
the connected components of $G'$ are only $G_1$ and $G_2$. Assume, on the 
contrary, that $G'$ has a connected component $G_3$ other than $G_1$ and 
$G_2$. Then, there exists a nation $w \in V-(C\cup V(G_3))$ which touches 
some nation $w'$ of $G_3$ in ${\cal M}$; otherwise, $G_3$ would be a connected 
component of $G-C$, a contradiction. But now, $(w, w')$ is a pair of adjacent 
nations of $G$ that belong to different connected components of $G'$, a 
contradiction. Thus, the connected components of $G'$ are only $G_1$ and $G_2$.
We may assume that $u \in V(G_1)$ and $v \in V(G_2)$. $G_1$ and $G_2$ only 
touch at $p$; otherwise, nations $a$ and $b$ would have to meet at a point 
other than $p$ in ${\cal M}$, a contradiction against the well-formedness of 
${\cal M}$. Hence, $\{u\} = V(G_1) \cap N_G(V(G_2))$ and $\{v\} = V(G_2) \cap 
N_G(V(G_1))$. 
\end{proof}

\subsection{Separating Quadruples}

\micsaid{Next lemma is OK, but referees might object here.}
Using Lemma~\ref{noS4-cycle}, 
we can modify the proof of Lemma~\ref{noStriple} to prove the following: 
\begin{lemma}\label{noSquadruple}
Suppose that $G$ has neither 
separating edge nor separating 4-cycle, but has a separating quadruple 
$\langle a,b,c,d\rangle$. Then, $G - \{a,b,c,d\}-{\cal E}[a,b]$ has 
exactly two connected components $G_1$ and $G_2$. Moreover, 
$\langle a, u, b, v\rangle$ is 
a correct 4-pizza, where $\{u\}=V(G_1)\cap N_G(V(G_2))$ and
$\{v\}=V(G_2)\cap N_G(V(G_1))$.
\end{lemma}

\begin{corollary}\label{cor:noSquadruple}
Suppose that $G$ does not have 
an MC$_5$ or a separating edge. Then, $G$ has a separating quadruple 
if and only if
for some induced 4-cycle $C$ in $G$, at most one pair of adjacent nations 
of $C$ weakly touch in ${\cal M}$.
\end{corollary}

\begin{proof}
The ``if" part is obvious.  For the ``only if" part, suppose $G$ 
has a separating quadruple $Q$. If $G$ has a separating 4-cycle $C$, 
then as observed in the proof of Lemma~\ref{noS4-cycle}, each pair of 
adjacent nations of $C$ strongly touch in ${\cal M}$. Otherwise, as observed 
in the modified proof of Lemma~\ref{noStriple} for Lemma~\ref{noSquadruple}, 
exactly one pair of adjacent nations of $Q$ weakly touch in ${\cal M}$. 
\end{proof}

\subsection{Separating Triangles}

\micsaid{pass22: Dropped mention of Corollary~\ref{cor:noStriangle},
and added text there instead about the two dropped assumptions.}
The previous results of this section allow our algorithm to simplify
$G$ whenever it contains a separating edge, triple, or quadruple; so, 
for the rest of this paper we assume:
\begin{assume}\label{ass:temp}
$G$ does not have a separating edge, triple, or quadruple.
\end{assume}


Suppose $G$ has a separating triangle $\langle a,b,c\rangle$. Let $C$ and 
$G'$ be as described in Definition~\ref{def:separators}(\ref{def:septria}). 
Our goal is to show that using $C$ and $G'$, our algorithm can proceed by 
finding correct 4-pizzas.  We begin with three preliminary claims.

\begin{claim}\label{aInAll}
If $\{u,v\}$ is an edge in $G-C$ but not in $G'$, then
$a\in N_G(u)\cap N_G(v)$.
Also, nations $u$, $v$, $b$, and $c$ cannot meet at a 4-point in
${\cal M}$.
\end{claim}

\begin{proof}
Since $\{u,v\}\in {\cal E}[a,b] \cup {\cal E}[a,c]$, either
$\{a,b,u,v\}$ or $\{a,c,u,v\}$ is an MC$_4$ of $G$.  In both cases,
$a\in N_G(u)\cap N_G(v)$.  For the last part, such a 4-point would
imply a 5-clique containing the MC$_4$, contradicting its maximality.
\end{proof}

\begin{claim}\label{CInAll}
For every connected component $K$ of $G'$, 
(i) $C \subseteq N_G(V(K))$ and
(ii) $G'$ has another connected component $J$
such that $V(K) \cap N_G(V(J)) \not= \emptyset$. 
\end{claim}

\begin{proof}
For (i), let $S=C \cap N_G(V(K))$.
  Since $G-C$ is connected, some edge in ${\cal E}[a,b] \cup {\cal E}[a,c]$
connects $K$ to an outside vertex, so to support the corresponding MC$_4$,
$S$ must contain either $\{a,b\}$ or $\{a,c\}$.
  If $|S|=2$, then $S$ would be a separating edge of $G$,
separating $K$ from the rest.
  So, $S=C$.

  For (ii), if on the contrary $V(K) \cap N_G(V(J)) = \emptyset$ for
every $J$, then $K$ would be a component of $G-C$, contradicting
Assumption~\ref{ass:4-con}.
\end{proof}

\begin{claim}\label{3point}
Let $Z$ be a subset of $V-C$. 
Suppose that a subset $\{u,v,w\}$ of $Z$ is a triangle of $G$ such that 
$u$ and $v$ belong to different connected components of $G'[Z]$. 
Then, the following hold:
\zzsaid{pass2: Used the ``enumerate" environment.}
\begin{enumerate}
\item Either (i) $C\subseteq N_G(u)$ and 
$\{C \cap N_G(v), C\cap N_G(w)\} = \{\{a,b\}, \{a,c\}\}$ or (ii) $C \subseteq 
N_G(v)$ and $\{C\cap N_G(u), C\cap N_G(w)\}=\{\{a,b\}, \{a,c\}\}$. 
\item There is no $x\in Z-\{u,v,w\}$ with $\{u,v,w\} \subseteq N_G(x)$.
\end{enumerate}
\end{claim}
Note that we typically use this claim with $Z=V-C$.

\micsaid{It seems the previous claim could be split into two.  In
particular the last part is essentially about what can happen to a
$K_4$ in $G-C$ (in particular, it stays connected).  Or in other
words, if a pair from a 4-clique is separated in $G'$, then it
intersects $C$ (in fact at $a$).}
\zzsaid{pass5: Not necessary, because I modified the rest of the paper
in such a way that whenever this claim is cited, I explicitly mention
which of the two statements is related.}

\begin{proof}
Since $u$ and $v$ are disconnected in $G'[Z]$, at least two of the
triangle edges are removed.  Claim~\ref{aInAll} applied to these
edges implies $\{u,v,w\} \subseteq N_G(a)$.
On the other hand, by Fact~\ref{fac:Eabcde} at most one triangle edge
is in each of ${\cal E}[a,b]$ and ${\cal E}[a,c]$, so in fact exactly
two edges are removed, and either edge $\{u,w\}$ or $\{v,w\}$ remains
in $G'$.

We suppose $\{v,w\}$ remains, the other case is similar (swap $u$ and $v$).
We also suppose $\{u,v\} \in {\cal E}[a,b]$ and $\{u,w\} \in {\cal E}[a,c]$,
the other case is similar (swap $b$ and $c$).
Then $\{a,b,c\} \subseteq N_G(u)$, $\{a,b\} \subseteq N_G(v)$, and
$\{a,c\} \subseteq N_G(w)$.  On the other hand, $G$ cannot have the
edges $\{v,c\}$ or $\{w,b\}$, since either would imply a 5-clique
containing an MC$_4$.  So, the first assertion holds.

For the second assertion, suppose on the contrary there is an $x\in
Z-\{u,v,w\}$ with $\{u,v,w\} \subseteq N_G(x)$. As above, we suppose
that both $\{a,b,u,v\}$ and $\{a,c,u,w\}$ are MC$_4$'s of $G$.  Then
neither $\{a,b\}$ nor $\{a,c\}$ is a subset of $N_G(x)$, since
otherwise $x$ would extend one of these MC$_4$'s to a 5-clique.  But
then the edges from $x$ would all survive in $G'[Z]$, contradicting the
disconnection of $u$ and $v$.
\end{proof}

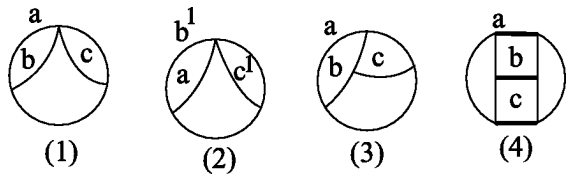
\fig[figure504]{adv4}{Possible displays of a separating triangle
 $\langle a,b,c \rangle$.}

Note that ${\cal M}|_C$ can have at most two holes. If ${\cal M}|_C$ has only 
one hole, then Figure~\ref{fig:adv4}(1), (2), or (3) displays it; otherwise, 
Figure~\ref{fig:adv4}(4) displays it.  
In the next three lemmas, we will show that in fact only 
Figure~\ref{fig:adv4}(4) is possible.

\begin{lemma}\label{not1}
Figure~\ref{fig:adv4}(1) does not display ${\cal M}|_C$. 
\end{lemma}

\begin{proof}
Assume, on the contrary, that Figure~\ref{fig:adv4}(1) 
displays ${\cal M}|_C$. Let $p$ be the point in ${\cal M}|_C$ at which nations 
$a$, $b$, and $c$ meet. Let $p_{a,b}$ (respectively, $p_{a,c}$) be the endpoint
 of the $(a,b)$-segment (respectively, $(a,c)$-segment) other than $p$ in 
${\cal M}$. There must exist a nation $d \in V-C$ which touches $p$ in ${\cal 
M}$. By the well-formedness of ${\cal M}$, nation $d$ touches $a$ only at $p$ 
and $\{a,d\}$ is not a marked edge in $G$. 
 Let $G'_d$ be the connected component of $G'$ containing $d$. 
 Let $K$ be a connected component of $G'$ 
other than $G'_d$ such that some nation $u$ of $G'_d$ touches some nation $v$ 
of $K$ in ${\cal M}$; $K$ exists by Claim~\ref{CInAll}.

\fig[figure505]{adv5}{Possible displays of $G[\{a,b,c,d,v,w\}]$.}

 We claim that nation $a$ touches some nation of $G'_d-\{d\}$ in ${\cal
M}$. Assume, on the contrary, that the claim is false.
 Clearly, $\{a,b,u,v\}$ or $\{a,c,u,v\}$ is an MC$_4$ of $G$. 
 Since no nation of $G'_d - \{d\}$ touches $a$ in ${\cal M}$, 
$u=d$. That is, $\{a,b,d,v\}$ or $\{a,c,d,v\}$ is an MC$_4$ of $G$. 
Since $\{a,b,c\}\subseteq N_G(d)$, we have $|N_G(v) \cap \{b,c\}| = 1$; 
otherwise, $\{a,b,c,d,v\}$ would be a 5-clique of $G$. We assume that $N_G(v) 
\cap \{b,c\} = \{b\}$; the other case is similar. Then, since nation $v$ 
cannot touch nation $c$ in ${\cal M}$ and ${\cal M}$ has no hole, there is a 
point in ${\cal M}$ at which nations $v$, $d$ and some $w\in V-\{a,b,c,d,v\}$ 
meet.
  By Claim~\ref{3point}, 
\micsaid {with $Z=V-C$}
$C\cap N_G(w)=\{a,c\}$ and there is 
no $x\in V-\{a,b,c,d,v,w\}$ such that $\{d,v,w\}\subseteq N_G(x)$. Now, we see 
that Figure~\ref{fig:adv5}(1) displays ${\cal M}|_{\{a,b,c,d,v,w\}}$. 
There is no $x\in V-\{a,b,c,d, v,w\}$ with $\{d,v\}\subseteq N_G(x)$; 
otherwise, $C\cap N_G(x)=\{a,c\}$ by Claim~\ref{3point}(1), which is impossible 
by Figure~\ref{fig:adv5}(1). Similarly, there is no $x\in V-\{a,b,c,d,v,w\}$ 
with $\{d,w\}\subseteq N_G(x)$. Thus, Figure~\ref{fig:adv5}(1) is 
transformable to Figure~\ref{fig:adv5}(2). By Figure~\ref{fig:adv5}(2) and 
the fact that $\{d,a\}$ is not a marked edge in $G$, $\langle b,c,w,v\rangle$ 
is a separating quadruple of $G$, a contradiction. So, the claim holds. 

\fig[figure506]{adv6}{Possible displays of $G[\{a,b,c,x,y,z\}]$.}

Next, we claim that for every connected component $K'$ of $G'$, there is no 
point $q$ in ${\cal M}$ at which two nations $x$ and $y$ of $K'$ together with 
two nations $w$ and $z$ of $V-V(K')$ meet cyclically in the order $x$, $w$, 
$y$, $z$. Assume, on the contrary, that such $q$ exists in ${\cal M}$. 
Then, by Claim~\ref{3point}(2), $C\cap \{w, z\} \not= \emptyset$. 
\micsaid{Argument that $C\cap\{w,z\}=\{a\}$ could be worked into
 Claim~\ref{3point}.}
\zzsaid{pass5: Not a big deal.}
By Figure~\ref{fig:adv4}(1), $q\not\in\{p, p_{a,b}, p_{a,c}\}$ and hence 
$|C\cap \{w,z\}|\leq 1$. So, $|C\cap \{w,z\}| = 1$. In turn, $C\cap \{w,z\} = 
\{a\}$; otherwise, by Claim~\ref{aInAll}, $\{x,y,a,w,z\}$ would be a 5-clique 
of $G$, a contradiction. 
We assume that $w=a$; the other case is similar. Now, by Claim~\ref{3point}(1), 
$\{C\cap N_G(x), C\cap N_G(y)\}=\{\{a,b\}, \{a,c\}\}$ and $C\subseteq N_G(z)$. 
We assume that $C\cap N_G(x)=\{a,b\}$ and $C\cap N_G(y)=\{a,c\}$; the other 
case is similar. In summary, Figure~\ref{fig:adv6}(1) displays 
${\cal M}|_{\{a,b,c,x, y, z\}}$. 
There is no $f\in V-\{a,b,c,x,y,z\}$ with $\{x,z\}\subseteq N_G(f)$; 
otherwise, by Claim~\ref{3point}(1), $C\cap N_G(f)=\{a,c\}$ which is impossible 
by Figure~\ref{fig:adv6}(1). Similarly, there is no $f\in V - \{
a,b,c,x,y,z\}$ with $\{y,z\}\subseteq N_G(f)$. So, Figure~\ref{fig:adv6}(1) 
is transformable to Figure~\ref{fig:adv6}(2). By the latter figure, 
$\langle a, z, b\rangle$ would be a separating triple of $G$, 
a contradiction. So, the claim holds. 

\fig[figure507]{adv7}{Possible displays of $G[\{a,b,c,u,v,w\}]$.}

By the above two claims and the fact (Claim~\ref{CInAll})
that $\{a,b,c\}\subseteq N_G(V(K))$, 
$p_{a,b}$ or $p_{a,c}$ is touched by both $G'_d$ and $K$ in ${\cal M}$. 
Suppose that $p_{a,b}$ is touched by both $G'_d$ and $K$; the other case 
is similar. Let $u$ (respectively, $v$) be the nation of $G'_d$ 
(respectively, $K$) touching $p_{a,b}$. Then, the boundaries of nations $u$ 
and $v$ in ${\cal M}$ share a curve segment $S$. One endpoint of $S$ is 
$p_{a,b}$. Let $q$ be the other endpoint of $S$. Neither nation $d$ nor $c$ 
touches $q$ in ${\cal M}$; otherwise, $\{u,v,a,b,d\}$ or $\{u,v,a,b,c\}$ 
would be a 5-clique of $G$. By the well-formedness of ${\cal M}$, 
it is impossible that nation $a$ or $b$ touches $q$.
 In turn, since 
${\cal M}$ has no hole, there is a nation $w \in V - \{a,b,c,d,u,v\}$ that 
touches $q$ in ${\cal M}$. Now, by Claim~\ref{3point}(1), 
$C\cap N_G(w)=\{a,c\}$ and either (i) $C \subseteq N_G(v)$ and 
$C\cap N_G(u)=\{a,b\}$ or (ii) $C\subseteq N_G(u)$ and 
$C\cap N_G(v)=\{a,b\}$. In case (i) holds, Figure~\ref{fig:adv7}(1) displays 
${\cal M}|_{\{a,b,c,u,v,w\}}$ and $\langle b, c, w, u\rangle$ would be a 
separating quadruple, a contradiction. So, (ii) holds and only 
Figure~\ref{fig:adv7}(2) can possibly display ${\cal M}|_{\{a,b,c,u,v,w\}}$. 
\micsaid{At this point, isn't $\langle a, w, u, b\rangle$ a separating
quadruple?} \zzsaid{pass2: No!} 
There is no $f\in V-\{a,b,c,u,v,w\}$ with $\{u,w\}\subseteq N_G(f)$; 
otherwise, by Claim~\ref{3point}(1), $C\cap N_G(f)=\{a,b\}$ which is impossible 
by Figure~\ref{fig:adv7}(2). By this, Figure~\ref{fig:adv7}(2) is 
transformable to Figure~\ref{fig:adv7}(3), by which $\langle a,u, c\rangle$ 
would be a separating triple of $G$, a contradiction. This completes the proof.
\end{proof}

\begin{lemma}\label{not2} 
Figure~\ref{fig:adv4}(2) does not display ${\cal M}|_C$. 
\end{lemma}

\begin{proof}
Assume, on the contrary, that Figure~\ref{fig:adv4}(2) 
displays ${\cal M}|_C$. We assume that $\langle b^1, c^1\rangle = \langle b ,c 
\rangle$ in the figure; the other case is similar. Define points $p$ and 
$p_{a,b}$, nation $d$ and $G'_d$ as in the proof of Lemma~\ref{not1}. By the 
well-formedness of ${\cal M}$, nation $d$ meets $b$ only at $p$ and $\{b,d\}$ 
is not a marked edge in $G$. Let $p_{b,c}$ be the endpoint of 
the $(b,c)$-segment other than $p$ in ${\cal M}$. 

\fig[figure508]{adv8}{Possible displays of $G[\{a,b,c,d,v,w\}]$.}

We claim that $G'_d-\{d\}$ touches nation $b$ in ${\cal M}$. Assume, on the 
contrary, that $G'_d-\{d\}$ does not touch nation $b$. Let $K$ be a connected 
component of $G'$ other than $G'_d$ such that some nation $u$ of $G'_d$ touches
some nation $v$ of $K$ in ${\cal M}$. By Claim~\ref{CInAll}, such $K$ exists.
Clearly, $\{a,b,u,v\}$ or $\{a,c,u,v\}$ is an MC$_4$ of $G$. 

\case{1} $u \not= d$.   Then $C \cap N_G(u)=\{a,c\}$
and $\{a,c,u,v\}$ is an MC$_4$ of $G$. Moreover, there is 
no $w \in V - \{a,b,c,u,v\}$ with $\{u,v\}\subseteq N_G(w)$; otherwise, since 
$C\cap N_G(u)=\{a,c\}$, we would have $C\subseteq N_G(v)$ and $C\cap N_G(w)=\{a,b\}$
by Claim~\ref{3point}(1), and in turn $w$ would be a vertex of $G'_d-\{d\}$ that 
touches nation $b$ in ${\cal M}$, a contradiction. So, by the fact that 
${\cal M}$ has no hole, the boundaries of nations $u$ and $v$ 
share a curve segment $S$ in ${\cal M}$, and both endpoints of $S$ are 3-points
one of which is touched by $a$ and the other is touched by $c$ in ${\cal M}$. 
By this, $S$ is a shrinkable segment in ${\cal M}$, $u$ and $v$ fall into 
different connected components of $G-\{a,c\}-{\cal E}[a,c]$, and $\{a,c\}$ would 
be a separating edge of $G$, a contradiction.

\case{2} $u=d$. Then 
$\{a,b,d,v\}$ or $\{a,c,d,v\}$ is an MC$_4$ of $G$. Since $\{a,b,c\}\subseteq 
N_G(d)$, we have $|N_G(v) \cap \{b,c\}| = 1$; otherwise, $\{a,b,c,d,v\}$ would 
be a 5-clique of $G$.  So we have two sub-cases.

\case{2.1} $N_G(v)\cap\{b,c\}=\{b\}$.
Then $C\cap N_G(v)=\{a,b\}$ and $\{a,b,d,v\}$ is an MC$_4$ of $G$. 
Moreover, since nation $v$ cannot touch nation $c$ in ${\cal M}$ and ${\cal 
M}$ has no hole, there is a point in ${\cal M}$ at which nations $v$, $d$ and 
some $w\in V-\{a,b,c,d,v\}$ meet. By Claim~\ref{3point}(1), 
$C\cap N_G(w)=\{a,c\}$ and there is no $x\in V-\{a,b,c,d,v,w\}$ such that 
$\{d,v,w\}\subseteq N_G(x)$. Now, we see that Figure~\ref{fig:adv8}(1) 
displays ${\cal M}|_{\{a,b,c,d,v,w\}}$. 
There is no $x \in V-\{a,b,c,d,v,w\}$ with $\{d,w\}\subseteq N_G(x)$; 
otherwise, $C\cap N_G(x)=\{a,b\}$ by Claim~\ref{3point}(1), 
which is impossible by Figure~\ref{fig:adv8}(1). Thus, 
Figure~\ref{fig:adv8}(1) is transformable to Figure~\ref{fig:adv8}(2). 
By Figure~\ref{fig:adv8}(2), $\langle b, v, w, c \rangle$ is a 
separating quadruple of $G$, a contradiction. 

\case{2.2} $N_G(v) \cap \{b,c\}=\{c\}$.
  If there is a 
$w\in V-\{a,b,c,d,v\}$ with $\{d,v\}\subseteq N_G(w)$, then similarly 
to Case 2.1, we can prove that $\langle b, 
w, v, c\rangle$ would be a separating quadruple of $G$, 
a contradiction. Otherwise, the boundaries of nations $d$ and $v$ share 
a curve segment $S$ in ${\cal M}$, and both endpoints of $S$ are 3-points 
one of which is touched by $a$ and the other is touched by $c$ in ${\cal M}$; 
by this, $S$ is a shrinkable segment in ${\cal M}$, $d$ and $v$ fall into 
different connected components of $G-\{a,c\}-{\cal E}[a,c]$, and $\{a,c\}$ 
would be a separating edge of $G$, a contradiction. 

Therefore, the claim holds: $G'_d-\{d\}$ touches $b$.

\fig[figure509]{adv9}{Possible displays of $G[\{a,b,c,x,y,z\}]$.}


Next, we claim that for every connected component $K'$ of $G'$, there is no 
point $q$ in ${\cal M}$ at which two nations $x$ and $y$ of $K'$ together with 
two nations $w$ and $z$ of $V-V(K')$ meet cyclically in the order $x$, $w$, 
$y$, $z$. Assume, on the contrary, that such $q$ exists in ${\cal M}$. 
Then, by Claim~\ref{3point}(2), $C\cap \{w, z\} \not= \emptyset$. 
By Figure~\ref{fig:adv4}(2), $q\not\in\{p, p_{a,b}, p_{b,c}\}$ and 
hence $|C\cap \{w,z\}|\leq 1$. So, $|C\cap \{w,z\}| = 1$. In turn, 
$C\cap \{w,z\}=\{a\}$; otherwise, by Claim~\ref{aInAll}, 
$\{x,y,a,w,z\}$ would be a 5-clique of $G$, a contradiction. 
We assume that $w=a$; the other case is similar. Now, by Claim~\ref{3point}(1), 
$\{C\cap N_G(x), C\cap N_G(y)\}=\{\{a,b\}, \{a,c\}\}$ and $C\subseteq N_G(z)$. 
We assume that $C\cap N_G(x)=\{a,b\}$ and $C\cap N_G(y)=\{a,c\}$; the other 
case is similar. In summary, Figure~\ref{fig:adv9}(1) displays 
${\cal M}|_{\{a,b,c,x, y, z\}}$. 
There is no $f\in V-\{a,b,c,x,y,z\}$ with $\{x,z\}\subseteq N_G(f)$; 
otherwise, by Claim~\ref{3point}(1), $C\cap N_G(f)=\{a,c\}$ which is impossible 
by Figure~\ref{fig:adv9}(1). Similarly, there is no $f\in V - \{a, b, c, x, 
y, z\}$ with $\{y,z\}\subseteq N_G(f)$. So, Figure~\ref{fig:adv9}(1) 
is transformable to Figure~\ref{fig:adv9}(2). By the latter figure, 
$\langle a, z, b\rangle$ would be a separating triple of $G$, a contradiction. 
So, the claim holds. 

\fig[figure510]{adv10}{Possible displays of $G[\{a,b,c,u,v,w\}]$.}

By the above two claims and the fact that $\{a,b,c\}\subseteq N_G(V(K))$, 
$p_{a,b}$ or $p_{b,c}$ is touched by both $G'_d$ and $K$ in ${\cal M}$. 
By Claim~\ref{aInAll}, $p_{b,c}$ cannot be touched by both $G'_d$ and 
$K$. So, $p_{a,b}$ is touched by both $G'_d$ and $K$. Let $u$ (respectively, 
$v$) be the nation of $G'_d$ (respectively, $K$) touching $p_{a,b}$. 
Then, the boundaries of nations $u$ and $v$ in ${\cal M}$ share a curve 
segment $S$. One endpoint of $S$ is $p_{a,b}$. Let $q$ be the other endpoint 
of $S$. Neither nation $d$ nor $c$ touches $q$ in ${\cal M}$; otherwise, 
$\{u,v,a,b,d\}$ or $\{u,v,a,b,c\}$ would be a 5-clique of $G$. 
By the well-formedness of ${\cal M}$, it is impossible that 
nation $a$ or $b$ touches $q$. 
So, there is a nation 
$w \in V - \{a,b,c,d,u,v\}$ that touches $q$ in ${\cal M}$. Now, by 
Claim~\ref{3point}(1), $C\cap N_G(w)=\{a,c\}$ and either (i) $C \subseteq 
N_G(u)$ and $C\cap N_G(v)=\{a,b\}$ or (ii) $C\subseteq N_G(v)$ and $C\cap 
N_G(u)=\{a,b\}$. In case (i) holds, Figure~\ref{fig:adv10}(1) displays 
${\cal M}|_{\{a,b,c,u,v\}}$; by the figure, it is impossible for nation 
$w$ to touch all of nations $v$, $a$, and $c$ in ${\cal M}$, a contradiction. 
So, only Figure~\ref{fig:adv10}(2) can possibly display ${\cal M}|_{\{a,b,c,u,
v,w\}}$. There is no $f\in V-\{a,b,c,u,v,w\}$ with $\{v,w\}\subseteq N_G(f)$; 
otherwise, by Claim~\ref{3point}(1), $C \cap N_G(f) = \{a,b\}$ which is impossible
by Figure~\ref{fig:adv10}(2). By this, Figure~\ref{fig:adv10}(2) is 
transformable to Figure~\ref{fig:adv10}(3), by which $\langle b,u,w,c\rangle$ 
would be a separating quadruple of $G$, a contradiction. 
This completes the proof.
\end{proof}

\begin{lemma}\label{not3} 
Figure~\ref{fig:adv4}(3) does not display ${\cal M}|_C$. 
\end{lemma}

\begin{proof}
Assume, on the contrary, that Figure~\ref{fig:adv4}(3) 
displays ${\cal M}|_C$. Define points $p$, $p_{a,b}$ and $p_{a,c}$ as in the 
proof of Lemma~\ref{not1}. Let $p_{b,c}$ be the endpoint of the $(b,c)$-segment
 other than $p$ in ${\cal M}$. Similarly to the proof of Lemma~\ref{not2},
 we can prove that for every connected component $K'$ of $G'$, there is no 
point $q$ in ${\cal M}$ at which two nations $x$ and $y$ of $K'$ together with 
two nations $w$ and $z$ of $V-V(K')$ meet cyclically in the order $x$, $w$, 
$y$, $z$. 

\fig[figure511]{adv11}{Possible displays of $G[\{a,b,c,u,v,w\}]$.}

Let $G'_1$ and $G'_2$ be two connected components of $G'$ such that $V(G'_1) 
\cap N_G(V(G'_2)) \not= \emptyset$. By the above claim, Claim~\ref{aInAll}, 
and Claim~\ref{CInAll}, $p_{a,b}$ or $p_{a,c}$ is touched by both 
$G'_1$ and $G'_2$ in ${\cal M}$. We assume that $p_{a,b}$ is touched by both 
$G'_1$ and $G'_2$ in ${\cal M}$; the other case is similar. 
Let $u$ (respectively, $v$) be the nation of $G'_1$ (respectively, $G'_2$) 
touching $p_{a,b}$. Similarly to the proof of Lemma~\ref{not1}, 
we can prove that there is a nation $w\in V - \{a, b, c, u, v\}$ such that 
only Figure~\ref{fig:adv11}(1) or (2) can possibly displays 
${\cal M}|_{\{a,b,c,u,v,w\}}$. If Figure~\ref{fig:adv11}(1) displays it, 
then $\langle b, u, w, c\rangle$ would be a separating quadruple 
(indeed, a separating 4-cycle) of $G$, a contradiction. So, suppose that 
Figure~\ref{fig:adv11}(2) displays it. Then, there is no $f\in V-\{a,b,c,u,v,
w\}$ with $\{u,w\}\subseteq N_G(f)$; otherwise, by Claim~\ref{3point}, $C \cap 
N_G(f) = \{a,b\}$ which is impossible by Figure~\ref{fig:adv11}(2). By this, 
Figure~\ref{fig:adv11}(2) is transformable to Figure~\ref{fig:adv11}(3). 
By the latter figure, $\langle b,v,w,c\rangle$ would be a separating quadruple 
of $G$, a contradiction. This completes the proof.
\end{proof}

By Lemmas~\ref{not1}, \ref{not2} and \ref{not3}, 
only Figure~\ref{fig:adv4}(4) can display ${\cal M}|_C$. 

\begin{lemma}\label{noSStriangle} 
Suppose that $C=\langle a,b,c \rangle$ is a strongly separating
triangle of $G$.
Let $d$ be the vertex that constitutes a connected component of $G'$. Then, 
there are exactly two vertices $x,y\in V-\{a,b,c,d\}$ such that 
$\{a,b,d,x\}$ and $\{a,c,d,y\}$ are MC$_4$ of $G$. Moreover, both 
$\langle a,d,b,x\rangle$ and $\langle a,d,c,y\rangle$ are correct 4-pizzas. 
\end{lemma}

\begin{proof}
Let ${\cal H}_1$ be one hole of ${\cal M}|_C$, 
and ${\cal H}_2$ be the other. Let $Z_1$ (respectively, $Z_2$) be the set 
of nations in $V-C$ that occupy hole ${\cal H}_1$ (respectively, ${\cal H}_2$) 
in atlas ${\cal M}$. Let $p_{a,b}$ be the point at which nations $a$ and $b$ 
together with some nation of $Z_1$ meet in ${\cal M}$. 
Define points $p_{a,c}$ and $p_{b,c}$ similarly.

\micsaid{I think the old version of this paragraph was weak.
 Please read this new version!}
\zzsaid{pass2: I checked your proof, and found an essential error.
I corrected it.}

  First, we observe that $C \subseteq N_G(V(K))$ for every connected
component $K$ of $G'[Z_1]$.
  If $K=Z_1$, then this is clear from Figure~\ref{fig:adv4}(4).
  Otherwise $G'[Z_1]$ contains some other component $K'$ adjacent to
$K$ in $G[Z_1]$, and now our argument resembles that for
Claim~\ref{CInAll}(i).
  That is, let $S=C \cap N_G(V(K))$.  Since an edge between $K$ and
$K'$ is absent in $G'$, $S$ contains either $\{a,b\}$ or $\{a,c\}$.
  Assume $S=\{a,b\}$; the $\{a,c\}$ case is similar.
  Then, in case $K$ is also a connected component of $G'$, it is clear 
  that $\{a,b\}$ would be a separating edge in $G$, separating $K$ from 
$K'$. In case $K$ is not a connected component of $G'$, there is exactly 
one edge $\{x_1,x_2\}\in E$ with $x_1\in V(K)$ and $x_2\in Z_2$; 
moreover, the four nations $a,x_1,b,x_2$ must meet at point $p_{a,b}$ 
in atlas ${\cal M}$ cyclically in this order (so, the $(a,b)$-segment in 
the layout in Figure~\ref{fig:adv4}(4) should be contracted to a point).
If $\{a,x_1,b,x_2\}$ is an MC$_4$ of $G$, then $K$ would be a connected 
component of $G'$, a contradiction. Otherwise, there is a 5-clique $C'$ 
containing $a,x_1,b,x_2$. The nation $x_3\in C'-\{a,x_1,b,x_2\}$ must 
belong to $Z_1$ and touch nation $c$, in order to touch $x_1$ and $x_2$. 
By this, edge $\{x_1,x_2\}$ remains in $G'$, and $x_3\in Z_1$, contradicting 
the fact that $K$ is a connected component of $G'[Z_1]$. So, $S=C$.

Similarly, we have $C \subseteq N_G(V(K))$ for every connected
component $K$ of $G'[Z_2]$.

\fig[figure512]{adv12}{Possible displays of $G[\{a,b,c,d,x,y\}]$.}

We assume that $d \in Z_1$; the other case is similar. We want to prove that 
$Z_1=\{d\}$. Towards a contradiction, assume that $Z_1\not=\{d\}$. Then, since 
${\cal M}$ has no hole, there is a connected component $K$ of $G'[Z_1]$ with 
$V(K) \cap N_G(d) \not= \emptyset$. First, we claim that $d$ and a nation of 
$K$ must meet at $p_{a,b}$, $p_{a,c}$, or $p_{b,c}$. Assume, on the contrary, 
that the claim does not hold. Then, since $C \subseteq N_G(V(K))\cap N_G(d)$ 
by the above observation, there must exist a point $q$ in ${\cal M}$ at which 
two nations $x$ and $y$ of $K$ together with $d$ and some $u\in C$ meet 
cyclically in the order $x$, $d$, $y$, $u$. Claim~\ref{3point} ensures that 
either (i) $C\cap N_G(x)=\{a,b\}$ and $C \cap N_G(y)=\{a,c\}$ or 
       (ii) $C\cap N_G(x)=\{a,c\}$ and $C \cap N_G(y)=\{a,b\}$. 
In either case, we have $u=a$. We assume that (i) holds; the other case is 
similar. Then, Figure~\ref{fig:adv12}(1) displays ${\cal M}|_{\{a,b,c,d,x,y\} 
}$. There is no $u\in Z_1-\{d,x,y\}$ with $\{x,d\}\subseteq N_G(u)$; 
otherwise, by Claim~\ref{3point}, $C\cap N_G(u)=\{a,c\}$ which is impossible 
by Figure~\ref{fig:adv12}(1). Similarly, there is no $u\in Z_1 - \{d, x, y\}$ 
with $\{y,d\}\subseteq N_G(u)$. So, Figure~\ref{fig:adv12}(1) is
transformable to Figure~\ref{fig:adv12}(2). By the latter figure and
Claim~\ref{aInAll}, $d$ and each of $b$ and $c$ strongly touch in ${\cal M}$.
In turn, by the fact that $\langle b, c, d\rangle$ is not
a separating triple of $G$, nations $b$, $c$ and $d$ meet at a point
$q_{b,c,d}$ in ${\cal M}$. By Claim~\ref{aInAll}, $q_{b,c,d}$ must be a
3-point; so, $b$ and $c$ strongly touch in ${\cal M}$.
\zzsaid{pass8: Added Figure 5.12(3), and modified next sentences.}
In summary, Figure~\ref{fig:adv12}(2) is transformable to
Figure~\ref{fig:adv12}(3). By the latter, $\langle x, y, c, b \rangle$ would
be a separating quadruple of $G$, a contradiction. So, the claim holds:
$d$ meets $K$ at $p_{a,b}$, $p_{a,c}$, or $p_{b,c}$.

\fig[figure513]{adv13}{Possible displays of $G[\{a,b,c,d,u,v\}]$.}

Next, we use the above claim to get a contradiction. By the above claim, $d$ 
and a nation $u$ of $K$ must meet at $p_{a,b}$, $p_{a,c}$, or $p_{b,c}$ in 
${\cal M}$. By Claim~\ref{aInAll}, $d$ and $u$ cannot meet at $p_{b,c}$. So, 
they meet at $p_{a,b}$ or $p_{a,c}$. We assume that they meet at $p_{a,b}$; 
the other case is similar. Then, the boundaries of nations $d$ and $u$ in 
${\cal M}$ share a curve segment $S$. One endpoint of $S$ is $p_{a,b}$. Let 
$q$ be the other endpoint of $S$. Nation $c$ cannot touch $q$ in ${\cal M}$; 
otherwise, $\{a,b,c,d,u\}$ would be a 5-clique of $G$. By the well-formedness 
of ${\cal M}$, it is impossible that only nations $a$ and $b$ touch $q$. 
So, there is a nation $v \in V - \{a,b,c,d,u\}$ 
that touches $q$ in ${\cal M}$. Now, by Claim~\ref{3point}, $C\cap N_G(v) = 
\{a,c\}$. Thus, Figure~\ref{fig:adv13}(1) or (2) displays 
${\cal M}|_{\{a,b,c,d,u,v\}}$. Actually, the former does not display 
${\cal M}|_{\{a,b,c,d,u,v\}}$ or else $\langle b,c,a\rangle$ 
would be a separating triple of $G$, a contradiction.
So, only Figure~\ref{fig:adv13}(2) can possibly display ${\cal M}|_{\{a,b,c,d,
u,v\}}$. There is no $w\in Z_1-\{d,u,v\}$ with $\{d,v\}\subseteq N_G(w)$; 
otherwise, by Claim~\ref{3point}, $C \cap N_G(w) = \{a,b\}$ which is impossible
by Figure~\ref{fig:adv13}(2). By this, Figure~\ref{fig:adv13}(2) is 
transformable to Figure~\ref{fig:adv13}(3). 
\zzsaid{pass8: Added Figure 5.13(4), and rewrote next sentences.}
The latter is further transformable to Figure~\ref{fig:adv13}(4), because 
each pair of nations in $\{b, c, d\}$ must strongly touch in ${\cal M}$ by 
Claim~\ref{aInAll} and the fact that $\langle b, c, d\rangle$ is not 
a separating triple of $G$. By Figure~\ref{fig:adv13}(4), 
$\langle u, b, c, v \rangle$ would be a separating quadruple of $G$, 
a contradiction. This completes the proof that $Z_1=\{d\}$. 

\fig[figure514]{adv14}{An extensible layout of $G[\{a,b,c\}]$.}

Now, $Z_1=\{d\}$. \zzsaid{pass6: Changed the next sentence.}
Thus, by Claim~\ref{aInAll} and Assumption~\ref{ass:temp}
($G$ has no separating triple), Figure~\ref{fig:adv14} displays ${\cal M}|_C$.
By the figure and the fact that $d$ constitutes a connected component of $G'$,
there are exactly two distinct nations $x$ and $y$ of $Z_2$ such that 
$\{x,d\}\in E$ and $\{y,d\}\in E$. By the figure, both $
\langle a,d,b,x\rangle$ and $\langle a,d,c,y\rangle$ are correct 4-pizzas. 
Since $Z_2=V-\{a,b,c,d\}$, finding $x$ and $y$ is easy. 
This completes the proof of Lemma~\ref{noSStriangle}.
\end{proof}

\begin{lemma}\label{noStriangle} 
Suppose that there is no strongly separating triangle of $G$. Further 
assume that $C=\langle a,b,c\rangle$ is a separating triangle of $G$. 
Then, $G'$ has exactly two connected components $G_1$ and $G_2$, and 
there are exactly two edges $\{u,v\}, \{x,y\}\in E$ with $\{u,x\}\subseteq 
V(G_1)$ and $\{v,y\}\subseteq V(G_2)$. Moreover, $\{a,b,u,v\}$ and 
$\{a,c,x,y\}$ are MC$_4$'s of $G$, and both $\langle a,u,b,v\rangle$ and 
$\langle a,x,c,y\rangle$ are correct 4-pizzas. 
\end{lemma}

\begin{proof}
Define sets $Z_1$ and $Z_2$ and points $p_{a,b}$, 
$p_{a,c}$, and $p_{b,c}$ as in Lemma~\ref{noSStriangle}. As in the proof of 
Lemma~\ref{noSStriangle}, we observe that $C \subseteq N_G(V(K))$ 
for every connected component $K$ of $G'[Z_1]$ or $G'[Z_2]$.

\fig[figure515]{adv15}{Possible displays of $G[\{a,b,c,x,y,z\}]$.}

We claim that for every connected component $K$ of $G'[Z_1]$, there is no 
point $q$ in ${\cal M}$ at which two nations $x$ and $y$ of $K$ together with 
two nations $w$ and $z$ of $(C\cup Z_1)-V(K)$ meet cyclically in the order 
$x$, $w$, $y$, $z$. Assume, on the contrary, that such $q$ exists in ${\cal 
M}$. Then, by Claim~\ref{3point} with $Z=Z_1$,
$C\cap \{w, z\} \not= \emptyset$. 
By Figure~\ref{fig:adv4}(4), $q\not\in\{p_{a,b},p_{a,c},p_{b,c}\}$ and 
hence $|C\cap \{w,z\}|\leq 1$. So, 
$|C\cap \{w,z\}| = 1$. In turn, $C\cap \{w,z\}=\{a\}$; 
otherwise, by Claim~\ref{aInAll}, $\{x,y,a,w,z\}$ would be a 5-clique of $G$, 
a contradiction. We assume that $w=a$; the other case is similar. Now, by 
Claim~\ref{3point}, $\{C\cap N_G(x), C\cap N_G(y)\}=\{\{a,b\}, \{a,c\}\}$ and 
$C\subseteq N_G(z)$. We assume that $C\cap N_G(x)=\{a,b\}$ and $C\cap N_G(y)=
\{a,c\}$; the other case is similar. In summary, Figure~\ref{fig:adv15}(1) 
displays $G[\{a,b,c,x,y,z\}]$. There is no $f\in Z_1-\{x,y,z\}$ with $\{x,
z\}\subseteq N_G(f)$; otherwise, by Claim~\ref{3point}, $C\cap N_G(f)=\{a,c\}$ 
which is impossible by Figure~\ref{fig:adv15}(1). Similarly, there is no $f 
\in Z_1 - \{x, y, z\}$ with $\{y,z\}\subseteq N_G(f)$. So, 
Figure~\ref{fig:adv15}(1) is transformable to Figure~\ref{fig:adv15}(2). 
The latter figure is further transformable to Figure~\ref{fig:adv15}(3), 
because (i) $\langle a,b,x\rangle$ and $\langle a,c,y\rangle$ are not 
separating triples of $G$ and (ii) both $\{a,b,x,z\}$ and 
$\{a,c,y,z\}$ are MC$_4$'s of $G$. By Figure~\ref{fig:adv15}(3), 
$\langle a,b,z\rangle$ would be a strongly separating triangle of 
$G$, a contradiction. So, the claim holds. 

\fig[figure516]{adv16}{Possible displays of $G[\{a,b,c,u,v,w\}]$.}

Next, we claim that $G'[Z_1]$ is connected. Assume, on the contrary, that 
$G'[Z_1]$ is disconnected. Then, since ${\cal M}$ has no hole, there are 
two distinct connected components $K$ and $K'$ of $G'[Z_1]$ such that $V(K) 
\cap N_G(V(K')) \not= \emptyset$. Since $C \subseteq N_G(V(K))$ and 
$C \subseteq N_G(V(K'))$, some nation $u$ of $K$ and some nation $v$ of $K'$ 
have to meet at $p_{a,b}$, $p_{a,c}$ or $p_{b,c}$ in ${\cal M}$, by the 
claim of the previous paragraph and Figure~\ref{fig:adv4}(4).
By Claim~\ref{aInAll}, $u$ and $v$ cannot 
meet at $p_{b,c}$ in ${\cal M}$. We assume that $u$ and $v$ meet at $p_{a,b}$ 
in ${\cal M}$; the other case is similar. Similarly to the proof of 
Lemma~\ref{not1}, we can prove that there is a nation $w\in Z_1 - \{u, v\}$ 
such that only Figure~\ref{fig:adv16}(1) or (2) can possibly 
display ${\cal M}|_{\{a,b,c,u,v,w\}}$. Actually, Figure~\ref{fig:adv16}(1) 
does not display it or else $\langle b, c, a\rangle$ would be a 
separating triple of $G$. So, only Figure~\ref{fig:adv16}(2) can 
possibly display it. Since $\langle a, w, v\rangle$ is not a 
separating triple of $G$, Figure~\ref{fig:adv16}(2) is 
transformable to Figure~\ref{fig:adv16}(3). By the latter figure, $\langle 
a,w,u\rangle$ would be a strongly separating triangle of $G$, 
a contradiction. So, the claim holds. Similarly, we can prove that 
$G'[Z_2]$ is connected. 

Since both $G'[Z_1]$ and $G'[Z_2]$ are connected, both are connected components
of $G'$ and $G'$ has no other connected component. So, by Claim~\ref{aInAll}, 
the figure obtained from Figure~\ref{fig:adv4}(4) by contracting the 
bold $(b,c)$-segment to a single point does not display ${\cal M}|_C$. 
In turn, the bold $(a,b)$-segment in Figure~\ref{fig:adv4}(4) should 
be contracted to a single point; otherwise, $\langle a,c,b\rangle$ would be 
a separating triple of $G$. Similarly,the bold $(a,c)$-segment 
in Figure~\ref{fig:adv4}(4) should be contracted to a single point. Thus, 
Figure~\ref{fig:adv14} displays ${\cal M}|_C$. Let $q_{a,b}$ (respectively, 
$q_{a,c}$) be the point where nations $a$ and $b$ (respectively, nations 
$a$ and $c$) meet in ${\cal M}|_C$. By the figure, a unique nation $u\in Z_1$
and a unique nation $v\in Z_2$ meet at $q_{a,b}$, and $\{a,b,u,v\}$ is an 
MC$_4$ of $G$. Similarly, a unique nation $x\in Z_1$ and a unique nation 
$y\in Z_2$ meet at $q_{a,c}$, and $\{a,c,x,y\}$ is an MC$_4$ of $G$. Moreover, 
both $\langle a,u,b,v\rangle$ and $\langle a,x,c,y\rangle$ are correct 
4-pizzas. By the figure, other than $\{u,v\}$ and $\{x,y\}$, there is no 
$\{w_1,w_2\}\in E$ with $w_1\in Z_1$ and $w_2\in Z_2$. 
\end{proof}

\micsaid{pass22: added this.}
Just for the next corollary, we temporarily drop
Assumptions~\ref{ass:4-con} and~\ref{ass:temp}.
\begin{corollary}\label{cor:noStriangle}
Suppose that $G$ does not have 
an MC$_5$, a separating edge, or a separating quadruple. Then, $G$ has 
a separating triangle if and only if
 for some 3-clique $C$ of $G$, (i) the nations of 
$C$ do not meet at a point in ${\cal M}$ and (ii) at least one pair of 
nations of $C$ strongly touch in ${\cal M}$.
\end{corollary}

\begin{proof}
 The ``if" part is obvious.  For the ``only if" part, suppose 
$G$ has a separating triangle $T$. If $G$ is not 4-connected, 
then by Lemma~\ref{lem:3-cut}, there is a 3-clique $C$ in $G$ such that 
the nations of $C$ do not meet at a point in ${\cal M}$ and every pair of 
nations of $C$ strongly touch in ${\cal M}$. So, we may assume that 
$G$ is 4-connected. Then, by the proof of Lemma~\ref{noStriple}, in case 
$G$ has a separating triple $C'$, the nations of $C'$ do not meet at a point 
in ${\cal M}$ and at most one pair of nations of $C$ weakly touch in ${\cal 
M}$. Thus, we may further assume that $G$ has no separating triple. 
Then, by the layouts found in Lemmas~\ref{not1} through \ref{noStriangle}, 
the nations of $T$ do not meet at a point in ${\cal M}$ and at least one 
pair of nations of $T$ strongly touch in ${\cal M}$. 
\end{proof}

By the reductions in this section, our algorithm may make progress
whenever $G$ has a separating edge, quadruple, or triangle.  Hereafter
we assume that all such reductions have been made:
\begin{assume}\label{ass:edgeQuadTri}
$G$ does not have a separating edge, quadruple, or triangle.
\end{assume}
In fact Assumption~\ref{ass:edgeQuadTri} implies
Assumptions~\ref{ass:4-con} (by Lemma~\ref{lem:3-cut}(1)) and
\ref{ass:temp} (by definition).  So this one statement summarizes the
effect of applying all the reductions in this section and the previous.
\micsaid{pass16: I reintroduced your implications, but with a bit
more justification.}
\zzsaid{pass17: Good!} 
\micsaid{pass12: I commented-out the previous sentence.  While it is true that
this assumption implies Assumption \ref{ass:4-con}, it is not obvious,
so it is just confusing.
  For example, suppose our nations are the (fattened) edges of the
dodecahedron.  Then this planar graph satisfies Assumptions
\ref{ass:3-con}, \ref{ass:big}, \ref{ass:temp}, and
\ref{ass:edgeQuadTri}, but it does not satisfy Assumption
\ref{ass:4-con}.  Of course the problem with this example is that it
has no atlas.}

\section{Removing Maximal 5-Cliques}\label{sec:five}

We assume that $G$ has an MC$_5$; our goal of this section is to
show how to remove MC$_5$'s from $G$. The idea behind the 
removal of an MC$_5$ $C$ from $G$ is to try to find and remove a correct 
center $P$ of $C$. By Fact~\ref{fac:center}, we make progress after 
removing $P$. 
\micsays{pass12: rewrote next two sentences (an option: delete them).}
\micsaid{pass18: commented out explicit list of violated assumptions,
 just used the one.}
  After removing $P$, the resulting $G$ may no longer satisfy
Assumption~\ref{ass:edgeQuadTri}; in that case, the algorithm must
therefore reapply the reductions of the previous
sections before considering another MC$_5$.
Also, not unexpectedly, our search for a correct center of $C$ may fail. 
In this case, we will be able to decompose $G$ into smaller graphs to 
make progress. 

\zzsaid{pass33: moved from Definition 3.4.}
For a positive integer $k$, two maximal cliques $C_1$ and $C_2$ are
{\em $k$-sharing} if $|C_1\cap C_2|=k$. 

\fig[figure601]{five1}
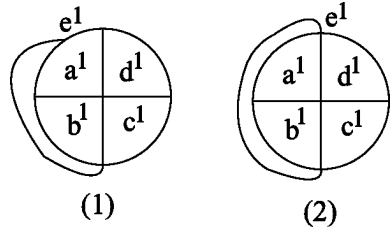
{Displays of an MC$_5$ $C$, 4-sharing with one other.}

\fig[figure602]{five2}
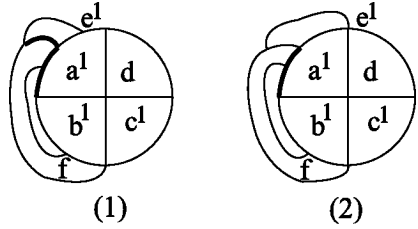
{Displays of 4-sharing MC$_5$'s $C$ and $C_1$.}

Every MC$_5$ $C$ of $G$ is 4-sharing with at most two other MC$_5$'s
$C'$ of $G$; this is because the center of $C'$ must be a 3-point
bordering a hole in ${\cal M}|_C$, and there are at most two such
points in the possible displays of Figure~\ref{fig:intro2}.
\micsaid{pass22: reworded previous.}
 We claim that 
at least one MC$_5$ of $G$ is 4-sharing with two other MC$_5$'s of $G$. 
Towards a contradiction, assume that the claim does not hold. 
Let $C=\{a,b,c,d,e\}$ be an MC$_5$ of $G$. When $C$ is 4-sharing with
no MC$_5$ of $G$, none of Figure~\ref{fig:intro2}(1) through (4)
displays ${\cal M}|_C$ or else either $V$ would equal $C$ or at least
one of $\langle e^1, a^1, b^1\rangle$, $\langle e^1, c^1, d^1\rangle$,
and $\langle e^1, a^1, d^1\rangle$ would be a separating triangle of
$G$, a contradiction.
So, consider the case where $C$ is 4-sharing with exactly one MC$_5$, say 
$C_1=\{a^1,b^1,c^1,e^1,f\}$, of $G$. 
In this case, by Assumption~\ref{ass:edgeQuadTri}
($G$ has no separating triangle),
Figure~\ref{fig:intro2}(1), (2), and (4) are 
transformable to Figure~\ref{fig:five1}(1),
and Figure~\ref{fig:intro2}(3) is 
transformable to Figure~\ref{fig:five1}(2). 
By Figure~\ref{fig:five1}(1) and (2), only Figure~\ref{fig:five2}(1) or (2) 
can possibly display ${\cal M}|_{\{a,\ldots,f\}}$. Actually, 
Figure~\ref{fig:five2}(2) does not display ${\cal M}|_{\{a,\ldots, f\}}$; 
otherwise, since $C_1$ is 4-sharing with no MC$_5$ of $G$ other than $C$, 
there is no $g\in V-\{a,\ldots, f\}$ with $\{a^1,b^1,e^1,f\} \subseteq N_G(g)$ 
and $\langle a^1, f, e^1\rangle$ would be a separating triangle of 
$G$, a contradiction. Similarly, Figure~\ref{fig:five2}(1) does not display 
${\cal M}|_{\{a,\ldots,f\}}$; otherwise, since $|V|\geq 9$, $\langle a^1, f, 
b^1\rangle$ or $\langle a^1, f, e^1\rangle$ would be a separating 
triple of $G$, a contradiction. Therefore, the claim holds. 

By the above claim, if $G$ has an MC$_5$, then it has an MC$_5$ that is 
4-sharing with two other MC$_5$'s of $G$. By our assumption that $G$ has 
an MC$_5$, $G$ has an MC$_5$ $C=\{a,b,c,d,e\}$ that is 4-sharing with 
two other MC$_5$'s, say $C_1=\{a, c, d, e, f\}$ and $C_2=\{a, b, c, e, g\}$, 
of $G$. Let $U = C \cup \{f, g\}$. We show how to find a correct center of 
$C$ below. First, we make a simple but useful observation. 

\begin{fact}\label{fac:crust}
Let $W$ be a subset of an MC$_5$ $C'$ of $G$ with $|W|\geq 3$. 
If all the edges in $E(G[W])$ are marked in $G$ or $G-C'$ has a vertex 
$x$ with $W=C'\cap N_G(x)$, then $W$ contains all correct crusts of $C'$.
\micsaid{pass30: we also use this specialization, is it worth stating?
``If $C$ and $C'$...''}
\zzsaid{pass31: added next sentence.}
In particular, if $C$ and $C'$ are MC$_5$'s with $|C\cap C'|\geq 3$,
then both crusts are in the intersection.
\end{fact}

\micsaid{pass4: in terms of the graph of 4-sharing MC$_5$'s, you have
shown: no isolated vertex, every degree 1 vertex is adjacent to a
degree-2 vertex, and no triangle. So, the graph is a set of paths of
length at least three.}
\zzsaid{pass5: Right.} 

\fig[figure603]{five3}
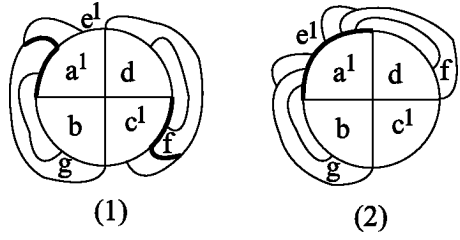
{Displays of MC$_5$ $C$, 4-sharing with $C_1$ and $C_2$.}

Vertices $f$ and $g$ are not adjacent in $G$; otherwise, only 
Figure~\ref{fig:intro2}(3) or (4) can display ${\cal M}|_C$, but after drawing 
nations $f$ and $g$ in the two figures, we see that the 4-connectedness of $G$ 
would force $V$ to equal $U$, contradicting Assumption~\ref{ass:big}. 
So, only Figure~\ref{fig:five3}(1) or Figure~\ref{fig:five3}(2) can 
display ${\cal M}|_U$. By the figures, a correct center of $C$ can be found 
from a correct crust immediately. So, it suffices to find out which one of 
$a$, $c$, and $e$ is a correct crust of $C$. 

Let $\alpha$ be the number of vertices $v \in \{a,c,e\}$ such that 
$N_G(v) \subseteq U$. $\alpha\leq 1$; otherwise, no matter which of 
Figure~\ref{fig:five3}(1) and (2) displays ${\cal M}|_U$, the 4-connectedness 
of $G$ would force $V$ to equal $U$, contradicting Assumption~\ref{ass:big}.
\micsaid{pass22: named preceding assumption.}
First, consider the case where $\alpha=0$. In this case, 
only Figure~\ref{fig:five3}(1) displays ${\cal M}|_U$.
\zzsaid{pass6: Changed the next sentence.} Moreover, by this figure 
and Assumption~\ref{ass:edgeQuadTri} ($G$ has no separating triple), 
there are a unique nation $h\in V-U$ with $\{a^1,b,e^1,g\} \subseteq N_G(h)$.
Similarly, there is a unique nation $i\in 
V-U$ with $\{c^1, d, e^1, f\} \subseteq N_G(i)$. 
So by Fact~\ref{fac:crust},
the unique nation in $N_G(h) \cap N_G(i)$ is a correct crust of $C$. 

Now, we may assume that $\alpha=1$. We may further assume that $c$ is the 
unique $u \in \{a,c,e\}$
\micsaid{pass22: was $\{b,c,e\}$}
such that $N_G(u)\subseteq U$. Then, we can delete 
$c$ from the permutable list $\langle a,c,e\rangle$
\micsaid{pass22: was $\langle b,c,e\rangle$}
in Figure~\ref{fig:five3}(2),
or more intuitively, we can let $c^1=c$ in the figure.
\micsaid{pass22: merged two old sentences into next.}
\zzsaid{pass23: Added next word:} Similarly, 
if Figure~\ref{fig:five3}(1) displays ${\cal M}|_U$, we can let
either $a^1=c$ or $c^1=c$ in the figure; this would imply either
$N_G(\{b,c,g\})\subseteq U$ or $N_G(\{c,d,f\})\subseteq U$,
respectively.
  No matter which 
of Figure~\ref{fig:five3}(1) and (2) displays ${\cal M}|_U$, if there is 
a $u\in \{a,e\}$ such that $\{u,d\}$ or $\{u,b\}$ is a marked edge in $G$, 
then the unique nation in $\{a,e\}-\{u\}$ is a correct crust of $C$. So, we 
may assume that none of $\{a,d\}$, $\{e,d\}$, $\{a,b\}$, and $\{e,b\}$ is 
a marked edge in $G$. It remains to consider three cases as follows.

\fig[figure604]{five4}
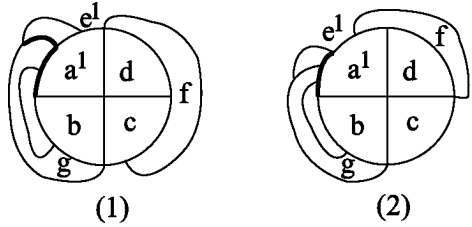
{Displays of $G[\{a,\ldots,g\}]$ in Case 1.}

\case{1} $N_G(\{c,d,f\})\subseteq U$. Then, Figure~\ref{fig:five3}(1) and 
(2) are transformable to Figure~\ref{fig:five4}(1) and (2), respectively. 

\fig[figure605]{five5}
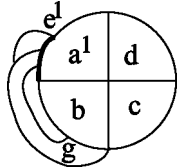
{A display of $G'[\{a,b,c,d,e,g\}]$ in Case 1.1.}

\case{1.1} Edge $\{c,f\}$ is not marked in $G$. Then, 
Figure~\ref{fig:five4}(1) is transformable to Figure~\ref{fig:five4}(2), and hence 
the latter displays ${\cal M}|_U$. Let $G'$ be the marked graph obtained from 
$G-\{f\}$ by marking the following edges: $\{b,c\}$, $\{c,d\}$, $\{a,e\}$, 
$\{a,d\}$, $\{e,d\}$. By Figure~\ref{fig:five4}(2), we can obtain a well-formed
 atlas ${\cal M}'$ of $G'$ from ${\cal M}$ by extending nation $e^1$ to occupy 
nation $f$. Figure~\ref{fig:five5} displays ${\cal M}'|_{\{a,\ldots,e,g\}}$. 
On the other hand, we claim that every well-formed atlas ${\cal M}''$ of $G'$ 
can be used to construct a well-formed atlas of $G$. To see this, first note 
that by Fact~\ref{fac:crust}, the crust of $C$ in ${\cal M}''$ must be either 
$a$ or $e$. Suppose that the crust is $e$; the other case is similar. Then, 
since edges $\{b,c\}$ and $\{c,d\}$ are marked in $G'$, the center of $C$ 
in ${\cal M}''$ must be $\langle a, b, c, d\rangle$. Moreover, since $N_{G'} 
(\{d\}) \subseteq C$, the four nations $a$, $c$, $d$, and $e$ must be 
related in ${\cal M}''$ as shown in Figure~\ref{fig:five5}. Thus, we can 
assign a suitable sub-region of $e$ to $f$ to obtain an atlas of $G$. 
This establishes the claim. 
\micsaid{pass22: I suppressed the following sentence: ``By the claim,
we only need to recursively find a correct crust of $C$.''  I see no
need for an mention recursion here, since we have already made
progress, by producing a simpler graph $G'$.}

\case{1.2} Edge $\{c,f\}$ is marked in $G$. Then, only 
Figure~\ref{fig:five4}(1) displays ${\cal M}|_U$. By the figure, at most one 
of edges $\{a,f\}$ and $\{e,f\}$ is marked in $G$. Moreover, if $\{a,f\}$ is 
marked in $G$, then $a$ is a correct crust of $C$. Similarly, if $\{e,f\}$ is 
marked in $G$, then $e$ is a correct crust of $C$. So, it remains to consider 
the case where neither $\{a,f\}$ nor $\{e,f\}$ is a marked edge in $G$. 
In this case, it suffices to construct a marked graph $G'$ as in Case~1.1. 

\case{2} $N_G(\{b,c,g\})\subseteq U$. Similar to Case~1, after relabeling.
\micsaid{pass22: we exchange $b$ with $d$ and $f$ with $g$, right?}
\zzsaid{pass23: right.}

\case{3}
Neither $N_G(\{b,c,g\})\subseteq U$ nor $N_G(\{c,d,f\})\subseteq U$.
Then as argued above, Figure~\ref{fig:five3}(2) displays $G[U]$. 
We consider three sub-cases as follows: 

\fig[figure606]{five6}
{A display of $G[\{a,\ldots,g\}]$ in Case 3.1.}
\fig[figure607]{five7}
{A display of $G'[\{a,\ldots,g\}]$ in Case 3.1.}

\case{3.1} There is no $v\in V-U$ such that $d\in N_G(v)$ and $N_G(v)\cap 
\{a, e\}\not=\emptyset$. Then, Figure~\ref{fig:five6} displays ${\cal M}|_U$. 
Let $G'$ be the marked graph obtained from $G-\{\{c,f\}\}$ by marking 
the following edges: $\{b,c\}$, $\{c,d\}$, $\{a,d\}$, $\{e,d\}$, $\{a,f\}$,
$\{e,f\}$, $\{d,f\}$. By Figure~\ref{fig:five6}, we can obtain a well-formed 
atlas ${\cal M}'$ of $G'$ by erasing the $(c,f)$-point in ${\cal M}$. 
Figure~\ref{fig:five7} displays ${\cal M}'|_{\{a, \ldots, g\}}$. 
By Figure~\ref{fig:five7} and Lemma~\ref{lem:3-cut}, 
\zzsaid{Change made: \{fac:3-cut\} $\rightarrow$ \{lem:3-cut\}.}
both $G'-\{a,d,f\}$ and $G'-\{e,d,f\}$ are connected. We claim that 
every well-formed atlas ${\cal M}''$ of $G'$ can be used 
to construct a well-formed atlas of $G$. To see this, first note that 
by Fact~\ref{fac:crust}, the crust of $C$ in ${\cal M}''$ must be either $a$ 
or $e$. We assume that the crust is $e$; the other case is similar. Then, 
since $\{b,c\}$ and $\{c,d\}$ are marked edges in $G'$, 
the center of $C$ in ${\cal M}''$ must be $\langle a, b, c, b\rangle$. 
Moreover, since $G'-\{a,d,f\}$ is connected, the marked edges $\{a,d\}$, 
$\{d,f\}$ and $\{f,a\}$ of $G'$ force nations $a$, $d$ and $f$ 
to meet at a 3-point in ${\cal M}''$. For a similar reason, nations $e$, 
$d$ and $f$ meet at a 3-point in ${\cal M}''$. Now, since $N_{G'}(c) 
\subseteq C$, the four nations $c$, $d$, $e$, and $f$ must be related in 
${\cal M}''$ as shown in Figure~\ref{fig:five7}. Thus, we can assign a suitable 
sub-region of $e$ to $f$ to obtain a well-formed atlas of $G$. 

\case{3.2} There is no $v\in V-U$ such that $b\in N_G(v)$ and $N_G(v) 
\cap \{a, e\}\not=\emptyset$. Similar to Case~3.1.

\case{3.3} There are nations $h$ and $i$ in $V-U$ such that $d\in N_G(h)$,
 $N_G(h)\cap \{a, e\}\not=\emptyset$, $b\in N_G(i)$, and $N_G(i)\cap \{a, e\} 
 \not=\emptyset$. By Figure~\ref{fig:five3}(2), no nation of $V-U$ can 
 touch both $b$ and $d$ in ${\cal M}$. So, $h$ and $i$ are distinct nations. 
 Moreover, if $|N_G(h)\cap \{a,e\}|=1$ 
 (respectively, $|N_G(i)\cap \{a,e\}|=1$), then the unique nation in 
 $\{a,e\}-N_G(h)$ (respectively, $\{a,e\}-N_G(i)$) must be a correct 
 crust and we are done. So, we assume that $\{a,e\}\subseteq N_G(h)$ and 
 $\{a,e\}\subseteq N_G(i)$. 
 Then, by Figure~\ref{fig:five3}(2), $\{a,d,e,f,h\}$ 
 and $\{a,b,e,g,i\}$ are MC$_5$'s in $G$.
 Let $U_h=U\cup\{h\}$. If $\{g,h\}$ 
 were an edge in $G$, then by Figure~\ref{fig:five3}(2), after drawing nation 
 $h$ in ${\cal M}|_U$, we see that the 4-connectedness of $G$ would force $V$ 
 to equal $U_h$, contradicting Assumption~\ref{ass:big}. 
 So, $\{g, h\}\not\in E$. Similarly, $\{f,i\}\not\in E$. Then, 
 Figure~\ref{fig:five8}(1) or Figure~\ref{fig:five8}(2) displays ${\cal M}|_{U_h}$.
 If edge $\{d,h\}$ is marked in $G$ or $N_G(d) - U_h \not= \emptyset$, 
 Figure~\ref{fig:five8}(2) displays ${\cal M}|_{U_h}$; otherwise, 
 Figure~\ref{fig:five8}(2) is transformable to Figure~\ref{fig:five8}(1). 
 So, we can decide whether Figure~\ref{fig:five8}(1) or (2) displays 
 ${\cal M}|_{U_h}$. 

\micsaid{pass30: broke the long paragraph of Case 3.3 into cases
3.3.1 and 3.3.2, corresponding to subfigures of
Figure~\ref{fig:five8}.  I changed the figure caption accordingly.}
\micsaid{pass30: slightly reworded last sentence.
  Since we are talking about a transformation, is it
  better to say $G[U_h]$, rather than a specific map ${\cal M}|_{U_h}$?
  Introduced two subcases below (it is still exactly your argument).}
\zzsaid{pass31: Here, the claim about each transformation is witnessed
	by the specific map ${\cal M}$. So, I think that mentioning
	the specific map helps the reader.} 
\micsaid{pass32: changed fig:five8 caption to use 
 ``${\cal M}|_{U_h}$'', for consistency.}
\fig[figure608]{five8}
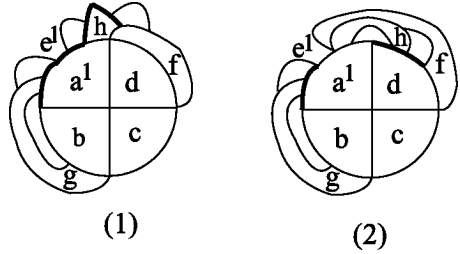
{Displays of 
 ${\cal M}|_{U_h}$ in Cases 3.3.1 and 3.3.2.}

\case{3.3.1}  Figure~\ref{fig:five8}(1) displays ${\cal M}|_{U_h}$.
  We further distinguish two cases as follows.

\case{3.3.1.1} There is no $v\in V-U_h$ such that $f\in N_G(v)$ and 
  $\{a,e\}\cap N_G(v)\not=\emptyset$. Let ${\cal D}$ be the figure obtained 
  from Figure~\ref{fig:five8}(1) by extending nation $h$ to occupy the hole  
  touched by $e^1$, $f$ and $h$. By Figure~\ref{fig:five8}(1), ${\cal D}$ 
  displays ${\cal M}|_{U_h}$ and so $N_G(f)\subseteq U_h$. Moreover, by figure 
  ${\cal D}$, if there is a $w\in\{a,e\}$ such that edge $\{w,f\}$ is marked 
  in $G$, then $w$ is a correct crust of $C$. So, we may assume that none of 
  the edges $\{a,f\}$ and $\{e,f\}$ is marked in $G$. Let $G'$ be the marked 
  graph obtained from $G-\{f\}$ by marking the following edges: $\{b,c\}$, 
  $\{c,d\}$, $\{a,d\}$, $\{e,d\}$, $\{a,h\}$, $\{e,h\}$, $\{d,h\}$. 
  By figure ${\cal D}$, we can obtain a well-formed atlas ${\cal M}'$ of 
  $G'$ from ${\cal M}$ by (i) erasing the $(c,f)$-point and further (ii) 
  extending nation $h$ to occupy $f$. Indeed, by renaming nation $f$ in 
  Figure~\ref{fig:five7} as $h$, we obtain a figure displaying 
  ${\cal M}'|_{\{a,\ldots,e,g,h\}}$. Moreover, similarly to Case~3.1, 
  we can prove that every well-formed atlas of $G'$ 
  can be used to construct a well-formed atlas of $G$. 

\micsaid{pass30: changed caption of Figure~\ref{fig:five9}}
\fig[figure609]{five9}
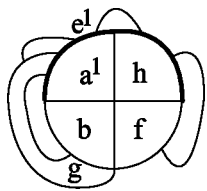
{Display of $G'[\{a,b,e,f,g,h\}]$ in Case 3.3.1.2.}

\case{3.3.1.2} There is a $j\in V-U_h$ such that $f\in N_G(j)$ and
$\{a,e\}\cap N_G(j)\not=\emptyset$. If $\{a,e\}\not\subseteq N_G(j)$,
then by Figure~\ref{fig:five8}(1), the unique nation in $\{a,e\}\cap
N_G(j)$ is a correct crust of $C$ and we are done. So, we assume that
$\{a,e\}\subseteq N_G(j)$.
Recall that $\{f,i\}\not\in E$. So, $j\not=i$. 
By Figure~\ref{fig:five8}(1), if there is a $w\in\{a,e\}$ such that
$\{w,c\}$ is a marked edge in $G$, then $w$ is a correct crust of
$C$. So, we may assume
that neither $\{a,c\}$ nor $\{e,c\}$ is a marked edge in $G$. Let $G'$ be 
the graph obtained from $G-\{c,d\}$ by adding the three edges $\{g,f\}$, 
$\{b,f\}$, and $\{h,b\}$ and further marking the two edges $\{b,f\}$ and 
$\{f,h\}$. By Figure~\ref{fig:five8}(1), we can obtain a well-formed atlas 
${\cal M}'$ of $G'$ from ${\cal M}$ by 
(i) erasing the $(d,e^1)$-point,
(ii) erasing the $(a^1,f)$-point,
(iii) extending nation $f$ to occupy nation $c$, and
(iv) extending nation $h$ to occupy nation $d$. 
Indeed, Figure~\ref{fig:five9} displays ${\cal M}'|_{\{a,e,b,f,g,h\}}$.
 We claim that every well-formed atlas ${\cal M}''$ of $G'$
can be used to construct a well-formed atlas of $G$. 
\micsaid{pass30: reworded next sentence, to make it clear
at the beginning that each set is an MC$_5$.}
To see this, first note that $G'$ contains the MC$_5$'s
$C'=\{a,e,b,f,h\}$, $C'_1=\{a,e,b,f,g\}$, $C'_2=\{a,e,f,h,j\}$, and 
$C'_3=\{a,e,b,g,i\}$.
These MC$_5$'s and Fact~\ref{fac:crust} ensure that the crust of $C'$ in 
${\cal M}''$ must be $a$ or $e$ and that the two nations $b$ and $h$ do not 
appear consecutively around the center of $C'$ in ${\cal M}''$. We assume that 
the crust of $C'$ in ${\cal M}''$ is $e$; the other case is similar. 
Then, the center of $C'$ in ${\cal M}''$ is $\langle a,b,f,h\rangle$. 
This together with the well-formedness of ${\cal M}''$ implies that the crust 
of $C'_1$ in ${\cal M}''$ is either $a$ or $f$. On the other hand, since 
$C'_1\cap N_G(i)=\{a,e,b,g\}$, $f$ is not a correct crust of $C'_1$ by 
Fact~\ref{fac:crust}. Thus, the crust of $C'_1$ in ${\cal M}''$ is $a$. 
Therefore, the centers of $C'$ and $C'_1$ are as shown in 
Figure~\ref{fig:five9}. {From} this, the claim follows immediately. 

\micsaid{pass30: the next text (now Case 3.3.2) used to be in the text
of Case 3.3, preceding what is now Case 3.3.1.  But I thought that
paragraph was simply too long, and also this presentation makes the
subfigure roles more obvious.}

\case{3.3.2} Figure~\ref{fig:five8}(2) displays ${\cal M}|_{U_h}$. 
 In this case, 
 we check if there is a $v\in V-U_h$ such that $d\in N_G(v)$ and 
 $N_G(v)\cap\{a, e\}\not=\emptyset$. If such $v$ exists, then $|N_G(v)\cap 
 \{a,e\}|=1$ and the unique nation in $\{a,e\}-N_G(v)$ is a correct crust of 
 $C$. If no such $v$ exists, then by Figure~\ref{fig:five8}(2) and the 
 4-connectedness of $G$, we have $N_G(\{d,f,h\})\subseteq U_h$ and so 
 Figure~\ref{fig:five8}(2) is transformable to a figure ${\cal D}$, where 
 ${\cal D}$ is obtained from Figure~\ref{fig:five8}(2) by extending nation 
 $h$ to occupy the two holes touched by $h$. By figure ${\cal 
 D}$, if there is a $w\in \{a,e\}$ such that edge $\{w,f\}$ is marked in 
 $G$, then $w$ is a correct crust of $C$. Similarly, if there is a $w\in 
 \{a,e\}$ such that edge $\{w,h\}$ is marked in $G$, then the unique nation 
 in $\{a,e\}-\{w\}$ is  a correct crust of $C$. So, we may assume that none of 
 the edges $\{a,f\}$, $\{e,f\}$, $\{a,h\}$ and $\{e,h\}$ are marked in $G$. 
 Let $G'$ be the marked graph obtained from $G-\{f,h\}$ by marking 
 the following edges: $\{b,c\}$, $\{c,d\}$, $\{a,e\}$, $\{a,d\}$, $\{e,d\}$. 
 By figure ${\cal D}$, we can obtain a well-formed atlas ${\cal M}'$ of $G'$ 
 from ${\cal M}$ by extending nation $e^1$ to occupy $f$ and $h$. On the other 
 hand, as in Case~1.1, we can prove that every well-formed atlas of $G'$ can be
 used to construct a well-formed atlas of $G$.

\section{Removing Maximal 4-Cliques}\label{sec:four}

Throughout this section, we assume that $G$ does not have an MC$_5$.
We further assume that 
$G$ has an MC$_4$; our goal of this section is to show how to remove 
MC$_4$'s from $G$. The idea behind the removal of an MC$_4$ $C$ from $G$ is 
to try to find and remove a correct 4-pizza via constructing an extensible 
layout of $C$. After the removal of a correct 4-pizza, the resulting $G$ may 
be not 4-connected and may have a separating 4-cycle, edge, triple, quadruple, 
or triangle. 
\micsaid{pass12: rewrote next sentence.}
To restore Assumption~\ref{ass:edgeQuadTri}, the algorithm reapplies the
reductions in Sections~\ref{sec:conn} and \ref{sec:adv} to the resulting $G$. 

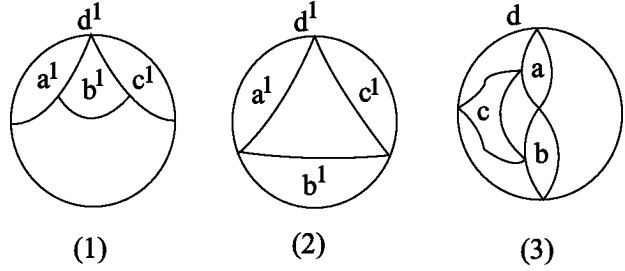
\fig[figure701]{four1}{Possible displays of MC$_4$ $\{a,b,c,d\}$.}

Suppose $C=\{a,b,c,d\}$ is an MC$_4$ of $G$; using
Corollary~\ref{cor:noStriangle} and $|V|>8$, we find that only
Figure~\ref{fig:four1}(1), (2) or (3) can possibly display ${\cal
M}_C$.
\zzsaid{Change made: \{fac:noStriangle\} $\rightarrow$ \{cor:noStriangle\}.}
\micsaid{pass34: reworded previous sentence, added next.}
Note these are a pizza, a pizza-with-crust, and a rice-ball, respectively.

\subsection{Finding Rice-Balls}\label{subsec:riceball}

Let $C=\{a,b,c,d\}$ be an MC$_4$ of $G$. For a subset $W$ of $C$, let 
${\cal E}[W]$ be the set of unmarked edges $\{u,v\}\in E$ such that 
$u\not\in W$, $v\not\in W$, and some MC$_4$ of $G$ consists of $u$, $v$, 
and two vertices in $W$. 

\micsaid{pass4: Here do you really want to include $\{u,v\}$ when the
 two vertices of $W$ span a marked edge?}
\zzsaid{pass5: I added another statement in the following lemma.} 
Let $G'=G-C-{\cal E}[C]$. 
A {\em 3-subset} of $C$ is a subset $S$ of $C$ with $|S|=3$. 
For each 3-subset $S$ of $C$, let $V_S=\cup_K V(K)$, where $K$ ranges over 
all connected components $K$ of $G'$ with $C \cap N_G(V(K)) = S$. 

\begin{lemma}\label{mc4:riceball} 
Figure~\ref{fig:four1}(3) displays ${\cal M}|_C$ if and only if
the following statements hold: 
\begin{enumerate}
\item No two vertices in $C$ are connected by a marked edge in $G$. 
\item \zzsaid{pass35: rewrote this item.}
    $V_{\{a,b,c\}}$, $V_{\{a,b,d\}}$, $V_{\{a,c,d\}}$, and $V_{\{b,c,d\}}$ 
    each are nonempty and induce a connected component of $G'$, and
    they together form a partition of $V-C$. 
\item For every pair of two distinct 3-subsets $S$ and $T$ of $C$, 
    $|V_S \cap N_G(V_T)|=1$, $|V_T \cap N_G(V_S)|=1$, and 
    $(S\cap T)\cup (V_S \cap N_G(V_T)) \cup (V_T \cap N_G(V_S))$ is 
    an MC$_4$ of $G$. 
\end{enumerate}
\end{lemma}
\begin{proof}
For the ``only if'' direction, suppose that Figure~\ref{fig:four1}(3) 
displays ${\cal M}|_C$. Then, ${\cal M}|_C$ has four holes, and each hole is 
touched by exactly three nations of $C$. For each 3-subset $S$ of $C$, let 
${\cal H}_S$ be the hole touched by the nations of $S$, and let $Z_S$ be the 
nations of $V-C$ that occupy ${\cal H}_S$ in atlas ${\cal M}$. We want to prove
that for each 3-subset $S$ of $C$, $Z_S=V_S$. To this end, first observe that 
for each connected component $K$ of $G'$, there is a 3-subset $S$ of $C$ with 
$V(K) \subseteq Z_S$. This observation follows from Figure~\ref{fig:four1}(3) 
immediately. 
\micsaid{pass32: changed next sentence, an equivalent claim.}
Consequently, $C\cap N_G(V(K)) \subseteq S$; we claim they are equal.
Towards a contradiction, assume that 
$G'$ has a connected component $K$ with $|C\cap N_G(V(K))| \leq 2$. 
Let $W = C\cap N_G(V(K))$. 
If $|W| \leq 1$, then $K$ would be a connected component of 
$G-W$, a contradiction. 
\micsaid{pass30:  next three sentences could be shorter...}
\zzsaid{pass31: shortened.}
If $|W| = 2$, then $K$ is a connected component of $G-W-{\cal E}[W]$,
and the vertices of $W$ define a separating edge, a contradiction.
So, the claim holds.
By this claim, the above observation and Figure~\ref{fig:four1}(3),
we have $Z_S=V_S$ for each 3-subset $S$ of $C$. 
In turn, by Figure~\ref{fig:four1}(3), Statements~1 through~3 hold. 

\zzsaid{pass13: Rewrote the next caption.}
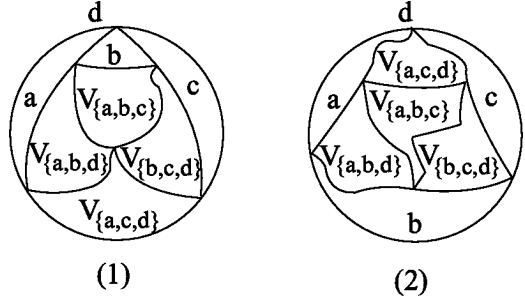
\fig[figure702]{four2}{Possible atlases of $G$.}

For the ``if'' direction, suppose that Statements 1 through 3 hold. 
We first prove that Figure~\ref{fig:four1}(1) does not display ${\cal M}|_C$. 
Towards a contradiction, assume that Figure~\ref{fig:four1}(1) displays 
${\cal M}|_C$. We may assume that $\langle a^1,b^1,c^1,d^1\rangle = \langle 
a, b, c, d\rangle$ in Figure~\ref{fig:four1}(1); the other cases are similar. 
Let $S$ be a 3-subset of $C$. We claim that there is no point in ${\cal M}$ at 
which two nations $u$ and $v$ of $V_S$ together with two nations $x$ and $y$ 
of $V-V_S$ meet cyclically in the order $u$, $x$, $v$, $y$. This claim holds; 
otherwise, $\{x,y\}\not\subseteq C$ by Figure~\ref{fig:four1}(1), so $x$ or $y$ 
belongs to $V_T$ for some 3-subset $T$ of $C$ other than $S$, and in turn 
$\{u,v\}$ would be a subset of $V_S\cap N_G(V_T)$, a contradiction against 
Statement 3 in the lemma. By this claim and Statement 2, 
the nations of $V_S$ form a disc homeomorph in ${\cal M}$.
\micsaid{pass34: to say that $V_S$ is a single disc homeomorph, we seem
 to need the statement that it has a single connected component.}
\zzsaid{pass35: rewrote previous sentence.}
Thus, by Statements 2 and 3, Figure~\ref{fig:four2}(1) displays ${\cal M}$.
By this figure, there is a 4-point $p$ in ${\cal M}$ such 
that for each 3-subset $S$ of $C$, exactly one nation $v_S\in V_S$ touches $p$.
Since the nations $v_S$ meet at $p$ but no two of them belong to the same 
connected component of $G'$, $C\cap N_G(v_S)=S$. 
In turn, by Figure~\ref{fig:four2}(1) and the 4-connectedness of $G$,
each $V_S$ would equal $\{v_S\}$, contradicting Assumption~\ref{ass:big}.
Therefore, Figure~\ref{fig:four1}(1) does not display ${\cal M}|_C$. 

We next prove that Figure~\ref{fig:four1}(2) does not display ${\cal M}|_C$. 
Towards a contradiction, assume that Figure~\ref{fig:four1}(2) displays 
${\cal M}|_C$.
\micsaid{pass34: shortened three sentences to the next.}
As in the last paragraph, we may assume $\langle
a^1,b^1,c^1,d^1\rangle = \langle a, b, c, d\rangle$, and we claim that
the nations of each $V_S$ form a disc homeomorph in ${\cal M}$.
Thus, by Statements 2 and 3, 
Figure~\ref{fig:four2}(2) displays ${\cal M}$. By this figure, there is 
a 4-point $p$ in ${\cal M}$ at which nation $a$, some $u\in V_{\{a,b,c\}}$, 
some $v\in V_{\{a,b,d\}}$, and some $w\in V_{\{a,c,d\}}$ meet. Since $u$, $v$ 
and $w$ meet at $p$ but no two of them belong to the same connected component 
of $G'$, $C\cap N_G(u)=\{a,b,c\}$, $C\cap N_G(v)=\{a,b,d\}$, and 
$C\cap N_G(w)=\{a,c,d\}$. In turn, by Figure~\ref{fig:four2}(2), $\{u\} = V_{\{ 
a,b,c\}}$. Moreover, nations $v$, $a$, $b$, $d$ meet at a point in ${\cal M}$, 
and nations $w$, $a$, $c$, $d$ meet at a point in ${\cal M}$. Thus, 
$V_{\{a,b,d\}}=\{v\}$ or else $\langle u, b, v\rangle$ would be a 
separating triple of $G$, a contradiction. Similarly, 
$V_{\{a,c,d\}}=\{w\}$. In a similar way, we can also prove that 
$|V_{\{b,c,d\}}|=1$. In summary, we have $|V|=8$, a contradiction. 
Therefore, Figure~\ref{fig:four1}(2) does not display ${\cal M}|_C$. 

Since both Figure~\ref{fig:four1}(1) and (2) do not display ${\cal M}|_C$, only 
Figure~\ref{fig:four1}(3) can display ${\cal M}|_C$. This completes the proof. 
\end{proof}

Since it is easy to check whether Statements 1 through 3 hold, 
we can easily decide whether $C$ has an extensible ``rice-ball'' layout. 
Once we know that $C$ has an extensible ``rice-ball'' layout, then by 
Figure~\ref{fig:four1}(3) and Statement 2, we can easily 
find and then remove six correct 4-pizzas from $G$.  
\micsaid{pass40: added next sentence.}
By examining all the MC$_4$'s in $G$, our algorithm can either find
one that is a rice-ball, and thus make progress; or else it can
establish that none of the MC$_4$'s 
is a rice-ball. \zzsaid{pass41: rewrote last sentence.}

\subsection{Distinguishing Pizzas and non-Pizzas}\label{subsec:dist}

\micsaid{pass40: small rewrite here.}
\zzsaid{pass41: rewrote next sentence.}
 By the previous discussion, we now suppose that our algorithm
reaches a point where none of the MC$_4$'s 
has a rice-ball layout.
Then all the remaining MC$_4$'s are either pizzas or
pizza-with-crusts.  Specifically, we have:
\begin{corollary}\label{cor:onlytwo}
 \micsaid{pass40: reworded.}
 For every MC$_4$ $C$ of $G$, 
 and for every well-formed atlas ${\cal M}$ of $G$,
 either Figure~\ref{fig:four1}(1) or (2) displays ${\cal M}_C$.
\end{corollary}

\micsaid{pass34: rewrote next paragraph to give more overview;
  incorporated material from old last paragraph of this subsection,
  pointing out relation to next subsection.}
Let $C=\{a,b,c,d\}$ be an MC$_4$ of $G$. Our goal in this section is
to give a
\micsaid{pass 37: added next} linear time
 decision procedure to decide which of
Figure~\ref{fig:four1}(1) and (2) displays ${\cal M}|_C$.
\micsaid{pass34: Is the following really true?
	 pass35: Yes. So, ZZ added it.}
\micsaid{pass40: changed next ``we'' to ``the procedure''.}
Moreover, the procedure
 always chooses \ref{fig:four1}(2) when both are possible.
Whenever Figure~\ref{fig:four1}(2) displays ${\cal M}|_C$, we will
have identified $d^1$ and therefore we immediately make progress by
removing three correct 4-pizzas.
\micsaid{pass34: a small worry: could it be incorrect to remove multiple
 pizzas at once, since it may violate our connectivity requirements?}
\zzsaid{pass35: Connectivity is not a problem here, as long as we 
 maintain the existence of a well-formed atlas. This existence is 
 always maintained no matter how many correct 4-pizzas you remove.
 Anyway, rewrote the last sentence to say only one 4-pizza is removed.} 
\micsaid{pass36: I changed back to remove all three 4-pizzas; this
   issue occurs at many points in the paper.  Rather than track them all
   down, I just added a relevant comment at end of lem:remove4.}
When Figure~\ref{fig:four1}(1) (the pizza) displays ${\cal M}|_C$, we
do nothing with this MC$_4$ $C$ and proceed to consider other
MC$_4$'s; this may eventually lead to a situation where all MC$_4$'s
in $G$ have to be pizzas, as considered in Section~\ref{subsec:pizzas}.

If Figure~\ref{fig:four1}(2) displays ${\cal M}|_C$, then there is no
$e\in V-C$ with $\{a^1,b^1,c^1\}\subseteq N_G(e)$ or else $V$ would
equal $\{a,b,c,d,e\}$ by Corollary~\ref{cor:onlytwo}, a contradiction.
\micsaid{pass34: reworded next sentence.}
Also, for Figure~\ref{fig:four1}(2) to possibly display ${\cal M}|_C$,
we must have 
(i) $C$ is 3-sharing with exactly three MC$_4$ $C_1$, $C_2$ and $C_3$
of $G$ and
(ii) the unique nation of $C_1\cap C_2\cap C_3$ is adjacent to no
nation of $V-(C\cup C_1\cup C_2\cup C_3)$ in $G$.
 We assume that $C_1\cap C_2\cap C_2=\{d\}$, $C_1 = \{a,b,d,e\}$,
$C_2=\{a,c,d,f\}$ and $C_3=\{b,c,d,g\}$; the other cases are
similar. Let $U=\{a, b, \ldots, g\}$. 
\zzsaid{pass35: added next sentence to reflect your next comment.}
\micsaid{pass38: added next ``the''}
By symmetry of the pizza, we may assume that
$d^1=d$ in Figure~\ref{fig:four1}(1).

\micsaid{pass34: in fig:four3(1), $d^1$ is not permutable.  Also,
  It is ok, but potentially confusing, that not all permutations
  allowed by the figure will display our cliques.  
  Also changed the caption (even though the old caption is later justified).}
\micsaid{pass40: change $d^1$ to $d$ in fig:four3(1)?}
\zzsaid{pass41: did it.}
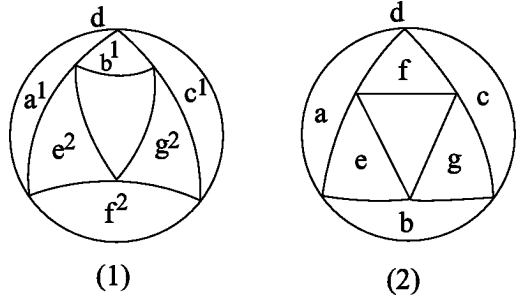
\fig[figure703]{four3} {Possible displays of $G[U]$ when $\{e,f,g\}$
is a clique.}

First, consider the case where $\{e, f, g\}$ is a clique in $G$.
In this case, by Corollary~\ref{cor:onlytwo},
Figure~\ref{fig:four3}(1) (or (2), respectively) displays ${\cal
M}|_U$ if and only if Figure~\ref{fig:four1}(1) (respectively, (2))
displays ${\cal M}|_C$.
 To distinguish the two figures, we check whether there is a nation
$h\in V-U$ with $\{e,f,g\} \subseteq N_G(h)$.  If no such $h$, then
Figure~\ref{fig:four3}(1) does not display ${\cal M}|_U$.
 If such $h$ exists, then Figure~\ref{fig:four3}(2) does not display
${\cal M}|_U$ because otherwise, $V$ would equal $\{a, b, \ldots, h\}$
according to Corollary~\ref{cor:onlytwo}.
 In summary, when $\{e,f,g\}$ is a clique of $G$, we know which of
Figure~\ref{fig:four1}(1) and (2) displays ${\cal M}|_C$.

So, in the sequel, we assume that $\{e, f, g\}$ is not a clique of $G$. 
\micsaid{pass34: rewrote from here...}
%
%
%
%
%
In case Figure~\ref{fig:four1}(1) displays ${\cal M}|_C$, a simple
inspection shows that one nation in $\{e,f,g\}$ (the one adjacent to
$a^1,c^1,d$) is adjacent to the other two. 
\zzsaid{pass35: previous $a^1,c^1,d$ was $\{a^1,c^1,d^1\}$.}
So we assume that only one edge is missing among $\{e,f,g\}$,
for otherwise Figure~\ref{fig:four1}(2) must display ${\cal M}|_C$.
We suppose the absent edge is $\{e,g\}$; the other cases are similar.
Note that $\{a,d,e,f\}$ and $\{c,d,f,g\}$ are MC$_4$'s in $G$.
\micsaid{pass34: ... to here.  Also changed fig:four4 caption.}
Moreover, by Corollary~\ref{cor:onlytwo}, Figure~\ref{fig:four1}(1) 
(or (2), respectively) displays ${\cal M}|_C$ if and only if
 Figure~\ref{fig:four4}(1) 
(respectively, (2)) displays ${\cal M}|_U$. 
\zzsaid{pass35: rewrote the rest of this paragraph.}
 Figure~\ref{fig:four4}(2) does not display ${\cal M}|_{U_i}$ if $\{d,f\}$
 is a marked edge. Also, if $\{d,b\}$ is a marked edge, then 
 Figure~\ref{fig:four5}(1) does not display ${\cal M}|_{U_i}$ and so 
 Figure~\ref{fig:four5}(2) displays ${\cal M}|_{U_i}$. 
 Thus, we may assume that neither $\{d,b\}$ nor $\{d,f\}$ is a marked edge.

\zzsaid{pass35: added paragraph break here.} 
 To distinguish Figure~\ref{fig:four4}(1) and (2), 
 we do a case-analysis as follows:

\fig[figure704]{four4}{Possible layouts of $G[U]$ when
$\{e,g\}\not\in E$.}

\case{1} There is no $h\in V-U$ with $\{a,b,e\}\subseteq N_G(h)$ or 
	there is no $i\in V-U$ with $\{b,c,g\}\subseteq N_G(i)$. 
   	Then, Figure~\ref{fig:four4}(1) does not display ${\cal M}|_U$. 
 \zzsaid{pass37: added next sentence.}
	Whether $h$ and $i$ exist can be decided in $O(1)$ time, because 
 	$|N_G(a)| = |N_G(c)| = 6$ by Figure~\ref{fig:four4}. 

\case{2} There are $h\in V-U$ and $i\in V-U$ such that $\{a,b,e\} 
 \subseteq N_G(h)$ and $\{b,c,g\}\subseteq N_G(i)$. Then, if $f\not\in N_G(h)$ 
 or $f\not\in N_G(i)$, Figure~\ref{fig:four4}(2) does not display ${\cal M}|_U$. 
 So, we may assume that $f\in N_G(h)$ and $f\in N_G(i)$. Then, $h\not=i$ 
 by Corollary~\ref{cor:onlytwo}, Figure~\ref{fig:four4}(1), and (2). 
 Let $U_i=U \cup \{h,i\}$. 

\micsaid{pass34: slight caption changes in figures four4 to four7.}

\fig[figure705]{four5} {Possible layouts of $G[U_i]$ when
$\{h,i\}\in E$.}

\case{2.1} $\{h,i\}\in E$. If $N_G(f)\subseteq U_i$, then 
 Figure~\ref{fig:four4}(1) does not display ${\cal M}_U$ 
 by Corollary~\ref{cor:onlytwo}. Similarly, if $N_G(b)\subseteq U_i$, 
 then Figure~\ref{fig:four4}(2) does not display ${\cal M}_U$. So, 
 we may assume that neither $N_G(b)\subseteq U_i$ nor $N_G(f)\subseteq U_i$.
 \zzsaid{pass6: Changed the next sentence.}
 Then, by Corollary~\ref{cor:onlytwo} and Assumption~\ref{ass:edgeQuadTri}
 ($G$ has no separating triple),
 Figure~\ref{fig:four5}(1) (respectively, (2)) displays ${\cal M}|_{U_i}$ 
 if and only if
 Figure~\ref{fig:four4}(1) (respectively, (2)) displays ${\cal M}_U$. 
\zzsaid{pass37: added next sentence and rewrote three subsequent sentences.}
 By Figure~\ref{fig:four5}, $|N_G(e)| = |N_G(g)| = 6$; let $j$ be 
 the nation in $N_G(e)-U_i$ and $k$ be the nation in $N_G(g)-U_i$. 
 In case $j$ or $k$ is not adjacent to $f$ in $G$, 
 Figure~\ref{fig:four5}(1) does not display ${\cal M}|_{U_i}$. 
 Similarly, in case $j$ or $k$ is not adjacent to $b$ in $G$, 
 Figure~\ref{fig:four5}(2) does not display ${\cal M}|_{U_i}$. So, we may 
 further assume that $j$ and $k$ are adjacent to both $f$ and $b$ in $G$.
 Then, by Corollary~\ref{cor:onlytwo}, Figure~\ref{fig:four5}(1) and (2), 
 we must have $j=k$ and $V=U_i\cup\{j\}$. Now, 
 Figure~\ref{fig:four5}(2) displays ${\cal M}|_{U_i}$ only if none of 
 $\{a,b\}$, $\{b,c\}$, $\{b,h\}$, $\{b,i\}$, $\{e,f\}$, $\{f,g\}$, $\{f,j\}$ 
 is a marked edge in $G$. On the other hand, if none of these edges is marked
 in $G$, then Figure~\ref{fig:four5}(1) is transformable to (2) and hence 
 the latter displays ${\cal M}|_{U_i}$. 

\fig[figure706]{four6}
{Possible layouts of $G[U_i]$ when $\{h,i\}\not\in E$.}

 \case{2.2} $\{h,i\}\not\in E$. Then, by Corollary~\ref{cor:onlytwo},
 Figure~\ref{fig:four6}(1) (respectively, (2)) displays ${\cal M}|_{U_i}$
 if and only if
 Figure~\ref{fig:four4}(1) (respectively, (2)) displays ${\cal M}_U$. 

   \case{2.2.1} There is no $j\in V-U_i$ with $\{e,f,h\}\subseteq N_G(j)$ 
     or there is no $k\in V-U$ with $\{f,g,i\}\subseteq N_G(k)$. 
   	Then, Figure~\ref{fig:four6}(1) does not display ${\cal M}|_{U_i}$. 
 \zzsaid{pass37: added next sentence.}
	Whether $j$ and $k$ exist can be decided in $O(1)$ time, because 
 	$|N_G(e)| = |N_G(g)| = 6$ by Figure~\ref{fig:four6}. 

   \case{2.2.2} There are $j\in V-U$ and $k\in V-U$ such that $\{e,f,h\} 
     \subseteq N_G(j)$ and $\{f,g,i\}\subseteq N_G(k)$. Then, if $b\not\in 
     N_G(j)$ or $b\not\in N_G(k)$, Figure~\ref{fig:four6}(2) does not display 
     ${\cal M}|_{U_i}$. So, we may assume that $b\in N_G(j)$ and $b\in N_G(k)$.
     Then, $j\not=k$ by Corollary~\ref{cor:onlytwo}, Figure~\ref{fig:four6}(1),
	and (2). Let $U_k=U_i \cup \{j,k\}$. 

   \case{2.2.2.1} $\{j,k\}\in E$. Then, similarly to Case~2.1 above, 
     we can distinguish which of Figure~\ref{fig:four6}(1) and (2) displays 
     ${\cal M}_{U_i}$. 

\fig[figure707]{four7}
{Possible layouts of $G[U_k]$ when $\{j,k\}\not\in E$.}

\zzsaid{pass37: fig:four8 was originally here.}

   \case{2.2.2.2} $\{j,k\}\not\in E$. Then, 
	by Corollary~\ref{cor:onlytwo},
     Figure~\ref{fig:four7}(1) (respectively, (2)) 
     displays ${\cal M}|_{U_k}$
     if and only if
     Figure~\ref{fig:four6}(1) (respectively, (2))
     displays ${\cal M}_{U_i}$. In case at least one of the edges $\{a,b\}$, 
     $\{e,f\}$, $\{c,b\}$, and $\{g,f\}$ is marked in $G$, 
     Figure~\ref{fig:four7}(2) does not display ${\cal M}|_{U_k}$. Moreover, 
     in case at least one of the edges $\{a,f\}$, $\{e,b\}$, $\{c,f\}$, and 
     $\{g,b\}$ is marked in $G$, Figure~\ref{fig:four7}(1) does not display 
     ${\cal M}|_{U_k}$. 
     \micsaid{pass34: rewrote list of edges as a product}
     So, we may assume that no pair in $\{a,c,e,g\} \times \{b,f\}$
     spans a marked edge of $G$. 

 \zzsaid{pass37: added the next two paragraphs to this case.} 
 Now, observe a resemblance between Figures~\ref{fig:four4} and 
 \ref{fig:four7}. We want to iterate the above case-analysis to 
 distinguish Figure~\ref{fig:four7}(1) and (2). To this end, first observe 
 that the above case-analysis is independent of nation $d$ and edge $\{a,c\}$.
 Moreover, the case-analysis can be viewed as a procedure 
 $CA(a, b, c, e, f, g)$ where the input parameters are nations of $G$ related
 as in Figure~\ref{fig:four4}(1) or (2) except for the possible absence of 
 edge $\{a,c\}$. Thus, to distinguish Figure~\ref{fig:four7}(1) and (2), 
 it suffices to call $CA(h, b, i, j, f, k)$. 

 There can be a linear number of subsequent calls of procedure $CA$. 
 Each call takes $O(1)$ time, so the overall time is linear.
 \micsaid{pass40: added end of previous sentence.}

\micsaid{pass36: That would be quadratic time to classify a single
  MC$_4$, which might then become a pizza (meaning no progress!),
  so we may take cubic time to make progress!
    This is still ok, since we can argue that we only go through this
  phase once (we don't manufacture any more MC$_4$'s in these reductions).
  But my point is that this would require some rewrite of the time analysis.
  Actually I think we can probably argue that the total time for deciding
  $C$ (with recursion) is still linear, since all the faces that come up
  (before we reach bottom) have bounded degree in $G$.
  But I'm not sure we need any of this, see my next pass36 remark, below.}

\zzsaid{pass37: [[ the rest was the original. ]]

     Now, let $G'$ be the marked graph obtained 
     from $G-\{a,e,c,g\}$ by adding the edges $\{h,d\}$, $\{j,d\}$, $\{h,i\}$, 
     $\{i,d\}$, and $\{k,d\}$. If Figure~\ref{fig:four7}(1) (respectively, (2))
     displays ${\cal M}|_{U_k}$, then we can obtain an atlas ${\cal M}'$ of 
     $G'$ from ${\cal M}$ by (i) extending nation $h$ to occupy nations $e$ and
     $a$ and extending nation $i$ to occupy nations $g$ and $c$, and then 
     (ii) contracting the $(h,f)$-segment (respectively, $(h,b)$-segment) to 
     	a single point and contracting the $(i,f)$-segment (respectively, 
	$(i,b)$-segment) to a single point. 
     Figure~\ref{fig:four8}(1) (respectively, (2)) displays ${\cal 
     M}'|_{\{b,d,f,h,i,j,k\}}$
     if and only if
     Figure~\ref{fig:four7}(1) (respectively, (2))
     displays ${\cal M}|_{U_k}$. 

\zzsaid{pass37: moved fig:four8 to here.} 
\fig{four8}{Possible layouts of $G'[\{b,d,f,h,i,j,k\}]$.}

\micsaid{pass36: Instead of your pass35 suggestion above, why don't we
  just stop here, having ``made progress''?  That is, we have produced
  a smaller graph $G'$, such that we can recover an atlas for $G$ from
  an atlas for $G'$ (in particular, the $C'$ in $G'$ won't be a rice-ball).
  It does not really matter that we have not classified the original
  $C$ (this will require a bit of rewrite of the goal for this
  section, though).  Also, although it doesn't really matter, I think
  we had a net loss of two MC$_4$'s ($\{e,f,h,j\}$ and
  $\{g,f,i,k\}$).}
\zzsaid{pass37: It may not be possible to recover an atlas for $G$ 
  from a given atlas for $G'$!!!} 

\micsaid{pass34: added paragraph break here.  The rest of this
  subsection is quite tricky.  In particular we should probably have
  something more on its time analysis, since we want this entire
  decision procedure to be linear time.}

     Moreover, by Figure~\ref{fig:four7}(1) and (2) 
     and Figure~\ref{fig:four8}(1) and (2), we can see that if one of the 
     following conditions (i) through (iv) does not hold, then its modification
     obtained by replacing $G'$ with $G$ and ${\cal M}'$ with ${\cal M}$ does 
     not hold: 

      (i) There is no shrinkable segment in ${\cal M}'$ whose ending nations 
	are adjacent in $G'$. 

      (ii) There is no induced 4-cycle $C'$ in $G'$ such that at most one pair 
	of adjacent nations of $C'$ weakly touch in ${\cal M}'$. 

      (iii) There is no 3-clique $C'$ of $G'$ such that the nations of $C'$ do 
	not meet at a point in ${\cal M}'$ and at least one pair of nations of 
	$C$ strongly touch in ${\cal M}'$. 

      (iv) For every MC$_4$ $C'$ of $G'$, the nations of $C'$ are related 
	in atlas ${\cal M}'$ in the same way as 
	either Figure~\ref{fig:four1}(1) or (2) shows. 

By Corollary~\ref{cor:onlytwo}, the modification of the above condition (iv) 
obtained by replacing $G'$ with $G$ and ${\cal M}'$ with ${\cal M}$ holds. 
Thus, (iv) holds. \zzsaid{pass6: Changed the next sentence.}
By Assumption~\ref{ass:edgeQuadTri},
the modifications of the above conditions (i), (ii) and (iii) obtained by 
replacing $G'$ with $G$ and ${\cal M}'$ with ${\cal M}$ hold. 
This follows from Corollaries~\ref{cor:noSedge}, \ref{cor:noSquadruple} 
and \ref{cor:noStriangle}. In turn, by the same corollaries, 
\zzsaid{Change made: \{fac:noSedge\} $\rightarrow$ \{cor:noSedge\}, 
\{fac:noSquadruple\} $\rightarrow$ \{cor:noSquadruple\}, and 
\{fac:noStriangle\} $\rightarrow$ \{cor:noStriangle\}.}
$G'$ has no separating edge, quadruple or triangle. Also observe that 
in the above case-analysis, we only need the existence of at least nine 
nations in $G$, the non-existence of a separating triangle in $G$, and 
the existence of a well-formed atlas ${\cal M}$ of $G$ such that for every 
MC$_4$ $C'$ of $G$, the nations of $C'$ are related in atlas ${\cal M}$ 
in the same way as either Figure~\ref{fig:four1}(1) or (2) shows. 
Therefore, when $G'$ has at least nine nations, 
the non-existence of a separating triangle in $G'$ and the existence of 
the atlas ${\cal M}'$ ensure that we can recursively do the same case-analysis 
to distinguish Figure~\ref{fig:four8}(1) and (2), as we did to distinguish 
Figure~\ref{fig:four4}(1) and (2). On the other hand, when $G'$ has at most 
eight nations, we can distinguish Figure~\ref{fig:four8}(1) and (2) 
by brute force. 
\micsays{pass12: something should be said about the time analysis of a
deep recursion of the previous kind, since it isn't explicitly
addressed elsewhere.}

}

\micsaid{pass34: merged last paragraph into first of subsection.}

\subsection{Removing Pizzas}\label{subsec:pizzas}

By the discussions in the last two subsections, we may assume that 
for every MC$_4$ $C=\{a,b,c,d\}$ of $G$, only Figure~\ref{fig:four1}(1) 
displays ${\cal M}|_C$. That is, the four nations of every MC$_4$ of $G$ 
meet at a point in ${\cal M}$. 

Fix an MC$_4$ $C=\{a,b,c,d\}$ of $G$. $C$ is 3-sharing with no MC$_4$ 
$C'$ of $G$ because otherwise, $C'$ would have a non-pizza layout. 
By Figure~\ref{fig:four1}(1), there are distinct nations $e$, $f$, $g$ and $h$ 
in $V-C$ such that $C\cap N_G(e)=\{a^1, b^1\}$, $C\cap N_G(f)= \{b^1, c^1\}$, 
$C\cap N_G(g) = \{c^1, d^1\}$ and $C\cap N_G(h)=\{d^1, a^1\}$, because 
${\cal M}$ has no hole. On the other hand, the existence of the nations $e$, 
$f$, $g$ and $h$ ensures that the nations of $C$ have to meet at a point in 
${\cal M}$ cyclically in the order $a^1$, $b^1$, $c^1$, $d^1$. Thus, 
by finding out nations $e$, $f$, $g$ and $h$, we can find and remove 
a correct 4-pizza from $G$. 

\zzsaid{pass41: added this paragraph.} 
\micsaid{pass42: removed ``dealt''.}
\micsaid{pass44: rewrote paragraph to refer to the multi-pizza argument.}
By this method we may identify a correct 4-pizza for every MC$_4$ in
$G$.  Since these 4-pizzas all exist in every well-formed atlas of
$G$, we may remove them all in one step by the remarks after
Lemma~\ref{lem:remove4}.

\section{Time Analysis}\label{sec:time} 

\micsaid{pass40: reworded}
\zzsaid{pass43: reworded next sentence.} 
Let $n$ and $m$ be the number of vertices and edges in the input graph $G$,
respectively.  Suppose this is not a base case; that is, $n\geq 9$ and $G$ 
has a 4-clique. Then we will show that 
the algorithm can always make progress in $O(n^2)$ time.
\micsaid{pass44: added next remark.}
In each case, the time needed to produce the subproblems from $G$
dominates the time needed to recover a solution from the subproblem
solutions, so we ignore the latter.

By Corollary~\ref{cor:edgebound} (with $k=4$) $G$ has $m=O(n)$ edges
and arboricity $\alpha(G)=O(1)$, so we can list its $O(n)$ maximal
cliques in linear time~\cite{CN85}.
\micsaid{pass44: JYP88 is $O(n^4)$, so rewrote the previous using CN85,
  and also moved the arboricity bound to sec:witness.}
{From} the listed MC$_4$'s, we can precompute the sets ${\cal E}[a,b]$
for all unmarked \zzsaid{pass17: Added ``unmarked" here}
edges $\{a,b\}$, again in linear time.

We claim that testing the existence of a separating triangle takes
$O(n^2)$ time.  Since $G$ has $O(n)$ maximal cliques and no 7-clique,
it has $O(n)$ 3-cliques and these can be found in linear time.
\micsaid{pass44: could use arboricity above, since a graph $G$ has
  at most $\alpha(G)(n-1)$ triangles.  In fact we could also use
  this to bound triangles in a planar graph, with $\alpha=3$.}
\zzsaid{pass45: just let us use the simplest idea.} 
For each 3-clique $C$, it takes $O(n)$ time to test 
whether some (ordered) list of the vertices in $C$ is a separating triangle. 
So, the claim holds. 
A similar analysis applies for finding a 3-cut (by
Lemma~\ref{lem:3-cut}(1)), a separating edge, or a separating triple.

In order to detect separating quadruples, we use an algorithm of Chiba
and Nishizeki~\cite{CN85} which implicitly lists all 4-cycles of $G$
in $O(m \cdot \alpha(G))=O(n)$ time.
\micsays{pass47: removed extra ``time'' in previous phrase.}
\micsaid{pass44: added ``$=O(n)$'' above, deleted arboricity definition.}
The algorithm produces a list of triples
$(u_i,v_i,S_i)$ with the following properties:
\begin{enumerate}
\item $u_i$ and $v_i$ are non-adjacent vertices of $G$.
\item $S_i$ is a set of vertices adjacent to both $u_i$ and $v_i$.
\item Every induced 4-cycle in $G$ occurs as
	$\langle u_i, x, v_i, y \rangle$
	for some choice of $i$ and $x,y \in S_i$.
\end{enumerate}
\micsaid{pass44: minor rewrite of next sentence.}
In particular, the sum of all $|S_i|$ is $O(n)$.

We claim that testing the existence of a separating quadruple takes
$O(n^2)$ time.  It suffices to show the following: for each triple
$(u_i,v_i,S_i)$, we can test whether there is a separating quadruple
$\langle u_i, x, v_i, y\rangle$ or $\langle v_i, x, u_i, y\rangle$
(with $x,y \in S_i$) in time $O(|S_i|n)$.  By similarity, it suffices
to show how to find those quadruples starting with~$u_i$.

For $x$ in $S_i$, let $G^x = G - \{u_i,v_i,x\} - {\cal E}[u_i,x]$.
In linear time we may compute $G^x$ and identify the set $S^x$ of all
cut vertices in $G^x$.  Now there is a separating quadruple of the
form $\langle u_i, x, v_i, y \rangle$ precisely if $S^x$ contains some
$y$ which is in $S$ but not adjacent to~$x$.  By repeating this for
every $x \in S_i$, we have the required time bound.

\micsaid{pass18: removed earlier mention of separating 4-cycle, added
next.}  A similar analysis applies for finding separating 4-cycles in
$O(n^2)$ time.

\zzsaid{pass37: rewrote this paragraph.} 
\micsaid{pass40: split next paragraph in two.}
The case analysis for eliminating an MC$_5$ in Section~\ref{sec:five}
may be executed in linear time.  In particular, we may identify an
MC$_5$ 4-sharing with two other MC$_5$'s in $O(n)$ time as
follows. First, for every MC$_5$ $C_i$ and for every $S\subseteq C_i$
with $|S|=4$, create a pair $(S,i)$. Next, bucket-sort all the pairs,
and use the result to count the number of 4-sharing MC$_5$'s with
each~$C_i$.

When the graph has no MC$_5$ but still has some MC$_4$'s, we make
progress in at most $O(n^2)$ time as follows.  First, we list the
$O(n)$ MC$_4$'s in some arbitrary order.  For each one, we test the
conditions of Lemma~\ref{mc4:riceball} in $O(n)$ time; if we find such
an MC$_4$, then we remove the identified 4-pizzas and we are done.
Otherwise, we go through the list again, this time applying the linear
time decision procedure of Section~\ref{subsec:dist}; if we determine
that some MC$_4$ is a non-pizza, then we remove the identified
4-pizzas and we are done again.  Otherwise, we have established that
all the MC$_4$'s are pizzas, and so we can remove a 4-pizza for each
MC$_4$ by the method in Section~\ref{subsec:pizzas}.

\micsaid{pass40: Since we have other reductions (separating 4-cycles
and quadruples) that take quadratic time, it does not matter that we
also use quadratic time to make progress here.  However, I think the
MC$_4$'s are not a quadratic bottleneck, and we can make progress in
linear time as follows.  First, look for a pair of 3-sharing MC$_4$'s.
If there is such a pair, then one of them is not a pizza; so we are
sure to make progress by applying both the rice-ball test and the
non-pizza test to each MC$_4$ in the pair.  Otherwise, if $C$ is an
MC$_4$ that is not 3-sharing with another, then we know it is not a
pizza-with-crust.  So, we first apply the rice-ball test; if that
fails, then it must be a pizza, and we can immediately apply the
method of subsec:pizzas.  That is, there is no need to establish that
all the other MC$_4$'s are pizzas before removing $C$.}
\zzsaid{pass41: This needs much more details and verification. 
So, I suggest we forget it this time.} 
\micsaid{pass42: ok.}

\zzsaid{pass39: Here, I meant the following: Each time our algorithm
begins to remove MC$_4$'s as discussed in Section~\ref{sec:four}, 
it first enumerates all MC$_4$'s, and then scan them in any order.
When it scans an MC$_4$ $C$, it spends linear time on deciding whether
$C$ has a non-pizza layout. If $C$ has, then it can make progress.
If $C$ does not have, then it proceeds to the next MC$_4$. In this way,
the algorithm may either quit or finish scanning all MC$_4$'s.
In the former case, it uses the found non-pizza layout to make progress
by removing a correct 4-pizza; this case takes $O(n^2)$ time.
In the latter case, it proceeds to the pizza-removal stage as 
discussed in Section~\ref{subsec:pizzas}; the pizza-removal stage 
is performed only once, and takes $O(n^2)$ time.}
\micsaid{pass38: I am confused by the above statement, I thought the
decision procedure was linear time?  I suppose the one quadratic thing
is the following: the final phase of removing all the accumulated
pizzas, which I suppose is quadratic overall, and does not follow the
``making progress'' analysis.  In fact, the decision procedure
(strictly speaking) does not make progress unless we count the number
of known pizzas as part of the ``size'' of our graph.  It all works of
course, as part of the base case.  Or we could rearrange to always
make progress as follows: if there are two 3-sharing 4-cliques, then
we make progress on at least one of the pair with (two invocations of?
) the decision procedure.  Otherwise, for a 4-clique that is not
3-sharing, it is a pizza and we handle it immediately.}
\zzsaid{pass39: What did you mean by ``decision procedure"? Did you mean
the process of deciding whether a given MC$_4$ has a non-pizza layout?
If you meant this, then the procedure indeed takes $O(n)$ time. }
\micsaid{pass40: Yes.}
\micsaid{pass32: removed remark about 4-sharing MC$_5$'s, in fact your
suggestion is linear time, with appropriate sorting.  Is there
anything else in those sections that might take more than linear time?}
\zzsaid{pass33: Yes, the last case in Section~\ref{subsec:dist}.}
 
Finally, if the algorithm reaches a base case, our graph $G$ either
has at most $8$ vertices, or no 4-clique.  In the former case we solve
the problem exhaustively in $O(1)$ time.
  Otherwise, $G$ should be planar; we finish in linear
time~\cite{HT74}, as described in Section~\ref{subsec:sketch}.
\micsaid{pass40: added this paragraph for completeness.}
\zzsaid{pass41: last sentence is incorrect in grammar.}
\micsaid{pass42: moved details to subsec:sketch.}

\zzsaid{pass43: rewrote this paragraph.}
\micsays{pass44: added some details to this paragraph.}
Let $N=n+m$ be the {\em size} of our input graph, and let $T(N)$ be
the maximum running time of the algorithm on any input of size~$N$.
We claim that there is a constant $c$ such that $T(N) \leq cN^3$. 
The claim is clearly true for the base cases, as argued above.
In all other cases, the algorithm makes progress in $c_1N^2$ time 
for some constant $c_1$. That is, the algorithm produces one or more 
smaller marked graphs whose total size is larger than that of $G$ by 
a constant $c_2$; the problem for $G$ is reduced to solving the problem 
for each of these smaller instances. 
More precisely, there are integers $n_1,\ldots,n_\ell\in\{1,\ldots,N-1\}$ such
that $\sum_{i=1}^\ell n_i\leq N+c_2$ and
$T(N) \leq \sum_{i=1}^\ell T(n_i) + c_1 N^2$. 
We prove our claim by induction.
For small $N$ ($N<c_2^2$), our claim is true simply by choosing $c$
large enough.
For larger $N$, we have $T(N) \leq \sum_{i=1}^\ell c n_i^3 +c_1 N^2$
by the inductive hypothesis.
Note that $\sum_{i=1}^\ell c n_i^3$ is maximized when $\ell=2$, 
$n_1=N-1$ and $n_2=c_2+1$.
\micsays{pass47: changed next ``at least'' to ``suffices''.}
Hence, by choosing $c$ large enough ($c_1+2$ suffices), we have $T(N)
\leq cN^3$.

\section{Concluding Remarks}\label{sec:conclude}

Our main algorithm is too complex. We would like to find a faster algorithm,
with simpler arguments. Perhaps such a simplification is possible using
some of Thorup's ideas. 

Naturally, we are very interested in polynomial-time algorithms for
recognizing (hole-free or not) $k$-map graphs with $k\geq 5$. In view
of the complication of our algorithm for hole-free 4-map graphs,
however, new insights seem to be needed in order to make progress in
this direction.

A natural and interesting question in connection with map graphs is to ask
whether $\lfloor 3k/2\rfloor$ colors suffice to color a $k$-map graph where
$k\geq 3$. Note that $\lfloor 3k/2\rfloor$ is the maximum clique size
in a $k$-map graph. In case $k=3$, the answer is positive because of the
famous Four Color Theorem. 
As Thorup observed~\cite{T98}, the answer is also positive for $k=4$:
the 4-map graphs are all 1-planar graphs, and 1-planar graphs are
known to be 6-colorable~\cite{B85}.
However, the answer is unknown when $k\geq 5$.

\micsaid{pass16: added this, do we have any tight bounds?}
\zzsaid{pass17: My edge number bound is tight for $k=4, 6$.}
\micsaid{pass18: Changed ``Chen'' to ``the first author''.}
Similarly, we are interested in tighter versions of the bounds in
Section~\ref{sec:fundamentals}; in particular the first
author~\cite{C99} has improved the edge bound in
Corollary~\ref{cor:edgebound} to~$kn-2k$.
\micsays{pass46: ``edge bound'' was ``upper bound''.}

The recognition problem of map graphs is just a special topological
inference problem, where each pair of regions either touch or are
disjoint.  One more general problem is obtained by allowing the
relation between certain pairs of regions to be left unspecified
(i.e., each such pair may touch or not touch). We conjecture that this
generalization is NP-complete.  Another generalization is obtained by
allowing a region to include another region as a subregion. We
conjecture that this generalization is polynomial-time solvable.
\micsaid{pass12: what are the partial results?}
\zzsaid{pass17: Rewrote the rest of this paragraph.}
  Note that the inclusion relations among the regions should induce a
rooted forest. The special case of this generalization where no four
leaf regions meet at a point and each non-leaf region is the union of
its descendant regions, can be solved by a nontrivial $O(n\log
n)$-time algorithm \cite{CH99}. In the real world, a non-leaf region
is usually not a closed disc homeomorph; this more general problem is
addressed in~\cite{CHK99}.


\micsaid {List of figures to review the captions. \listoffigures}

\end{document}